\documentclass[10pt]{article}
\usepackage{array}

\usepackage{graphicx}
\usepackage{amssymb}
\usepackage{amsmath}
\usepackage{amsthm}
\usepackage{enumitem}
\usepackage{undertilde}
\usepackage{subfigure}
\usepackage{color}      % use if color is used in text
\usepackage{multirow}
\usepackage{booktabs}
\usepackage{natbib}
\usepackage{indentfirst}
\usepackage[left=2.5cm,right=2.5cm,top=2.8cm,bottom=2.8cm,letterpaper]{geometry}
\usepackage{multirow}
\usepackage{booktabs}
\usepackage{array}
\usepackage{tabularx}
\usepackage{adjustbox}
\usepackage{bm}
\usepackage{titlesec}
\usepackage{url}

\newlength{\bibitemsep}
\newlength{\bibparskip}
\let\oldthebibliography\thebibliography
\renewcommand\thebibliography[1]{%
  \oldthebibliography{#1}%
  \setlength{\parskip}{\bibitemsep}%
  \setlength{\itemsep}{\bibparskip}%
}

\titleformat{\section}
    {\centering\normalfont\Large\bfseries}{\thesection.}{1em}{}

\newlength\Origarrayrulewidth

\def\thesection{\arabic{section}}
\begin{document}

%%%%%%%%%%%%%%%%%%%%%%%%%%%%%%%%%%%%%%%%%%%%%%%%%%%%%%%%%%%%%%%%%%%%%%%%%%%%%%%%%%%%%%%%%%%%%%%%%%%%%%%%%%%%%%%%%%%%%%%%%%%%
%%%%%%%%%%%%%%%%%%%%%%%%%%%%%%%%%%%%%%%%%%%%%%%%%%%%%%%%%%%%%%%%%%%%%%%%%%%%%%%%%%%%%%%%%%%%%%%%%%%%%%%%%%%%%%%%%%%%%%%%%%%%

\renewcommand{\baselinestretch}{1.2}

\markboth{\hfill{\footnotesize\ Taewoon Kong and Brani Vidakovic} \hfill}
{\hfill {\footnotesize\rm Non-decimated Quaternion Wavelet Spectral Tools with Applications} \hfill}

\renewcommand{\thefootnote}{}
$\ $\par

%%%%%%%%%%%%%%%%%%%%%%%%%%%%%%%%%%%%%%%%%%%%%%%%%%%%%%%%%%%%%%%%%%%%%%%%%%%%%%%%%%%%%%%%%%%%%%%%%%%%%%%%%%%%%%%%%%%%%%%%%%%%

\fontsize{10.95}{14pt plus.8pt minus .6pt}\selectfont
\vspace{0.8pc}
\centerline{\large\bf Non-decimated Quaternion Wavelet Spectral Tools with Applications}
\vspace{2pt}
\centerline{\large\bf }
\vspace{.4cm}
\centerline{Taewoon Kong$^*$,  \url{twkong@gatech.edu}}
\centerline{Brani Vidakovic, \url{brani@gatech.edu}}
\vspace{.2cm}
\centerline{\it H. Milton Stewart School of Industrial \& Systems Engineering}
\centerline{\it Georgia Institute of Technology, Atlanta, USA}
\vspace{.55cm}
\fontsize{9}{11.5pt plus.8pt minus .6pt}\selectfont

\begin{abstract}

Quaternion wavelets are redundant wavelet transforms generalizing complex-valued non-decimated wavelet transforms.
In this paper we propose a matrix-formulation for non-decimated quaternion wavelet transforms and define spectral tools for use in machine learning tasks.
Since quaternionic algebra is an extension of complex algebra, quaternion wavelets bring redundancy in the components that proves beneficial in wavelet based tasks.
Specifically, the wavelet coefficients in the decomposition are quaternion-valued numbers that define the modulus and three phases.

The novelty of this paper is definition of non-decimated quaternion wavelet spectra based on the modulus and phase-dependent statistics as low-dimensional summaries for 1-D signals or 2-D images.
A structural redundancy in non-decimated wavelets and a componential redundancy in quaternion wavelets are linked to extract more informative features.
In particular, we suggest an improved way of classifying signals and images based on their scaling indices in terms of spectral slopes and information contained in the three quaternionic phases.
We show that performance of the proposed method significantly improves when compared to the standard versions of wavelets including the complex-valued wavelets.

To illustrate performance of the proposed spectral tools we provide two examples of application on real-data problems: classification of sounds using scaling in high-frequency recordings over time and monitoring of steel rolling process using the fractality of captured digitized images. The proposed tools are compared with the counterparts based on standard wavelet transforms.

\vspace{9pt}
\par
\end{abstract}\par

\fontsize{10.95}{14pt plus.8pt minus .6pt}\selectfont

%\newpage
%
%\tableofcontents
%
%\newpage

%------------------------------------------
\section{Introduction}{\label{sec-Intro}}
In recent decades the traditional real-valued discrete wavelet transform (DWT) has  been utilized as powerful mathematical tool in signal and image processing, in tasks of denoising, segmentation, compression, classification, and so on \citep{Rajini2016}.
The traditional real-valued orthogonal discrete wavelet transforms (DWT) feature elegant, parsimonious, and informative
representations, but have two shortcomings.
The first is that DWT is not shift-invariant.
Even a small shift of a signal results in complete change of wavelet coefficients, which causes problems in efficient   computation and feature extraction in real-time.
The second is that no phase information is encoded, unlike the Fourier representations \citep{Chan2008}.
To accommodate the phase information, the complex wavelet transform ($\text{WT}_\text{\large{c}}$) was proposed in \citet{Lina1997}.

%To accommodate the phase information and shift-invariance %%%TAEWOON LINA's COMPLEX WAVELETS
% ARE ORTHOGONAL, THEY ARE NOT SHIFT INVARIANT

We denote this transform as ($\text{WT}_\text{\large{c}}$) where c refers to ``complex'' instead of CWT that usually stands for the continuous wavelet transform.
The $\text{WT}_\text{\large{c}}$  is orthogonal, symmetric, and have decomposing atoms of compact size \citep{Lina1997, Gao2010}.
Most notable characteristic is the phase information that $\text{WT}_\text{\large{c}}$ additionally provides compared to real-valued wavelet decompositions.
This phase information enables the $\text{WT}_\text{\large{c}}$ to pack more information about the signal or image that it represents.
Because of these merits, $\text{WT}_\text{\large{c}}$ has been exploited in various wavelet-based tasks \citep{MacGibbon1997, Remenyi2014, Jeon2014, Kong2019}.

As an extension of the $\text{WT}_\text{\large{c}}$, the quaternion wavelet transform (QWT) provides a richer scale-space analysis by taking into account the axioms of the quaternion algebra \citep{Billow1999, Gai2015}. This transform leads to quaternion-valued wavelet coefficients in the form of a vector of one modulus and three phases that possess symmetry properties and near shift-invariance, according to Billow's results \citep{Billow1997}. The modulus reflects the outline of signal or image while the three phases encode local image shifts and represent subtle information such as cusps, boundaries, and texture structure. Preserving the benefits of $\text{WT}_\text{\large{c}}$, the QWT can provide a more extensive redundancy with its three phases. Based on these merits, the QWT has been utilized in image denoising \citep{Gai2015}, texture classification \citep{Soulard2011}, image segmentation \citep{Subakan2011}, face recognition \citep{Jones2006}, image fusion \citep{Zheng2016}, etc. Note that these tasks usually have been performed with constructing quaternion wavelets utilizing four real-valued DWT: the first corresponding to the real part of the quaternion and the other three linked with the first via Hilbert transform correspond to the three imaginary parts of the quaternion wavelets. This transform possesses approximate shift invariance, abundant phase information, and limited redundancy while retaining the traditional wavelet time-frequency localization ability \citep{Rajini2016}. However, this original so-called QWTs were really DWT or $\text{WT}_\text{\large{c}}$ in disguise \citep{Fletcher2017}. In fact, their filter coefficients are real-valued, which means that it was technically wrong to name them QWT. In recent years, several studies have been conducted to construct a bonafide QWT that is not a conglomerate of DWTs and $\text{WT}_\text{\large{c}}$  \citep{Carre2006, Hogan2012, Ginzberg2012}.
 \citet{Ginzberg2012} suggested  true quaternion matrix-valued wavelets with quaternion-valued filter coefficients. Of the provided filters, for the analysis in this paper, we selected  non-trivial quaternion scaling and wavelet filters of length $L=10$ and with five vanishing moments $(A=5)$ as a compromise between locality and smoothness. The selected filters correspond to non-trivial symmetric quaternion wavelet functions with compact support via a matrix-based implementation. Selected quaternion basis can address some critical problems from which other established but not fully quaternion wavelet designs had suffered. In addition, full quaternion approach leads to meaningful uniqueness and selective existence for filters of only certain lengths and numbers of vanishing moments. More details can be found in \citet{Ginzberg2012}.

Another popular version of wavelet transform is a non-decimated wavelet transform (NDWT) which is shift-invariant.
It represents a discrete approximation to continuous wavelet transforms by equispaced sampling in time and log-scale.
Operationally the redundancy of non-decimated wavelets is due to their repeated filtering in Mallat's algorithm with a minimal shift (or with a maximal sampling rate) which remains the same at all dyadic scales.
As a result, each decomposition level retains the same number of wavelet coefficients as the size of original data.
Even though the total size of decomposition obtained by NDWT is larger than that of the orthogonal transform, this redundancy is often preferred by practitioners in many fields. More details can be found in excellent monographs \citet{percival2006} and \citet{Mallat2009}.

Note that the redundancy in $\text{WT}_\text{\large{c}}$ and the NDWT is of different nature.
\citet{Kong2019} focused on  non-decimated complex wavelet transforms ($\text{NDWT}_\text{\large{c}}$) that combine $\text{WT}_\text{\large{c}}$ and NDWT.
Specifically, $\text{NDWT}_\text{\large{c}}$ produces redundant wavelet coefficients both as complex numbers and by NDWT as oversampled. The authors suggested a way of building phase-based statistics as variables for classification problems and showed significant increase in precision of classification, compared to other existing wavelet-based methods. Furthermore, the $\text{NDWT}_\text{\large{c}}$ is more flexible than the decimated wavelet transforms because of the matrix-based implementation proposed in \citet{Kang2016}.
The decimated wavelet transform methods including $\text{WT}_\text{\large{c}}$ and even convolution-based NDWT can be routinely applied to signals and squared images dyadic sizes \citep{Lina1999, percival2006}.
However, a real-world data typically do not have such sizes and need pre-processing prior to application. The matrix-based NDWT enables us to directly analyze 1-D signals of an 2-D images of arbitrary size. More details can be found in \citet{Kong2019}.

Given the useful characteristics of the QWT and matrix-based NDWT, we propose a non-decimated quaternion wavelet transform (NDQWT) and its wavelet spectra defined by quaternionic modulus and the three phases. Since the QWT is an extension of the $\text{WT}_\text{\large{c}}$, we expect that classification accuracy would improve with QWT-defined
spectral tools.
The modulus of the QWT behaves as the wavelet spectra in a conventional wavelet transform,
 so that our main focus is on the contribution by the three phases.
Several researches including \citet{Billow1999} and \citet{Soulard2010} also focused on exploiting phases of quaternion wavelets. However, as far as we know, there is no proposal of quaternionic phase-based levelwise statistics for use in classification problems, and wider, for machine learning.
Also, as we pointed out, existing use of phase depended on the artificially constructed QWT.
Taken together with non-decimated nature of the underlying transform, the proposed spectral
tools are novel in terms of providing new modalities for classification problems.
And finally, the goal of this study is to demonstrate superiority of the proposed method over several
 competing wavelet-based methods through applications in real-data: in classifying sound signals and
 in tasks of anomaly detection in rolling bar images.

The paper is organized as follows.
Section \ref{sec-QuatAlgebra} briefly reviews quaternion algebra focusing on calculation of a modulus and the three phases.
Next, Section \ref{sec-NDQWT} explains the NDQWT for 1-D and 2-D cases, respectively. For the 2-D case, we present the scale-mixing version of 2-D NDQWT.
Section \ref{sec-NDQwavespec} describes the non-decimated quaternion wavelet-based spectra focusing on the modulus information of the coefficients, while Section \ref{sec-phase} suggests construction of a three phase-based statistics as new covariates in discriminatory analysis.
In Section \ref{sec-app}  the proposed tools are applied on 1-D and 2-D real-data, and finally, concluding remarks and directions for future study are given in  Section \ref{sec-Conc}.

\section{Quaternion Algebra}{\label{sec-QuatAlgebra}}
Sir William Hamilton in 1843 first developed a quaternion algebra; the notation $\mathbb{H}$ for the field of all quaternions, is proposed after him \citep{Giirlebeck1997}.
In a four-dimensional (4-D) algebra, the elements of $\mathbb{H}$ are given as linear combinations of a real scalar and three orthogonal imaginary units $i, j,$ and $k$ with real coefficients as
\begin{equation}\label{quatexpression}
\mathbb{H} = \{q = q_0 + q_1 i + q_2 j + q_3 k \; | \;  q_0, q_1, q_2, q_3 \in \mathbb{R} \},
\end{equation}
where the three imaginary units ($i,j,k$) satisfy the following non-commutative Hamilton's multiplication rules as
\begin{equation*}\label{quatrule}
ij=-ji=k, \; jk=-kj=i, \; ki=-ik=j, \; i^2=j^2=k^2=ijk=-1.
\end{equation*}

The conjugate of a quaternion $q$ can be written as
\begin{equation}\label{quatconj}
\overline{q} = q_0 - q_1 i - q_2 j - q_3 k,
\end{equation}
and some useful properties for the conjugate as follows:
\begin{equation*}
\overline{\overline{q}} = q, \; \overline{q+p} = \overline{q} + \overline{p}, \;
\overline{qp} = \overline{q}~\overline{p}, \;\;\;  \forall \; q, p \in \mathbb{H}.
\end{equation*}

Since the product of a quaternion $q$ and its conjugate $\overline{q}$ in Equation (\ref{quatconj}) is
\begin{equation*}
q\overline{q} = q_0^2 + q_1^2 + q_2^2 + q_3^2,
\end{equation*}
the modulus $|q|$ of a quaternion $q$ is correspondingly defined as
\begin{equation*}
|q| = \sqrt{q\overline{q}} = \sqrt{q_0^2 + q_1^2 + q_2^2 + q_3^2}.
\end{equation*}

In a manner similar to complex numbers, the expression of quaternion $q$ in Equation (\ref{quatexpression}) can have an alternative representation in polar form as
\begin{equation}\label{quatexpression2}
q = |q| e^{i\phi} e^{j\theta} e^{k\psi},
\end{equation}
where $(\phi, \theta, \psi) \in [-\pi, \pi] \times [-\frac{\pi}{2}, \frac{\pi}{2}] \times [-\frac{\pi}{4}, \frac{\pi}{4}]$.
For a unit quaternion, $|q|=q\overline{q}=1$, their corresponding three phases can be evaluated as follows:
\begin{enumerate}[label=\emph{\hspace{0\marginparsep} \arabic*}., leftmargin=*]
\item First, compute $\psi$ as
\begin{equation*}\label{quatpsi}
\psi = -\frac{1}{2}\mbox{arcsin} \Big(2(q_1 q_2 - q_0 q_3)\Big).
\end{equation*}

\item If $\psi \in (-\frac{\pi}{4}, \frac{\pi}{4})$, then
\begin{equation*}\label{quatphases1}
\left \{
\begin{array}{l}
\phi = \frac{1}{2}\arctan2 \big( 2(q_0 q_1 + q_2 q_3), q_0^2 - q_1^2 + q_2^2 -q_3^2 \big), \\
\theta = \frac{1}{2}\arctan2 \big( 2(q_0 q_2 + q_1 q_3), q_0^2 + q_1^2 - q_2^2 -q_3^2 \big).
\end{array} \right.
\end{equation*}

\item If $\psi = \pm \frac{\pi}{4}$, then select either
\begin{equation*}\label{quatphases2}
\left \{
\begin{array}{l}
\phi = \frac{1}{2}\arctan2 \big( 2(q_0 q_1 - q_2 q_3), q_0^2 - q_1^2 - q_2^2 + q_3^2 \big), \\
\theta = 0.
\end{array} \right.
\end{equation*}
or
\begin{equation*}\label{quatphases3}
\left \{
\begin{array}{l}
\phi = 0, \\
\theta = \frac{1}{2}\arctan2 \big( 2(q_0 q_2 - q_1 q_3), q_0^2 - q_1^2 - q_2^2 + q_3^2 \big) \big).
\end{array} \right.
\end{equation*}

\item If $e^{i\phi} e^{j\theta} e^{k\psi} = -q$ and $\phi \geq 0$, then $\phi \rightarrow \phi - \pi$.

\item If $e^{i\phi} e^{j\theta} e^{k\psi} = -q$ and $\phi < 0$, then $\phi \rightarrow \phi + \pi$.

\end{enumerate}

\section{Non-decimated Quaternion Wavelet Transform}{\label{sec-NDQWT}}
The quaternion scaling and wavelet functions in \citet{Bayro2005} and \citet{Chan2008} satisfy
\begin{eqnarray}\label{quatfunctions}
\phi(x) &=& \sum_{k \in \mathbb{Z}} h_k \sqrt{2} \phi(2x-k) = w_0(x) + i \cdot w_1(x)  + j \cdot w_2(x)  + k \cdot w_3(x) , \\
\psi(x) &=& \sum_{k \in \mathbb{Z}} g_k \sqrt{2} \phi(2x-k) = v_0(x) + i \cdot v_1(x)  + j \cdot v_2(x)  + k \cdot v_3(x) ,
\end{eqnarray}
where $h_k$ denotes the low pass filter and $g_k$ is the high pass filter.

As we mentioned, \citet{Ginzberg2012} suggested a true quaternion matrix-valued wavelets with quaternion-valued filter coefficients.
 From the provided filters, for applications in this paper, we selected the non-trivial quaternion
 scaling and wavelet filters of length $L=10$ and with five vanishing moments $(A=5)$.

\begin{figure}[!ht]
\centering
\includegraphics [scale=0.37 , clip]{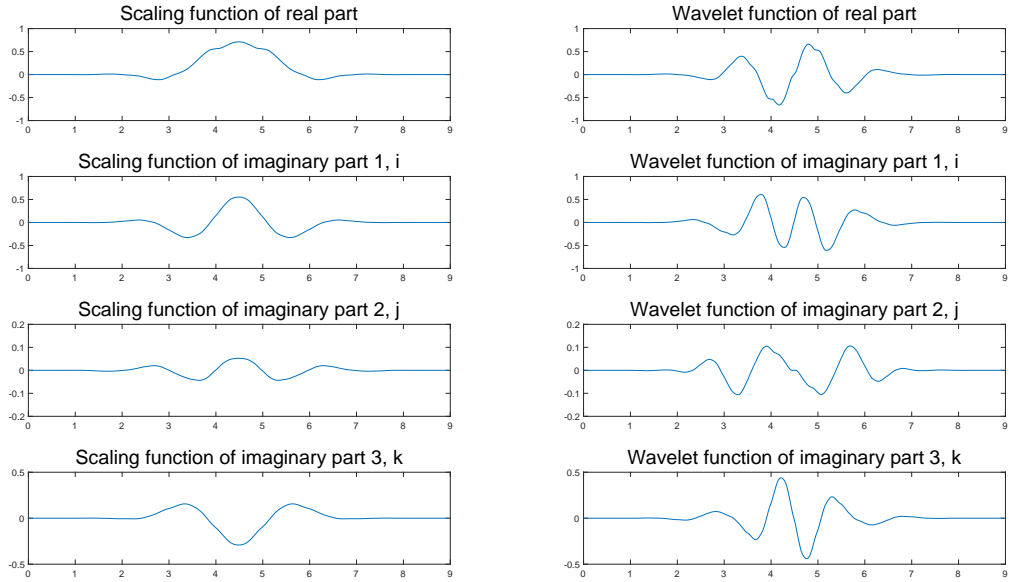}
\caption{Quaternion scaling and wavelet functions for $L = 10$ and $A = 5.$}
\label{fig:quaternionfunctions}
\end{figure}

Specifically, for $L=10$ and $A=5$, the set of design equations is given as
\begin{eqnarray*}\label{quatdesignequations}
\sum_{k=0}^{9} \mathbf{H}_k &=& \sqrt{2} \mathbf{I}_4, \nonumber \\
\sum_{k=0}^{9} (-1)^k k^d \mathbf{H}_k &=& \mathbf{0}_{4} \;\;\;\; \mbox{for} \;\;\;\; d = 0, 1, 2, 3, 4, \\
\sum_{k=0}^{9-2m} \mathbf{H}_k \mathbf{H}_{k+2m}^T &=& \delta_{m, 0} \mathbf{I}_4 \;\;\;\; \mbox{for} \;\;\;\;  m = 1, 2, 3, 4, \nonumber
\end{eqnarray*}
where the $\mathbf{H}_k$ each denote $4 \times 4$ matrix representations of quaternion.
By solving these equations, one obtains the wavelet filters as
\begin{eqnarray*}\label{quatwavefunc}
h_0 &=& h_9 = C_2 i, \nonumber \\
h_1 &=& h_8 = -5 C_1 + C_2 k, \nonumber \\
h_2 &=& h_7 = -7 C_1 -7 C_2 i +3 C_2 k,  \\
h_3 &=& h_6 = 35 C_1 - 5 C_2 i + C_2 k, \nonumber \\
h_4 &=& h_5 = 105 C_1 + 11 C_2 i - 5 C_2 k, \nonumber
\end{eqnarray*}
where $C_1 = \frac{\sqrt{2}}{256}$ and $C_2 = \frac{\sqrt{35}}{256}$.
The corresponding antisymmetric scaling filters are also described as
\begin{eqnarray*}\label{quatscalfunc}
g_0 &=& -g_9 = C_3 (89\sqrt{35}i + 35\sqrt{2}j - 35\sqrt{35}k ), \nonumber \\
g_1 &=& -g_8 = C_3 (-480\sqrt{2} + 35\sqrt{35}i - 175\sqrt{2}j + 79\sqrt{35}k ), \nonumber \\
g_2 &=& -g_7 = C_4 (84\sqrt{2} - 91\sqrt{35}i +35\sqrt{2}j + \sqrt{35}k ),  \\
g_3 &=& -g_6 = C_5 (35\sqrt{2} + 5\sqrt{35}i - \sqrt{35}k ), \nonumber \\
g_4 &=& -g_5 = C_6 (-5040\sqrt{2} + 577\sqrt{35}i -245\sqrt{2}j + 5\sqrt{35}k ), \nonumber
\end{eqnarray*}
where $C_3 = \frac{1}{24576}, C_4 = \frac{1}{3072}, C_5 = \frac{1}{256}$ and $C_6 = \frac{1}{12288}$.
Note that we renamed Ginzberg's $\mathbf{G}$ for $\mathbf{H}$ and $g$ for $h$ in accordance with  conventional notations of $\mathbf{H}$ for low pass filter and $\mathbf{G}$ for high pass filter.
The scaling and wavelet functions for all real and imaginary parts are presented in Figure \ref{fig:quaternionfunctions}.

 Next, we define the non-decimated quaternion wavelet transform (NDQWT) separately for 1-D and 2-D cases by connecting the quaternion low- and high-pass filters with a non-decimation property.

\subsection{1-D case}{\label{sec-1dNDQWT}}
Given a specified a multiresolution framework and a data vector $\bm{\mathit{y}} = (y_0, y_1, \dots, y_{m-1}$) of size $m$, the discrete data vector $\bm{\mathit{y}}$ can be connected to a function $f$ which is a linear combination
 of shifts of the scaling function at some decomposition level $J$,
\begin{eqnarray}\label{NDQWTfunc}
f(x) &=& \sum_{k=0}^{m-1} y_k \phi_{J,k}(x)
\end{eqnarray}
where $J-1 < \log_2m  \le  J$, i.e. $J = \lceil \log_2 m \rceil$ and
\begin{gather}\label{NDQWTscalefunc1}
\phi_{J,k}(x) = 2^{\frac{J}{2}} \phi(2^J (x - k) ).
\end{gather}
For the NDQWT, the scaling functions in Equation (\ref{NDQWTfunc}) and (\ref{NDQWTscalefunc1}) will be the quaternion-valued scaling functions from Equation (\ref{quatfunctions}).

\begin{figure}[!ht]
\centering
\includegraphics [scale=0.57 , clip]{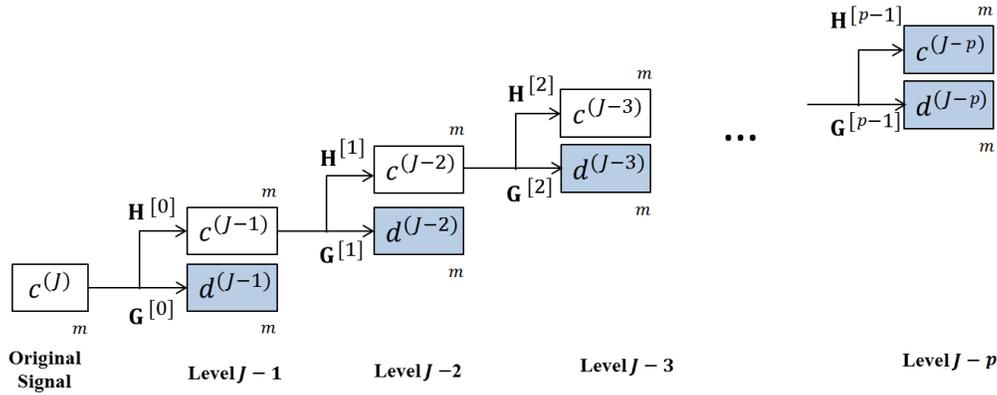}
\caption{Graphical illustration of the  Mallat algorithm. The NDQWT decomposes the original signal of size $m$ to $p+1$ multiresolution subspaces comprising of $p$ levels of detail coefficients and one level of coarse coefficients. The
The shaded boxes represent the transformation, ${\bm d}^{(J-1)}, \bm{d}^{(J-2)}, \dots,  \bm{d}^{(J-p)},$ and $\bm c^{(J-p)}$ .}
\label{fig:NDWTG}
\end{figure}

Alternatively,  the data interpolating function $f$ also can be represented in terms of wavelet coefficients as follows:
\begin{eqnarray}\label{NDQfx}
f(x)&=& \sum_{k=0}^{m-1} c_{J_0,k} \phi_{J_0,k}(x) + \sum_{j=J_0}^{J-1} \sum_{k=0}^{m-1}  d_{j,k}\psi_{j,k}(x),
\end{eqnarray}
where
\begin{eqnarray}\label{NDQref}
\phi_{J_0,k}(x) &=& 2^{\frac{J_0}{2}} \phi\left( 2^{J_0} (x - k) \right), \nonumber\\
\psi_{j,k}(x) &=& 2^{\frac{j}{2}}  \psi\left(2^{j} (x - k) \right),
\end{eqnarray}
and $J_0$ is the coarsest decomposition level.
Note that $2^J (x - k)$ is used inside of the scaling function in Equation (\ref{NDQWTscalefunc1}) and (\ref{NDQref}) instead of $2^J x - k$ for the traditional decimation in wavelet domain in order to make this wavelet decomposition non-decimated.
By using $2^J (x - k)$, the shift indicator $k$ remains constant at all levels, and this corresponds to
levelwise sampling rate that results in the non-decimation. In comparison, for DWT case, the shifts are level dependent as $2^{-j}k$.
After performing the NDQWT on the vector $\bm{\mathit{y}}$, one obtains a vector of smooth coefficients as
\begin{equation}\label{NDQcoarsetsvector}
\bm{\mathit{c}}_{(J_0)} = (c_{J_0,0}, c_{J_0,1}, \dots, c_{J_0,m-1})
\end{equation}
which corresponds to the coarsest approximation of $\bm{\mathit{y}}$.
Likewise, the vectors of detail coefficients  are given as
\begin{equation}\label{NDQdetailvector}
\bm{\mathit{d}}_{(j)} = ( d_{j,0}, d_{j,1}, \dots, d_{j,m-1}), \;\; j=J_0, \dots, J-1,
\end{equation}
which carry fine-scale information within the input $\bm{\mathit{y}}$.
Of course, the number of coefficients in the vectors $\bm{\mathit{c}}^{(J_0)}$ and $\bm{\mathit{d}}^{(j)}$ is always $m$ and this is due to the non-decimated property of NDQWT.
As a result, we  obtain total $(p+1) \times m$ wavelet coefficients,  with $p m$  details and $p$ coarse coefficients.
The Mallat type of algorithm for forward NDQWT is graphically illustrated in Figure \ref{fig:NDWTG}.

Next we focus on each wavelet coefficient  in Equation (\ref{NDQcoarsetsvector}) and (\ref{NDQdetailvector}). Since the scaling and wavelet functions in Equation (\ref{NDQref}) are quaternion-valued, the non-decimated quaternion wavelet coefficients $c_{J_0,k}$ and $d_{j,k}$ in Equation (\ref{NDQfx}) have one real and three imaginary parts as
\begin{eqnarray}\label{NDQcoeffi}
c_{J_0,k} &=& \text{Re}(c_{J_0,k}) + i \cdot \text{Im}^{i}(c_{J_0,k}) + j \cdot \text{Im}^{j}(c_{J_0,k}) + k \cdot \text{Im}^{k}(c_{J_0,k}), \nonumber\\
d_{j,k} &=& \text{Re}(d_{j,k}) \hspace{0.16cm} + i \cdot \text{Im}^{i}(d_{j,k}) \hspace{0.16cm} + j \cdot \text{Im}^{j}(d_{j,k}) \hspace{0.16cm} + k \cdot \text{Im}^{k}(d_{j,k}),
\end{eqnarray}
where $j=J_0, \dots, J-1$ and $\text{Im}^{i}(q)=q_1$, $\text{Im}^{j}(q)=q_2$, and $\text{Im}^{k}(q)=q_3$.
These quaternion-valued wavelet coefficients would be considered in the later sections for construction
of a NDQWT-based spectra, as well as level-dependent phase summaries.

Note that the multiresolution levels and location parameters are traditionally denoted as $j$ and $k$; this is the same
notation for the second and third imaginary unit in the quaternion algebra. This overlap in notation should not cause a confusion when representing  the quaternion-valued wavelet coefficients because
the different context is clear. In the sequel we denote both multiresolution level and second imaginary unit by $j$, and both location parameter and third imaginary unit by $k$.

\begin{figure}[!ht]
\centering
\includegraphics [scale=0.48 , clip]{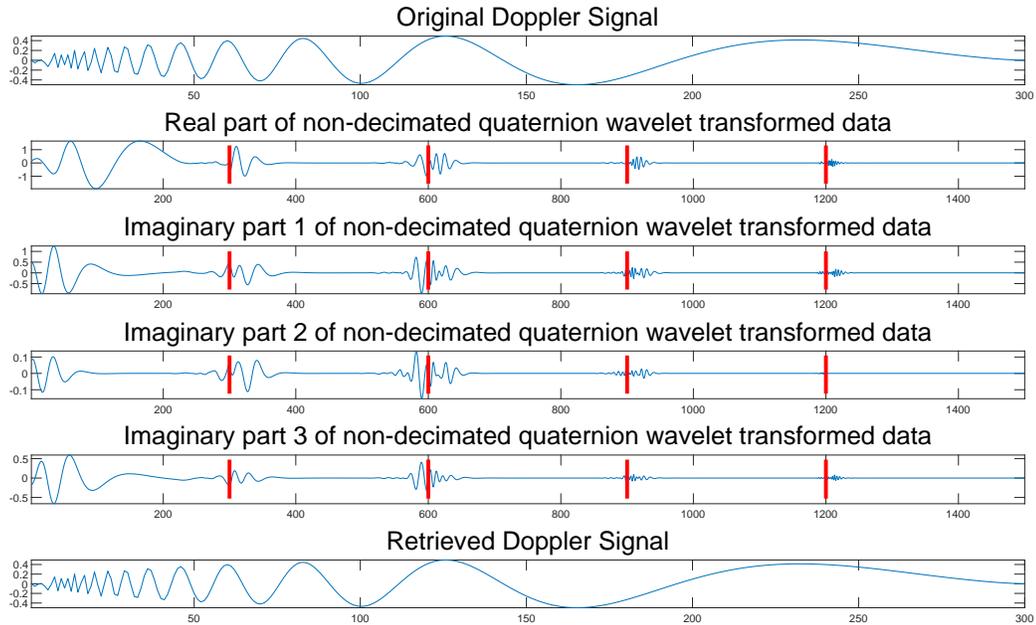}
\caption{An example of matrix-based NDQWT of a Doppler signal of length 300.}
\label{fig:NDQWTdoppler}
\end{figure}

Unlike standard convolution-based approach, the matrix-based NDWT
 can provide several additional features. First, the matrix-formulation allows us to use any non-dyadic size signal. Due to typical sizes of the signals and images
processed,
the matrix based transform does not significantly increase practical computational complexity.
Thus, we incorporate the quaternion scaling and wavelet filters in Equation (\ref{quatfunctions}) into the matrix formulation of NDWT in order to utilize its convenient properties. We obtain a non-decimated quaternion wavelet matrix, $W_m^{(p)}$, that is formed directly from quaternion wavelet filter coefficients with $p$ detail levels and $m$ size of input data. Details for constructing $W_m^{(p)}$ are explained in \citet{Kang2016}.
To obtain a non-decimated quaternion wavelet transformed vector $\boldsymbol{d}$ with depth $p$ from a 1-D signal  $\boldsymbol{y}$ of size $m \times 1$, we multiply $\boldsymbol{y}$ by $W_m^{(p)}$ as
\begin{eqnarray*}\label{eq:NDQWT1dconst}
\boldsymbol{d}&=& {W}_{m}^{(p)} \cdot \boldsymbol{y},
\end{eqnarray*}
where $p$ and $m$ are arbitrary.
For the reconstruction from $\boldsymbol{d}$ to $\boldsymbol{d}$, we need an additional weight matrix for $W_m^{(p)}$ as ${T}_{m}^{(p)}$ that is defined as
\begin{equation}\label{eq:NDQWT1dweight}
 {T}_{m}^{(p)}=\mbox{diag}(\overbrace{1/2^p, \dots,1/2^{p}}^\text{$2m$},\overbrace{1/2^{p-1},\dots,1/2^{p-1}}^\text{$ m$},\dots, \overbrace{1/2, \dots,1/2}^\text{$m$}).
\end{equation}
Utilizing the weight matrix, ${T}_{m}^{(p)}$, we can perform the perfect reconstruction as
\begin{eqnarray*}\label{eq:NDQWT1dcorrectreconst}
\boldsymbol{y} &=&  ({W}_{m}^{(p)})^{\dagger} \cdot {T}_{m}^{(p)} \cdot \boldsymbol{d}
\end{eqnarray*}
where the $W^{\dagger}$ denotes a Hermitian transpose matrix of $W$.
Graphical illustrations of matrix-based NDQWT is displayed in Figure \ref{fig:NDQWTdoppler}.

\subsection{2-D case}{\label{sec-2dNDQWT}}
The real power of matrix implementation of wavelet transforms can be seen in 2-D cases, where the so called
scale-mixing property is utilized. The scale-mixing transforms typically have lower entropy compared to
traditional 2-D transforms which is beneficial in tasks of wavelet shrinkage.
For the scaling analysis, scale mixing transforms enable definition of a range of spectra along the hierarchies of
scale-mixing spaces. In this paper we focus only on diagonal hierarchy, but emphasize that the
spectral tools can be further generalized.

In this section, the 1-D NDQWT from Section \ref{sec-1dNDQWT} is extended to a scale-mixing 2-D NDQWT of
$f(x,y)$ where $(x,y) \in  \mathbb{R}^2$. The decomposition
 has one scaling function and three wavelet functions defined as tensor product
 of functions Equations (\ref{quatfunctions}):
\begin{eqnarray}\label{eq:2dsNDQwavscafunctions}
\phi(x,y) &=& \phi(x) \phi(y)  \nonumber \\
&=& \mu(x,y) + i  \alpha(x,y) + j  \beta(x,y) + k  \gamma(x,y),  \nonumber\\
\psi^{(h)}(x,y) &=& \phi(x) \psi(y)  \nonumber\\
&=& \xi^{(h)}(x,y) + i \zeta^{(h)}(x,y) + j \gamma^{(h)}(x,y) + k \omega^{(h)}(x,y), \\
\psi^{(v)}(x,y) &=& \psi(x) \phi(y) \nonumber \\
&=& \xi^{(v)}(x,y) + i \zeta^{(v)}(x,y) + j \gamma^{(h)}(x,y) + k \omega^{(h)}(x,y),  \nonumber\\
\psi^{(d)}(x,y) &=& \psi(x) \psi(y) \nonumber \\
&=& \xi^{(d)}(x,y) + i \zeta^{(d)}(x,y) + j \gamma^{(h)}(x,y) + k \omega^{(h)}(x,y), \nonumber
\end{eqnarray}
where symbols $h,v,d$ denote the horizontal, vertical, and diagonal directions, respectively.

\subsubsection{Scale-Mixing 2-D Non-decimated Quaternion Wavelet Transform}{\label{sec-sm2dNDQWT}}
The various versions of the 2-D WT with appropriate tessellations of the detail spaces have been considered in 2-D wavelet literature. Here we focus on the scale-mixing 2-D wavelet transform  because of its remarkable flexibility,
 compressibility, and ease of computation \citep{Ramirez2013}.  From the scaling and wavelet functions in Equations (\ref{eq:2dsNDQwavscafunctions})  the wavelet atoms of the scale-mixing 2-D NDQWT can be represented as
\begin{eqnarray}\label{eq:2dscalemixtransNDQfunctions}
\phi_{J_{01},J_{02},k_1,k_2}(x,y) &=& \mu_{J_{01},k_1,k_2}(x,y) \;+ \nonumber \\
& &   i \cdot \alpha_{J_{02},k_1,k_2}(x,y) \;+\; j \cdot \beta_{J_{02},k_1,k_2}(x,y) \;+\; k \cdot \gamma_{J_{02},k_1,k_2}(x,y), \nonumber \\
\psi_{J_{01},j_2,k_1,k_2}^{(h)}(x,y) &=& \xi^{(h)}_{J_{01},k_1,k_2}(x,y) \;+ \nonumber \\
& &  i \cdot \zeta^{(h)}_{j_2,k_1,k_2}(x,y) \;+\; j \cdot \gamma^{(h)}_{j_2,k_1,k_2}(x,y) \;+\; k \cdot \omega^{(h)}_{j_2,k_1,k_2}(x,y),  \nonumber \\
\psi_{j_1,J_{02},k_1,k_2}^{(v)}(x,y) &=& \xi^{(v)}_{j_1,k_1,k_2}(x,y) \;+ \\
& &   i \cdot \zeta^{(v)}_{J_{02},k_1,k_2}(x,y) \;+\; j \cdot \gamma^{(v)}_{j_2,k_1,k_2}(x,y) \;+\; k \cdot \omega^{(v)}_{j_2,k_1,k_2}(x,y), \nonumber \\
\psi_{j_1,j_2,k_1,k_2}^{(d)}(x,y) &=& \xi^{(d)}_{j_1,k_1,k_2}(x,y) \;+ \nonumber \\
& &   i \cdot \zeta^{(d)}_{j_2,k_1,k_2}(x,y) \;+\; j \cdot \gamma^{(d)}_{j_2,k_1,k_2}(x,y) \;+\; k \cdot \omega^{(d)}_{j_2,k_1,k_2}(x,y), \nonumber
\end{eqnarray}
where $k_1 = 0, \dots, m-1$, $k_2 = 0, \dots, n-1$, $j_1 = J_{01}, \dots, J-1$, $j_2 = J_{02}, \dots, J-1$, and $J = \lceil \log_2 \min(m,n) \rceil$. Notice that $J_{01}$ and $J_{02}$ indicate the coarsest decomposition levels of rows and columns, respectively. Using these definitions, we can express any function $f \in L_2(\mathbb{R}^2)$  via wavelet decomposition as
\begin{eqnarray*}\label{eq:2dscalemixtransNDC}
f(x, y) &= & \sum_{k_1}  \sum_{k_2} c_{J_{01}, J_{02}, k_1,k_2} \phi_{J_{01}, J_{02}, k_1,k_2}(x,y)  \nonumber \\
&+ &  \sum_{j_2>J_{02}} \sum_{k_1}  \sum_{k_2} d_{J_{01}, j_2, k_1,k_2}^{(h)} \psi_{J_{01}, j_2, k_1,k_2}^{(h)}(x,y) \nonumber\\
&+ &  \sum_{j_1>J_{01}} \sum_{k_1}  \sum_{k_2} d_{j_1, J_{02}, k_1,k_2}^{(v)} \psi_{j_1, J_{02}, k_1,k_2}^{(v)}(x,y) \nonumber \\
&+ & \sum_{j_1>J_{02}} \sum_{j_2>J_{01}} \sum_{k_1}  \sum_{k_2} d_{j_1, j_2, k_1,k_2}^{(d)} \psi_{j_1, j_2, k_1,k_2}^{(d)}(x,y).
\end{eqnarray*}
This defines a scale-mixing 2-D NDQWT.
Unlike the standard 2-D NDQWT using a single scale denoted by $j$,
here we denote by pair $(j_1, j_2)$ a mixture of two scales.
The coefficients corresponding to these scale-mixed atoms in the decomposition
capture the local ``energy flux" between scales $j_1$ and $j_2.$

Finally, we can obtain the scale-mixing non-decimated quaternion wavelet coefficients as
\begin{eqnarray}
c_{ J_{01},J_{02}, k_1, k_2 } &=&   \iint   f(x,y) \overline{\phi}_{J_{01},J_{02}, k_1, k_2} (x, y)\; dxdy \nonumber \\
&=& \text{Re}(c_{J_{01}, J_{02}, k_1,k_2}) + i \cdot \text{Im}^{i}(c_{J_{01}, J_{02}, k_1,k_2}) \;+ \nonumber\\
&& j \cdot \text{Im}^{j}(c_{J_{01}, J_{02}, k_1,k_2}) + k \cdot \text{Im}^{k}(c_{J_{01}, J_{02}, k_1,k_2}) \nonumber\\
d_{ J_{01},j_{2},k_1, k_2 }^{(h)} &=&  \iint   f(x,y) \overline{\psi}_{J_{01},j_{2}, k_1, k_2}^{(h)}(x, y)\; dxdy \nonumber\\
&=& \text{Re}(d_{J_{01}, j_2, k_1,k_2}^{(h)}) + i \cdot \text{Im}^{i}(d_{J_{01}, j_2, k_1,k_2}^{(h)}) \;+ \nonumber\\
&& j \cdot \text{Im}^{j}(d_{J_{01}, j_2, k_1,k_2}^{(h)}) + k \cdot \text{Im}^{k}(d_{J_{01}, j_2, k_1,k_2}^{(h)}) \\
d_{ j_{1},J_{02},k_1, k_2 }^{(v)} &=&  \iint  f(x,y) \overline{\psi}_{j_{1},J_{02}, k_1, k_2}^{(v)}(x, y)  \; dxdy \nonumber \\
&=& \text{Re}(d_{j_1, J_{02}, k_1,k_2}^{(v)}) + i \cdot \text{Im}^{i}(d_{j_1, J_{02}, k_1,k_2}^{(v)}) \;+ \nonumber\\
&& j \cdot \text{Im}^{j}(d_{j_1, J_{02}, k_1,k_2}^{(v)}) + k \cdot \text{Im}^{k}(d_{j_1, J_{02}, k_1,k_2}^{(v)}) \nonumber\\
d_{ j_1,j_2,k_1, k_2}^{(d)} &=&   \iint   f(x,y) \overline{\psi}_{j_1,j_2, k_1, k_2}^{(d)}(x, y) \; dxdy \nonumber \\
&=& \text{Re}(d_{j_1, j_2, k_1,k_2}^{(d)}) + i \cdot \text{Im}^{i}(d_{j_1, j_2, k_1,k_2}^{(d)}) \;+ \nonumber\\
&& j \cdot \text{Im}^{j}(d_{j_1, j_2, k_1,k_2}^{(d)}) + k \cdot \text{Im}^{k}(d_{j_1, j_2, k_1,k_2}^{(d)}), \nonumber
\label{eq:2dscalemixtransNDQcoef}
\end{eqnarray}
where $\overline{\phi}$ denotes the quaternion conjugate of $\phi$ defined in Equation (\ref{quatconj}).
As we can see, the non-decimated quaternion wavelet coefficients in Equation (\ref{eq:2dscalemixtransNDQcoef}) contain one real and three imaginary parts as quaternion numbers.

Matrix formulation can be used to perform the scale-mixing 2-D NDQWT for images of any size without a preprocessing work.
Transformation of a 2-D image $\boldsymbol{A}$ of size $m \times n$ into a non-decimated quaternion wavelet transformed matrix $\boldsymbol{B}$ with depth $p_1$ and $p_2$ is implemented as
\begin{equation*}\label{eq:NDQWT2dcorrectconst}
\boldsymbol{B}=   {W}_{m}^{(p_1)}  \cdot \boldsymbol{A} \cdot ({W}_{n}^{(p_2)})^{\dagger}
\end{equation*}
where $p_1,p_2,m$, and $n$ are arbitrary. Also, the $W_m^{(p_1)}$ and $W_n^{(p_2)}$ with $p_1$, $p_2$ detail levels and $m$, $n$ size of input data, respectively, are constructed from the quaternion scaling and wavelet filters in Equation (\ref{eq:2dsNDQwavscafunctions}). Then, the resulting transformed matrix $\boldsymbol{B}$ has a size of $(p_1 + 1) m \times (p_2 + 1) n$ and represents a finite-dimensional implementation of the Equation (\ref{eq:2dscalemixtransNDQcoef}) for $f(x)$ sampled in a matrix form.
To correctly reconstruct the original image $\boldsymbol{A}$ of size $m \times n$, we need two weight matrices ${T}_{m}^{(p_1)}$ and ${T}_{n}^{(p_2)}$ with $p_1$- and $p_2$-level weight matrices, which are equally obtained as Equation (\ref{eq:NDQWT1dweight}) with different $m,n,p_1,p_2$. Then the reconstruction can be implemented using the weight matrices as
\begin{equation*}\label{eq:NDQWT2dcorrectreconst}
\boldsymbol{A} =  {W}_{m}^{(p_1)} \cdot  {T}_{m}^{(p_1)} \cdot  \boldsymbol{B} \cdot  {T}_{n}^{(p_2)}\cdot  ({W}_{n}^{(p_2)})^{\dagger}.
\end{equation*}
More rigorous details on these matrix formulation for real-valued wavelets can be found in \citet{Kang2016}.

\section{Non-decimated Quaternion Wavelet Spectra}{\label{sec-NDQwavespec}}
Wavelet-based spectra is an efficient tool to estimate Hurst exponent in analyzing self-similar processes,
such as fractional Brownian motion. Any hierarchy of multiresolution spaces can lead to definition of spectra.
Especially important is that the multiscale analysis is generated by orthogonal filters because of
energy preservation and resulting unbiased spectra.
The literature on different approaches to defining a spectra based on wavelets is vast.

In recent study, \citet{Kong2019} suggested the non-decimated complex wavelet spectra and demonstrated that consideration of redundancy and phase information were beneficial in the tasks of signal and image classification. Here, we extend the complex-valued method into the quaternion-valued wavelet spectra retaining the non-decimation and in 2-D case scale-mixing decomposition.
First, we will explain this method for 1-D case and then expand its to 2-D case, by considering the scale-mixing 2-D transforms.

A real-valued stochastic process $\{ X(t), t \in \mathbb{R} \}$ is said to be self-similar with the Hurst exponent $H$ if
\begin{equation}\label{selfsimilar}
X(\lambda t) \stackrel{d}{=} \lambda^H X(t) \;\; \mbox{for any} \;\; \lambda \in \mathbb{R},
\end{equation}
where $\stackrel{d}{=}$ indicates equality in all joint finite-dimensional distributions.
Given Equation (\ref{selfsimilar}), the wavelet coefficient, $d_{j,k}$, can be represented as
\begin{equation}\label{NDQWTdetail}
d_{j,k} \stackrel{d}{=} 2^{-j(H+\frac{1}{2})}d_{0,k}
\end{equation}
under $L_2$ normalization in a real-valued wavelet transform at fixed dyadic scale $j$.
In the NDQWT, we need to use a modulus $|d_{j,k}|$ instead of $d_{j,k}$. The $|d_{j,k}|$ is defined as
\begin{equation*}
|d_{j,k}| = \sqrt{Re(d_{j,k})^2 +  \text{Im}^{i}(d_{j,k})^2 + \text{Im}^{j}(d_{j,k})^2 + \text{Im}^{k}(d_{j,k})^2}, \;\; j=J_0, \dots, J-1.
\end{equation*}
Then, we can re-state Equation (\ref{NDQWTdetail})as
\begin{equation*}\label{NDQWTdetail1}
|d_{j,k}| \stackrel{d}{=} 2^{-j(H+\frac{1}{2})}|d_{0,k}|, \;\; j=J_0, \dots, J-1.
\end{equation*}
where the numbers of $k$ are all same for each $j$ because of the non-decimation property of NDQWT.
Here the notation $\stackrel{d}{=}$ means the equality in all finite-dimensional distributions.
When $X(t)$ shows a stationary increment, $E(|d_{0,k}|)=0$ and $E(|d_{0,k}|^q)=E(|d_{0,0}|^q)$. This leads to
\begin{equation}\label{NDQWTdetail2}
E(|d_{j,k}|^q) = C 2^{-jq(H+\frac{1}{2})}, \;\; j=J_0, \dots, J-1
\end{equation}
where $C=E(|d_{0,0}|^q)$.
Power $q$ is usually $2,$ corresponding to ``power spectrum, or energy spectrum''
and this would be used in this paper.
 By taking logarithms on both sides of the Equation (\ref{NDQWTdetail2})
  we obtain a basis for wavelet-based estimation of $H$, as
\begin{equation*}\label{NDQWTwavespectra1d}
S(j) = \log_2 (E(|d_{j,k}|^2)) = -j(2H+1) + C', \;\; j=J_0, \dots, J-1.
\end{equation*}
With all considered scaling levels as $j \in \mathbb{Z}$, a set of $S(j)$ represents a wavelet-based spectra.
It describes a transition of energies along the scales. If a signal has a regular scaling, the energies would regularly decay,
 the plot of log-energy against the log-scale is a straight line.
 The rate of energy decay, that is, the slope of  the regression line, measures self-similarity of a given signal.

\begin{figure}[!ht]
\centering
\includegraphics [width=\linewidth,height=\textheight,keepaspectratio]{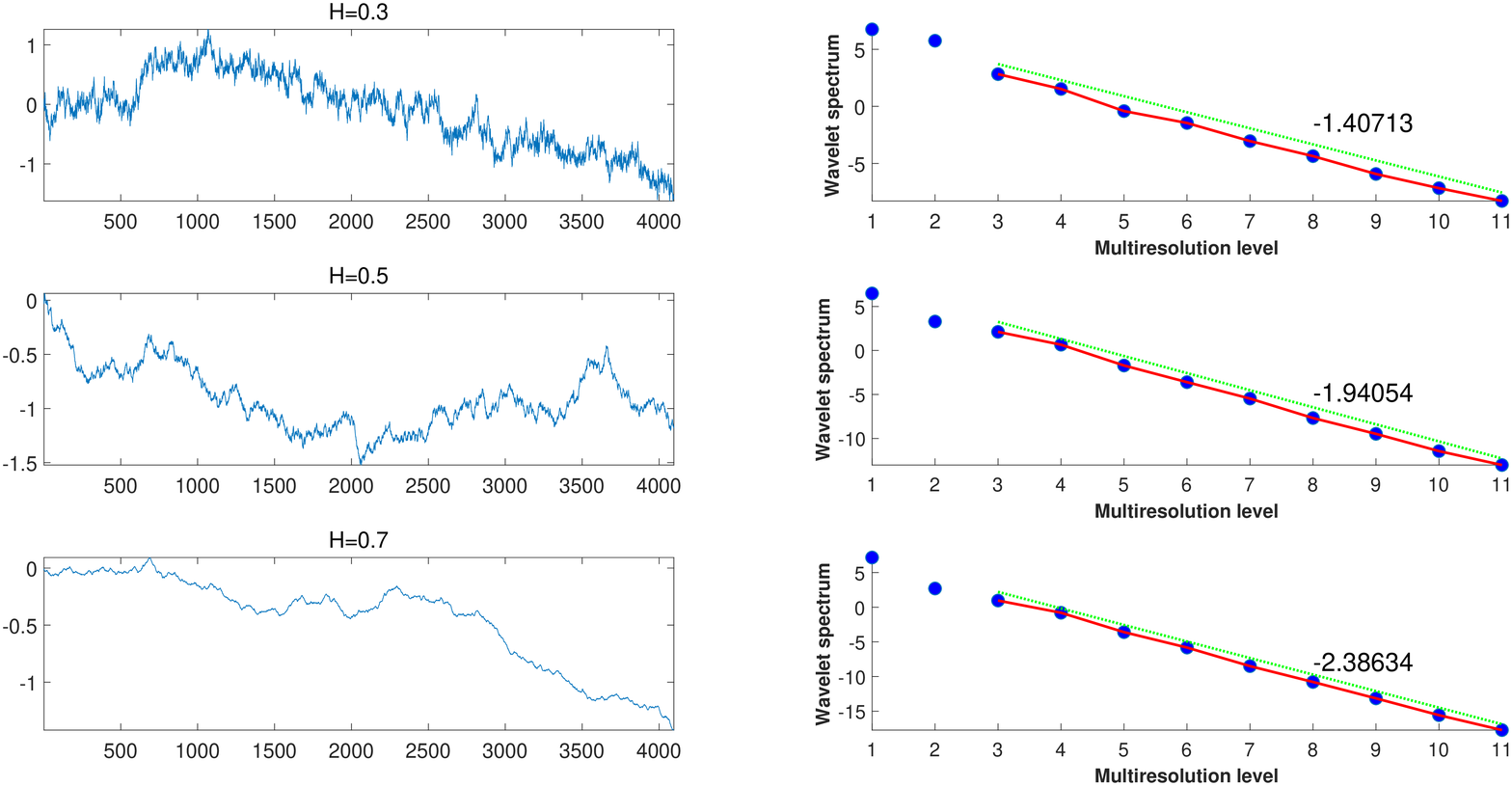}
\caption{Examples of non-decimated quaternion wavelet spectra using the modulus of coefficients. The slopes are -1.40713, -1.94054, and -2.38634 corresponding to estimator $\hat{H}= 0.2035, 0.4703, \; \mbox{and} \; 0.6932$. The original 4096-length signals were simulated as a fBm with Hurst exponent 0.3, 0.5, and 0.7.}
\label{exwaveletspectramod}
\end{figure}

Operationally, we empirically estimate the wavelet-based spectrum, $S(j)$, as
\begin{equation*}\label{NDQWTwavespectra1dempi}
\hat{S}(j) = \log_2 \frac{1}{m} \sum_{k=1}^{m} |d_{j,k}|^2 = \log_2 \overline{|d_{j,k}|^2}, \;\; j=J_0, \dots, J-1
\end{equation*}
where $m$ is the number of given data. Then, we can plot a $2$nd order Logscale Diagram (2-LD) that is a set of $\hat{S}(j)$ against $j$ as $\big( j, \; \hat{S}(j) \big)$ as displayed in Figure \ref{exwaveletspectramod}. Finally, we can measure the slope of energy decay by regression methodology (an ordinary, weighted, or robust regression) and calculate the Hurst exponent $H$ based on the slope as $H = -(\mbox{slope} +1)/2$. More rigorous proof and explanation of wavelet-based spectra  and its applications can be found in \citet{Veitch1999}, \citet{Mallat2009}, and \citet{Ramirez2013}.

\subsection{Scale-Mixing 2-D Non-decimated Quaternion Wavelet Spectra}{\label{sec-scNDQwavespec}}
A 2-D fractional Brownian motion (fBm) in two dimensions, $B_H(\mathbf{u})$ for $\mathbf{u} \in [0,1] \times [0,1]$ and $H \in (0,1)$, will be used as a model to explain a scale-mixing 2-D non-decimated complex wavelet spectra. The 2-D fBm, $B_H(\mathbf{u})$, is a
self-similar process
\begin{equation*}
B_H(a\mathbf{t}) \stackrel{d}{=} a^H B_H(\mathbf{t}) \;\; \mbox{for any} \;\; a \in \mathbb{R},
\end{equation*}
with stationary zero-mean Gaussian increments.
When 2-D fBm is decomposed by a scale-mixing non-decimated quaternion transform,
the wavelet detail coefficients are
\begin{equation*}\label{detailfBm}
d_{(j_1,j_2+s,k1,k2)}=2^{\frac{1}{2}(j_1+j_2+s)}\int B_{H}(\mathbf{u})
\overline{\psi} \left( 2^{j_1}(u_{1}-k_{1}),2^{j_2+s}(u_{2}-k_{2})  \right)  d\mathbf{u},
\end{equation*}
where $\overline{\psi}$ is the quaternion conjugate of $\psi^{(d)}$ defined in Equation (\ref{eq:2dscalemixtransNDQfunctions}).
We will  focus on the main diagonal hierarchy where the 2-D scale indices coincide,
we will use notation   $j=j_1 = j_2$ and $J_0=J_{01} = J_{02} $ in the sequel.

As in the 1-D case, we need to consider a modulus, $|d_{(j,j+s,k1,k2)}|$, instead of quaternion-valued $d_{(j,j+s,k1,k2)}$
defined as:
\begin{gather*}
|d_{(j,j+s,k_1,k_2)}|  = \\
\sqrt{Re(d_{(j,j+s,k_1,k_2)})^2 +  Im^i(d_{(j,j+s,k_1,k_2)})^2 + Im^j(d_{(j,j+s,k_1,k_2)})^2 + Im^k(d_{(j,j+s,k_1,k_2)})^2}, \\
j=J_0, \dots, J-1.
\end{gather*}
Next, we can calculate an average of squared modulus of the coefficients as
\begin{eqnarray*}\label{mixd}
& &\mathbb{E}\left[ |d_{(j,j+s,k_1,k_2)} |^{2}\right] =
 2^{2j+s}\int \psi\left(2^j (u_{1}-k_{1}),2^{j+s}(u_{2}-k_{2}) \right) \nonumber \\
& &~~~~~~~\times \overline{\psi}\left(2^j (v_{1}-k_{1}),2^{j+s}(v_{2}-k_{2}) \right)  \mathbb {E}\left[ B_{H}(\mathbf{u})B_{H}(\mathbf{v}) \right]d\mathbf{u} \ d\mathbf{v},
\end{eqnarray*}
which be expressed as
\begin{equation}\label{spectrumfbm}
\mathbb{E}\left[ |d_{(j,j+s,k_1,k_2)} |^{2}\right] = 2^{-j(2H+2)}\  V_{\psi,s}(H).
\end{equation}
This was proven in \citet{Jeon2014} for complex wavelets and in \citet{Kang2016} for non-decimated wavelets.
Here, the $V_{\psi,s}(H)$ can be treated as constant with respect to scale $j$ but it depends on $\psi$, $H$ and $s$.
Finally, we can obtain the scale-mixing 2-D non-decimated quaternion wavelet spectrum by taking logarithms on both sides of the Equation (\ref{spectrumfbm}) as following:
\begin{equation*}\label{wavespectra2d}
S(j, j+s) = \log_2 (\mathbb{E}(|d_{j,j+s,k_1,k_2}|^2)) = -j(2H+2) + C', \;\; j=J_0, \dots, J-1.
\end{equation*}
The empirical counterpart of $S(j, j+s) $ is
\begin{equation*}\label{wavespectra2dempi}
\hat{S}(j, j+s) = \log_2 \frac{1}{mn} \sum_{k_1=1}^{m}\sum_{k_2=1}^{n} |d_{j,j+s,k_1,k_2}|^2 = \log_2 \overline{|d_{j,j+s,k_1,k_2}|^2}, \;\; j=J_0, \dots, J-1
\end{equation*}
where $m$ is a row length and $n$ is a column length.
To estimate Hurst exponent $H$ we use  the spectral slope as in the 1-D case except that in 2-D case  $\hat H = -(\mbox{slope} +2)/2$ instead of $\hat H = -(\mbox{slope} +1)/2$.

\section{Phase-based Statistics for Classification Analysis}{\label{sec-phase}}
Importance of properly utilizing phase information that is not available for the real-valued wavelets
was exemplified in \citet{Jeon2014} and \citet{Kong2019} for the complex-valued wavelets.
Although the spectra based on the phase information cannot be used to estimate the Hurst exponent,
here we suggest  the use of  phase-based modalities to improve performance in classification tasks.
Given the  three phases in quaternion decompositions, we expect that the discriminatory
power of summaries that include phase modalities would significantly increase.

First, we need to calculate three phases of non-decimated quaternion wavelet coefficient defined in Equation (\ref{NDQcoeffi}). For 1-D case, substituting the four coefficients, $\text{Re}(d_{j,k}), \text{Im}^{i}(d_{j,k}), \text{Im}^{j}(d_{j,k}),$ and $\text{Im}^{k}(d_{j,k})$ for the $q_0, q_1, q_2,$ and $q_3$, we obtain the three phases $\phi_{d_{j,k}}, \theta_{d_{j,k}},$ and $\psi_{d_{j,k}}$ as explained in Section \ref{sec-QuatAlgebra}. It is the same for 2-D case after replacing the $d_{j,k}$ with $d_{(j,j+s,k1,k2)}$.

Then, the three phase averages at level $j$ can be obtained as
\begin{gather}\label{avgphase}
\phi_{j} = \frac{1}{m} \sum_{k=1}^{m} \phi_{d_{j,k}}, \;\; \theta_{j} = \frac{1}{m} \sum_{k=1}^{m} \theta_{d_{j,k}}, \;\;  \psi_{j} = \frac{1}{m} \sum_{k=1}^{m} \psi_{d_{j,k}}, \\
\phi_{j} = \frac{1}{mn} \sum_{k_1=1}^{m}\sum_{k_2=1}^{n} \phi_{d_{(j,j+s,k1,k2)}}, \;\; \theta_{j} = \frac{1}{mn} \sum_{k_1=1}^{m}\sum_{k_2=1}^{n} \theta_{d_{(j,j+s,k1,k2)}}, \;\;  \psi_{j} = \frac{1}{mn} \sum_{k_1=1}^{m}\sum_{k_2=1}^{n} \psi_{d_{(j,j+s,k1,k2)}},  \nonumber \\
j=J_0, \dots, J-1 \nonumber
\end{gather}
for 1-D and 2-D cases, separately.

\begin{figure}[h!tb]
  \centering
  \subfigure[]{\includegraphics[width=1.8in, height=1.6in]{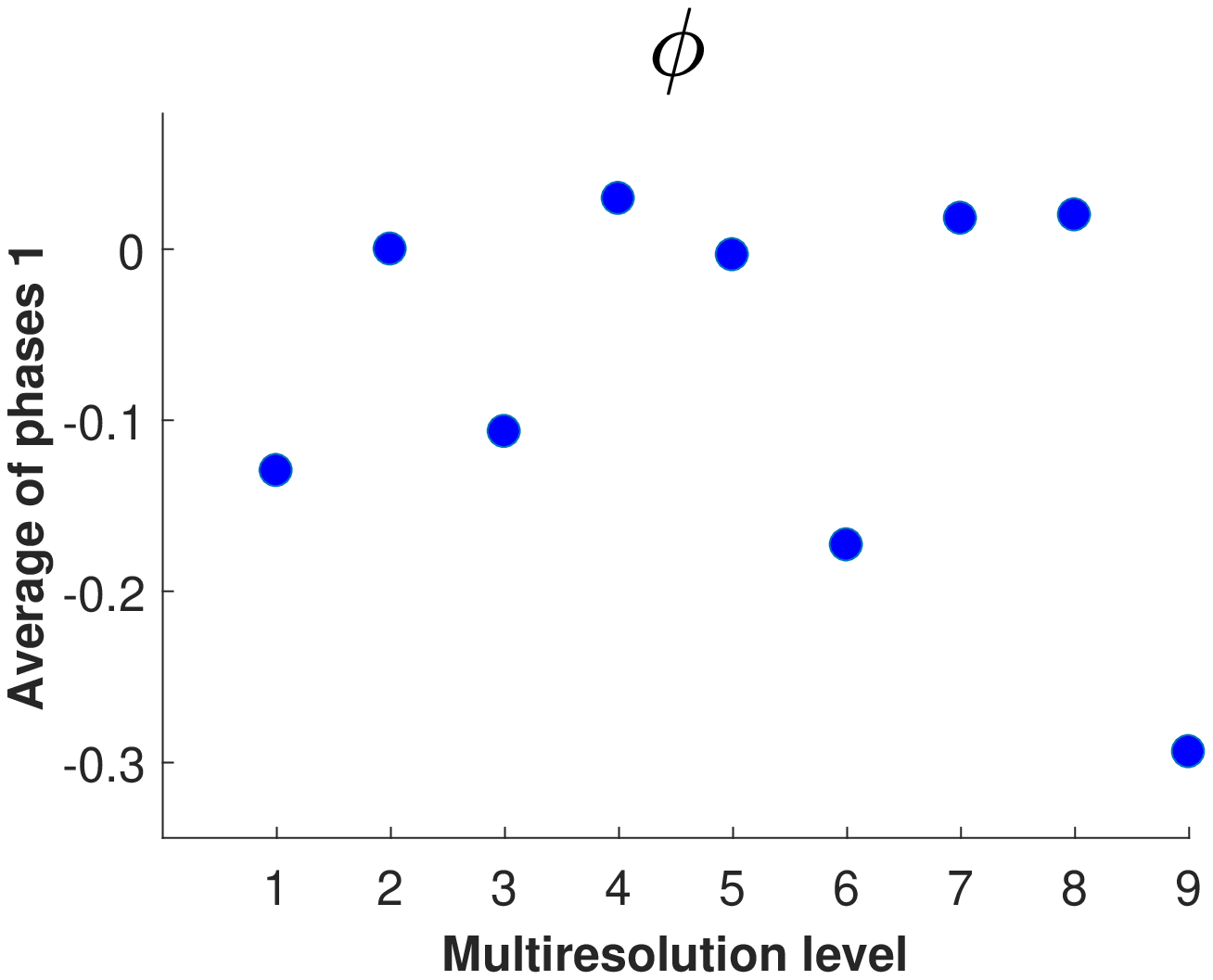}} \qquad
  \subfigure[]{\includegraphics[width=1.8in, height=1.6in]{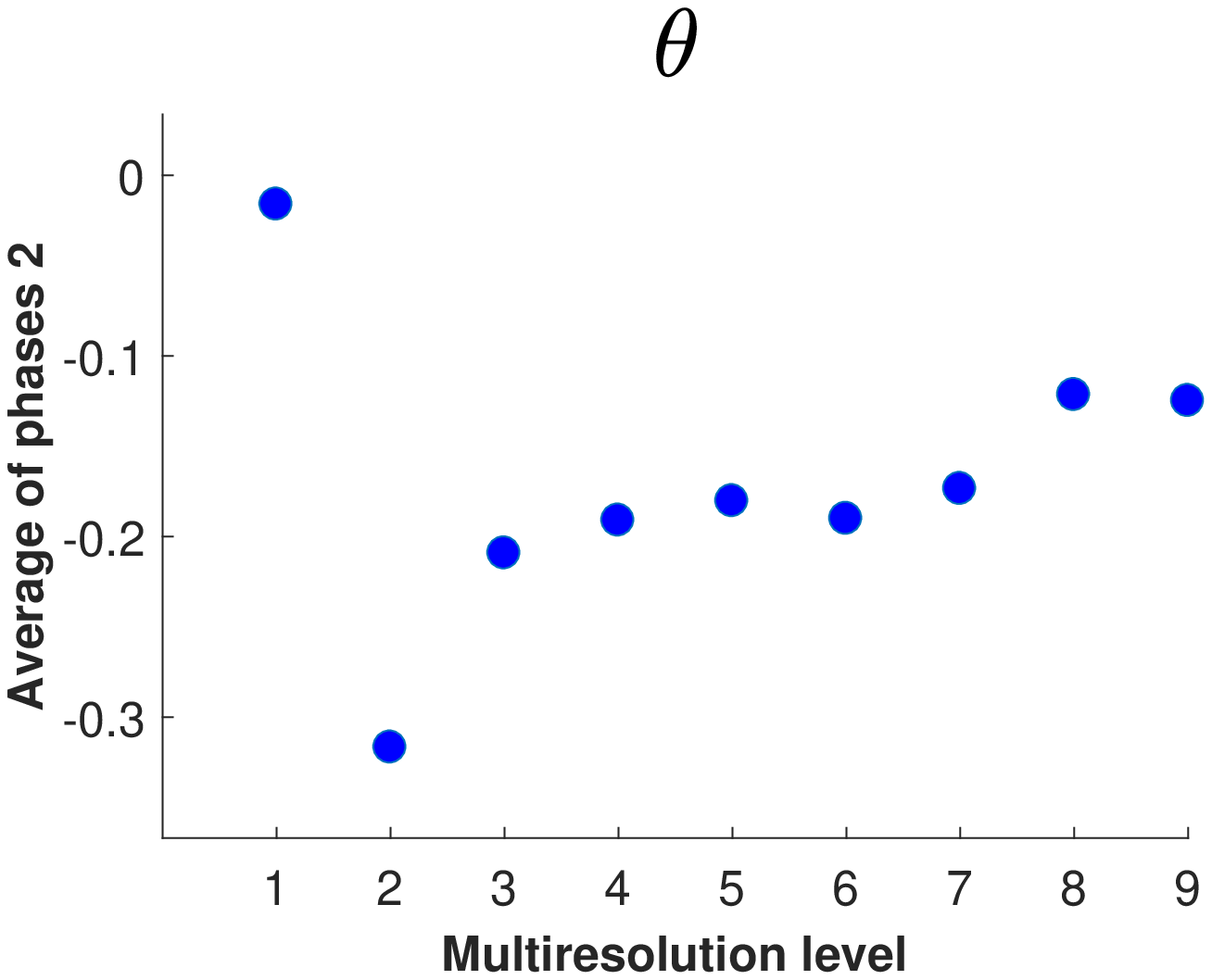}} \qquad
  \subfigure[]{\includegraphics[width=1.8in, height=1.6in]{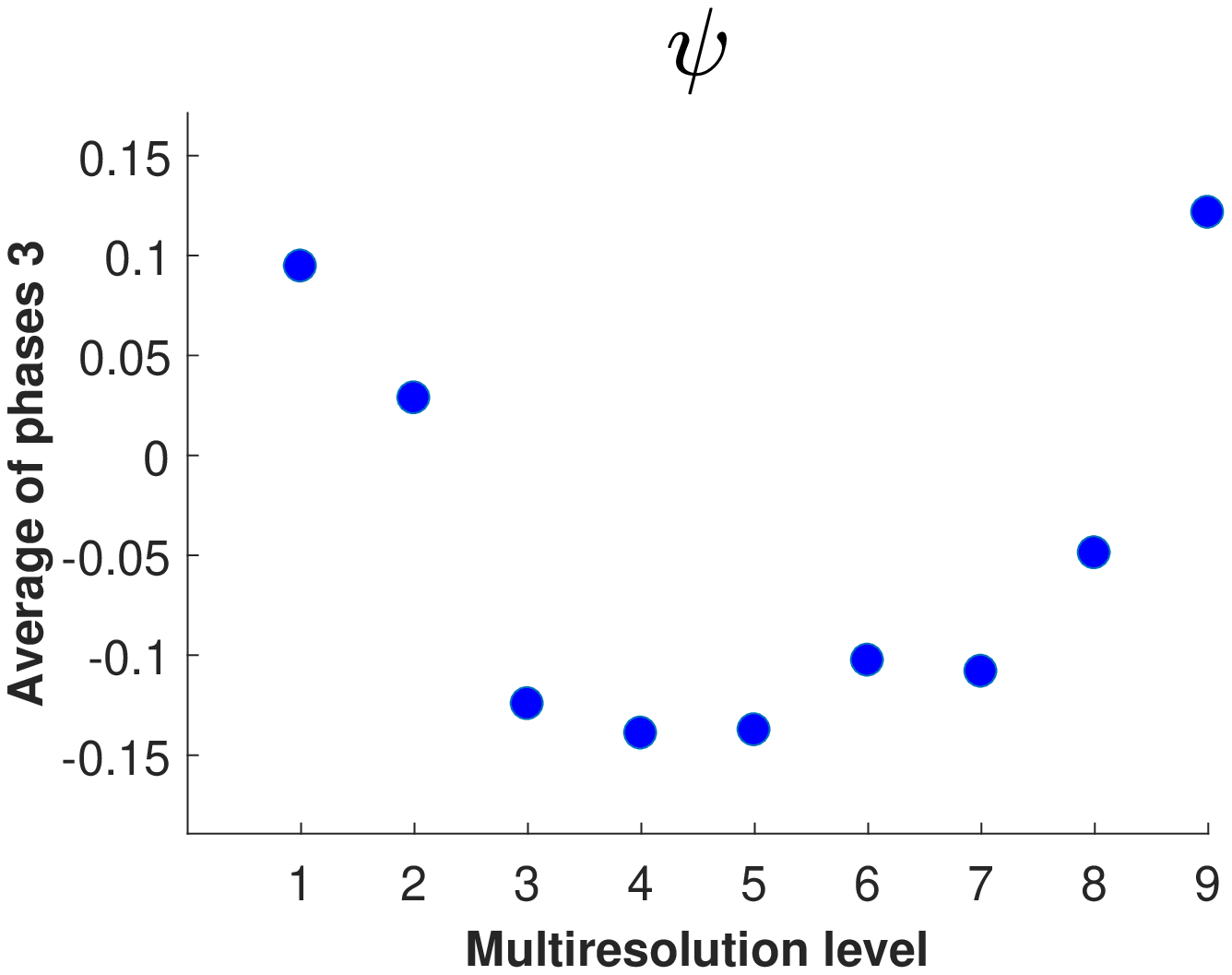}} \qquad
  \caption{Visualization of three phase averages ($\phi, \theta, \psi$) at all multiresolution levels.}
  \label{exwaveletquaternionspectraphase}
\end{figure}

While the phases do not indicate any scaling regularity as explained at the beginning of this section and as displayed in Figure \ref{exwaveletquaternionspectraphase}, the  three phase averages defined in Equation (\ref{avgphase}) would improve a power of classification if used with the wavelet-based spectra described in section \ref{sec-NDQwavespec}. In Section \ref{sec-app}
we demonstrate this and show that the new modalities have surpassed the traditional wavelet-based spectra method
which is based on the modulus of wavelet coefficients.

\section{Applications}{\label{sec-app}}

To illustrate the proposed methodology, we consider two applications in tasks of supervised learning: classification of sounds and
quality control in industrial production.

\subsection{Application in Classifying Sounds data}{\label{sec-sound}}
Nowadays the air conditioners (AC) are quieter than ever. High-efficiency AC utilizes sound-dampening technology and two-stage compressors to keep noise levels below 55 decibels. So if unusual sounds come from an air conditioner during
the course of normal operation,
one should not ignore them as this could be a sign of malfunction or wearout.
Ignoring unusual noises from AC can turn minor issues into major expenses because these noises could indicate a specific problem. The sooner we can find and resolve the cause of the noise, the better. Therefore, if  an automatic noise analyzing system is available, it could increase AC's reliability and maintainability.

To develop an automatic noise analyzer, we are given three sound signals that AC (from unnamed company) could make. Since the normal sound signal is not provided, our task is to build a classification model for the three noises named as air, sha, and water, which would be used as a prototypes.
Descriptions of the noise sounds and what they may signify are as follows.

`Air' indicates hissing noise sound and implies a possible leak. So if there is a hissing sound coming from AC, it is likely either a ductwork issue or a refrigerant leak.
There may be a leak in ducts allowing air or refrigerant to escape. When air is leaking the system is not running efficiently, while if the refrigerant is leaking, users may be exposed to a dangerous chemical.

`Sha' represents buzzing noise sound. Buzzing is almost always a sign of an electrical issue.
If buzzing only occurs when triggering certain settings through a control panel, it is likely just because of a faulty part. But   the constant buzzing is more likely to indicate a problem with the wiring, like a loose or exposed wire, causing electricity to spark within the unit.
Slight humming is common and usually does not mean anything serious. On the other hand, if air conditioner is making loud buzzing, it could be a sign of loose parts or motor problems.

`Water' relates to bubbling noise sound.  Although bubbling noise is not common, the problem is not
likely to be a serious issue.
Bubbling noise usually occurs when the condensing pump malfunctions. As condensate builds up within the AC, it drips to the bottom of the air handler where it empties into a drain pan via either gravity or a condensing pump.  Water accumulation and pump malfunctioning can lead to such noises.

Since a variety of noises including the air, sha, and water require different types of professional attention, it is important task to classify them. To achieve this goal the extraction of informative features from the sound signals is critical.
It is notable that trends are not significant because they usually relate to volume of the sounds. Alternatively,
focusing on the scaling information may be discriminatory since these sound signals digitized at a high frequency are typically self-similar in nature.
Here we propose a classification model based on the wavelet spectra method described in Section \ref{sec-NDQwavespec} with phase-based statistics suggested in Section \ref{sec-phase}.

\subsubsection{Description of Data}{\label{sec-sounddata}}
Data on sound signals are recorded in the system at a rate of 0.0333MHz.
The original dataset consists of three long sound signals: air, sha, and water, of unequal sizes.
We segmented the signals to make the dataset convenient for our analysis.
For each signal, we take subsequent non-overlapping 1024-length pieces.
For instance, we obtain a total 4 dataset (segments) of length 1024 from a signal of 4096 length.
We emphasize that pieces of arbitrary length can be selected to form the data set,
but because of comparisons with decimated transforms, we selected length which is a power of 2.

Table \ref{sounddata} summarizes the finalized dataset according to this segmentation
concept and finally the total number of samples is 1341.

\begin{table}[h!tb]
\begin{center}
\begin{adjustbox}{max width=\textwidth}

\begin{tabular}{c|c|c}
  \specialrule{1.3pt}{1pt}{1pt}
   Group & Original length  & Number of samples  \\\hline \hline
  Air & 491520  & 479    \\
   Sha & 655360 & 639    \\
   Water & 229376   & 223    \\
  \specialrule{1.3pt}{1pt}{1pt}
\end{tabular}

\end{adjustbox}
\end{center}
\caption{Group characterization summary.}\label{sounddata}
\end{table}

\subsubsection{Classification}{\label{sec-soundclassification}}

In this section, we explain how to classify the sound signals based on the proposed NDQWT. First, we performed the proposed 1-D NDQWT to the segmented signals described in Section \ref{sec-sounddata} using the quaternion filter.
Additionally, DWT, NDWT, $\text{WT}_\text{\large{c}}$, $\text{NDWT}_\text{\large{c}}$, and QWT are implemented for comparison.

\begin{figure}[h!tb]
  \centering
  \subfigure[]{\includegraphics[width=1.5in, height=1.5in]{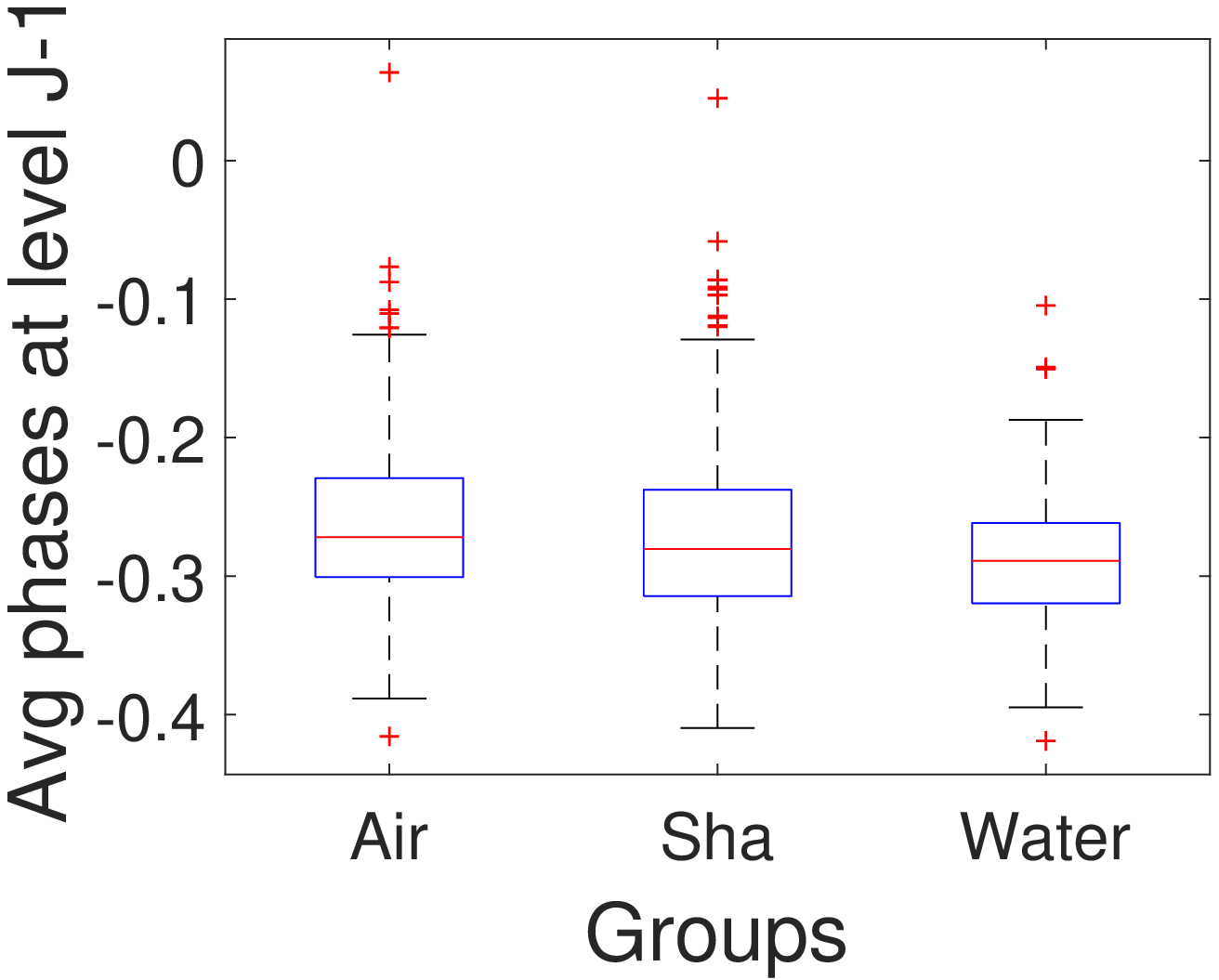}} \qquad
  \subfigure[]{\includegraphics[width=1.5in, height=1.5in]{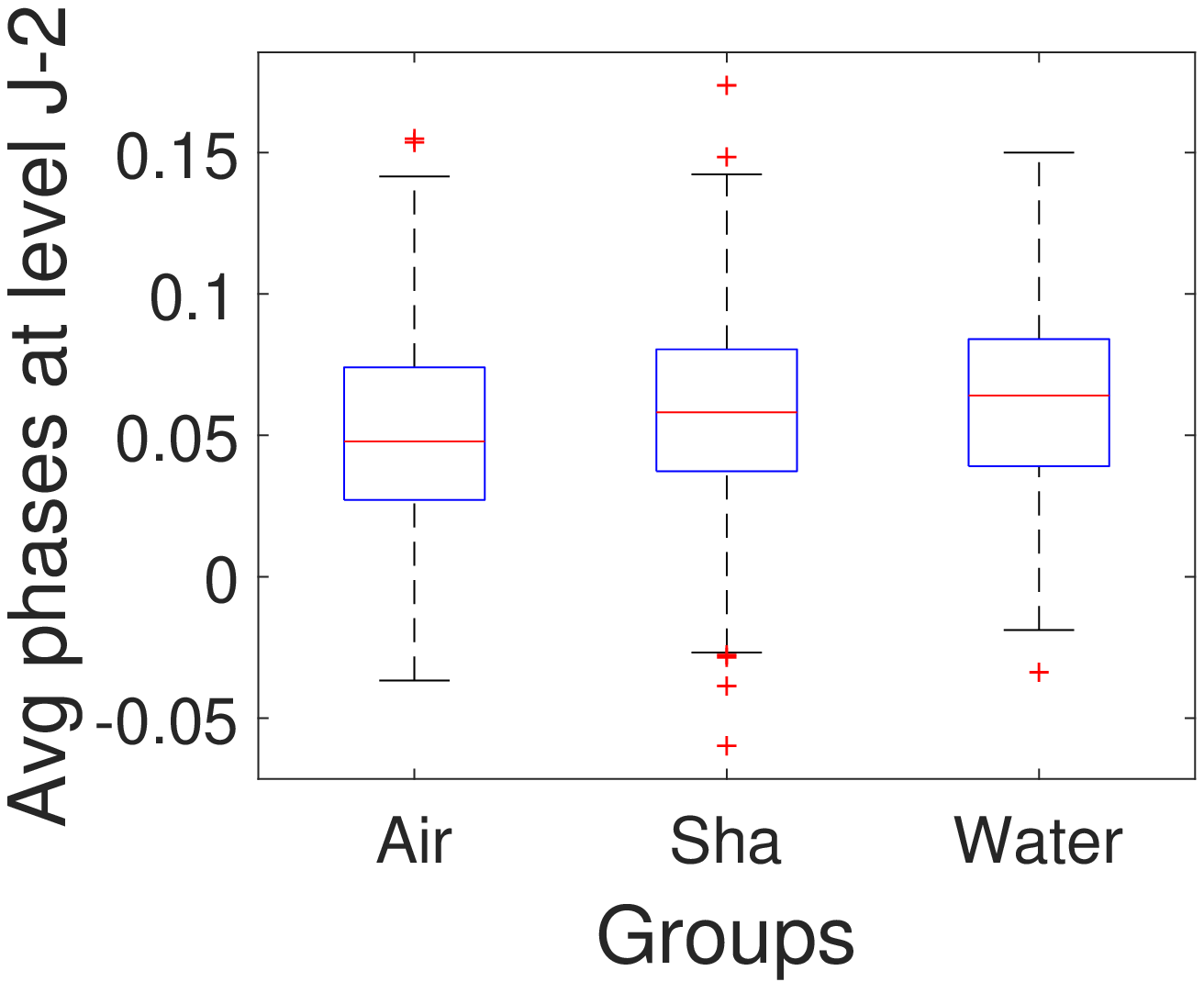}}  \qquad
  \subfigure[]{\includegraphics[width=1.5in, height=1.5in]{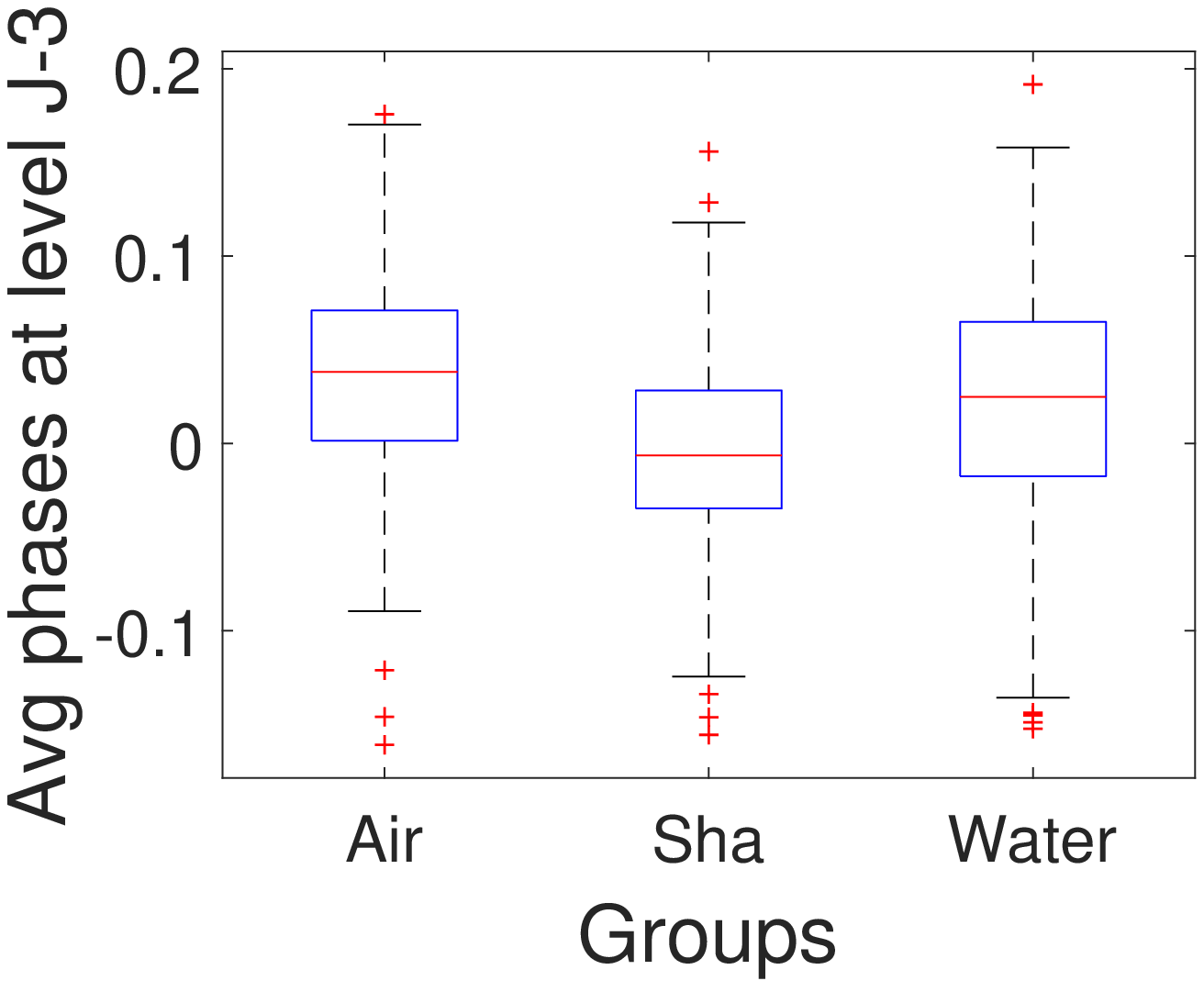}} \\
  \subfigure[]{\includegraphics[width=1.5in, height=1.5in]{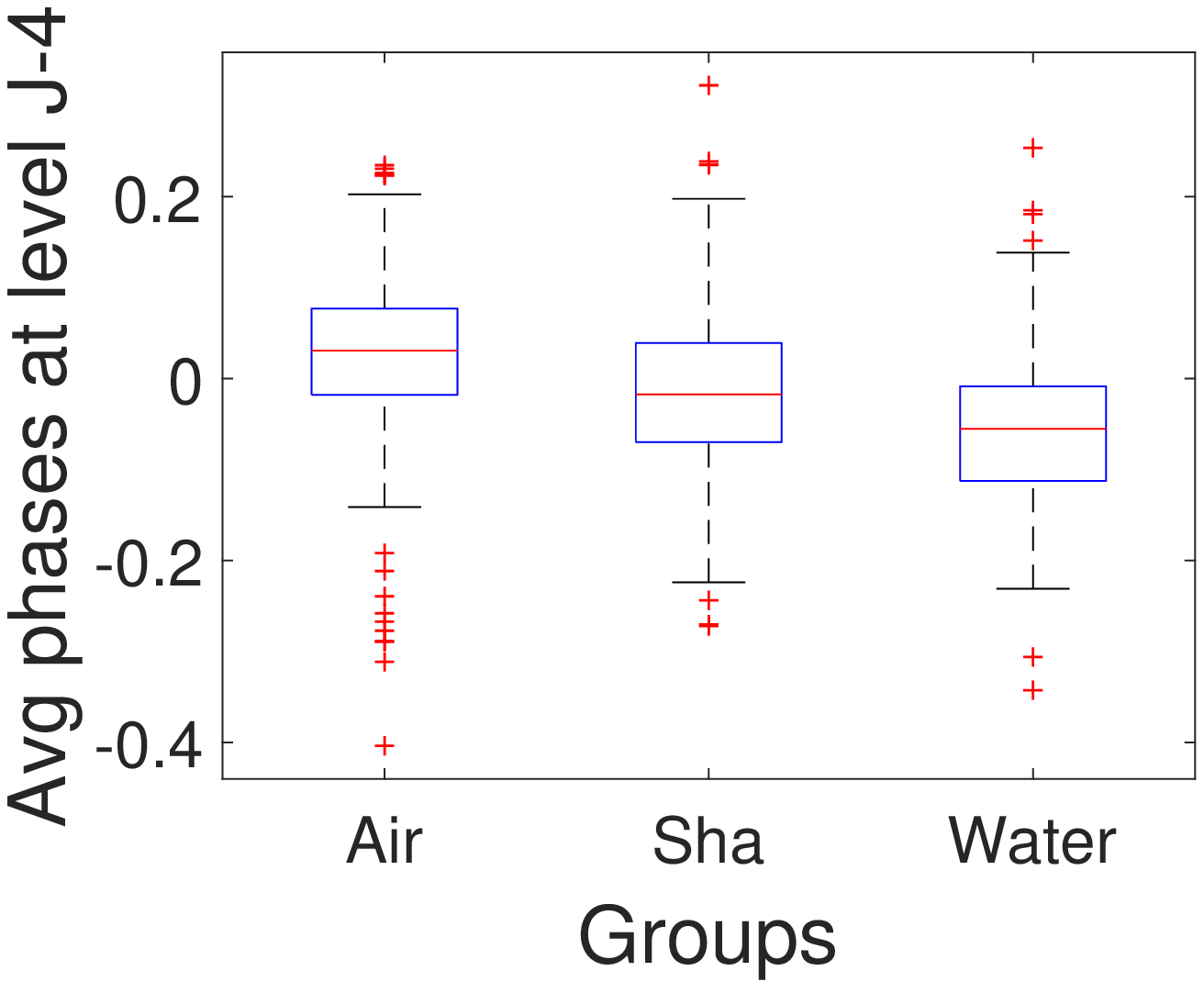}} \qquad
  \subfigure[]{\includegraphics[width=1.5in, height=1.5in]{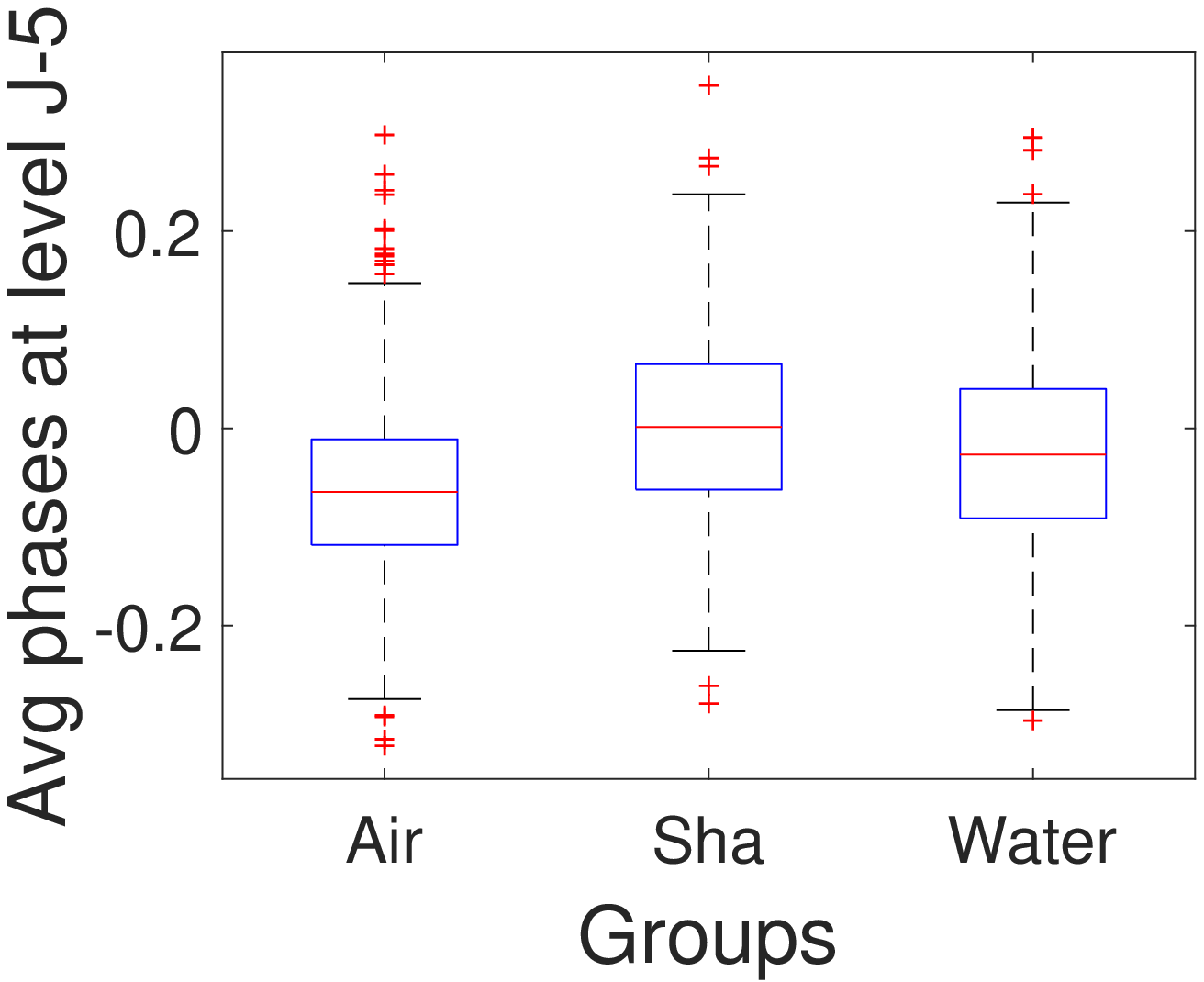}}  \qquad
  \subfigure[]{\includegraphics[width=1.5in, height=1.5in]{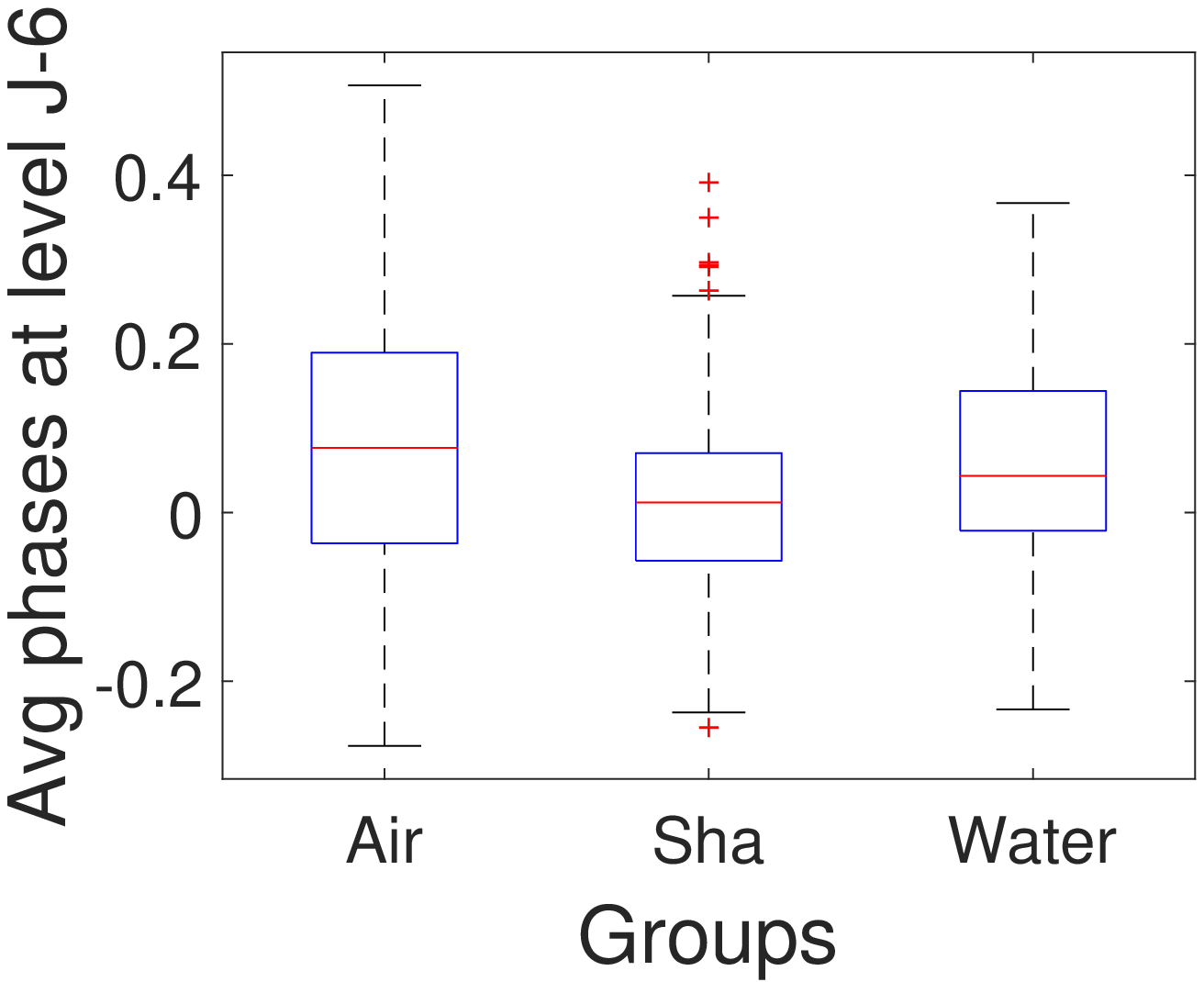}} \\
  \subfigure[]{\includegraphics[width=1.5in, height=1.5in]{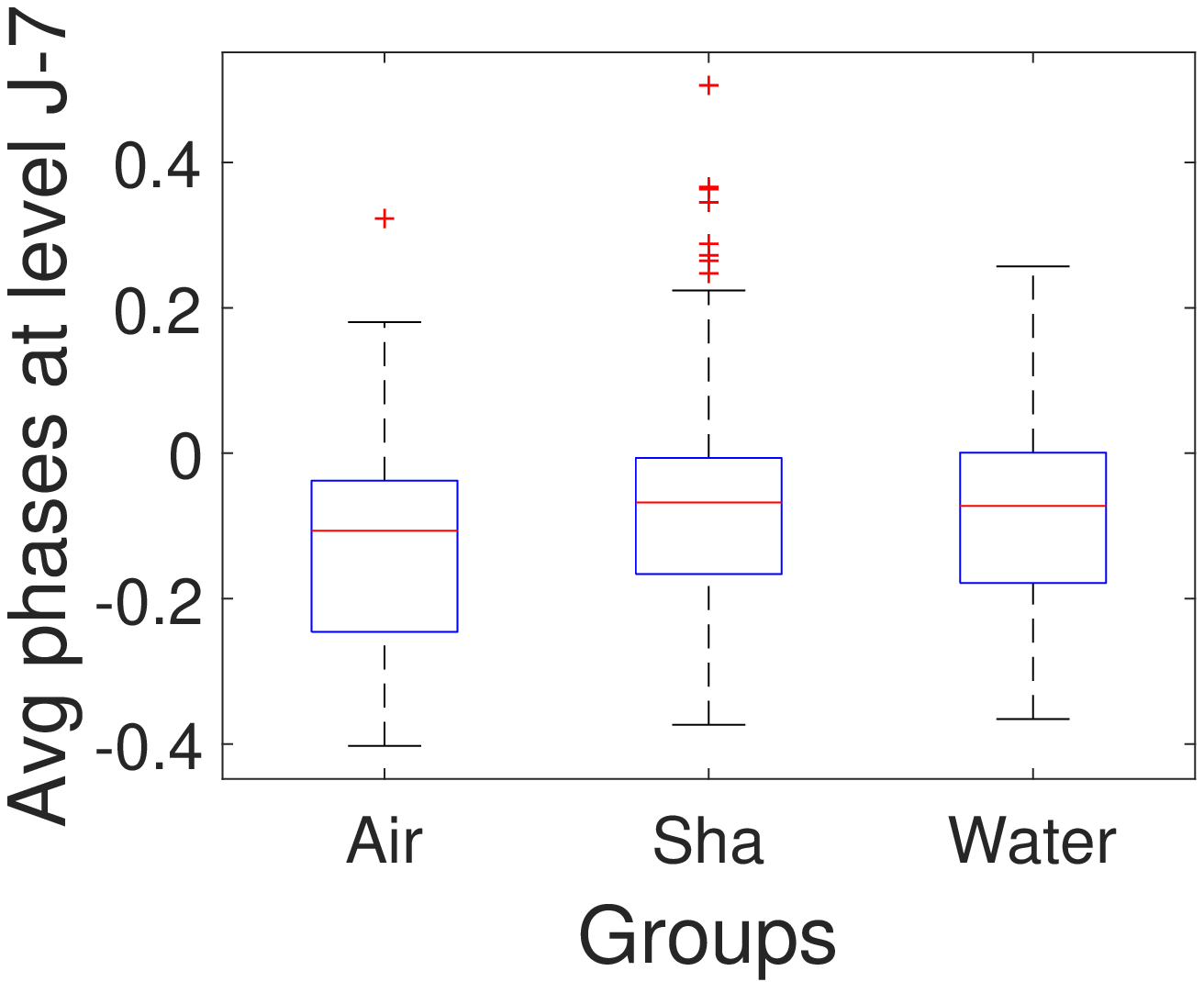}} \qquad
  \subfigure[]{\includegraphics[width=1.5in, height=1.5in]{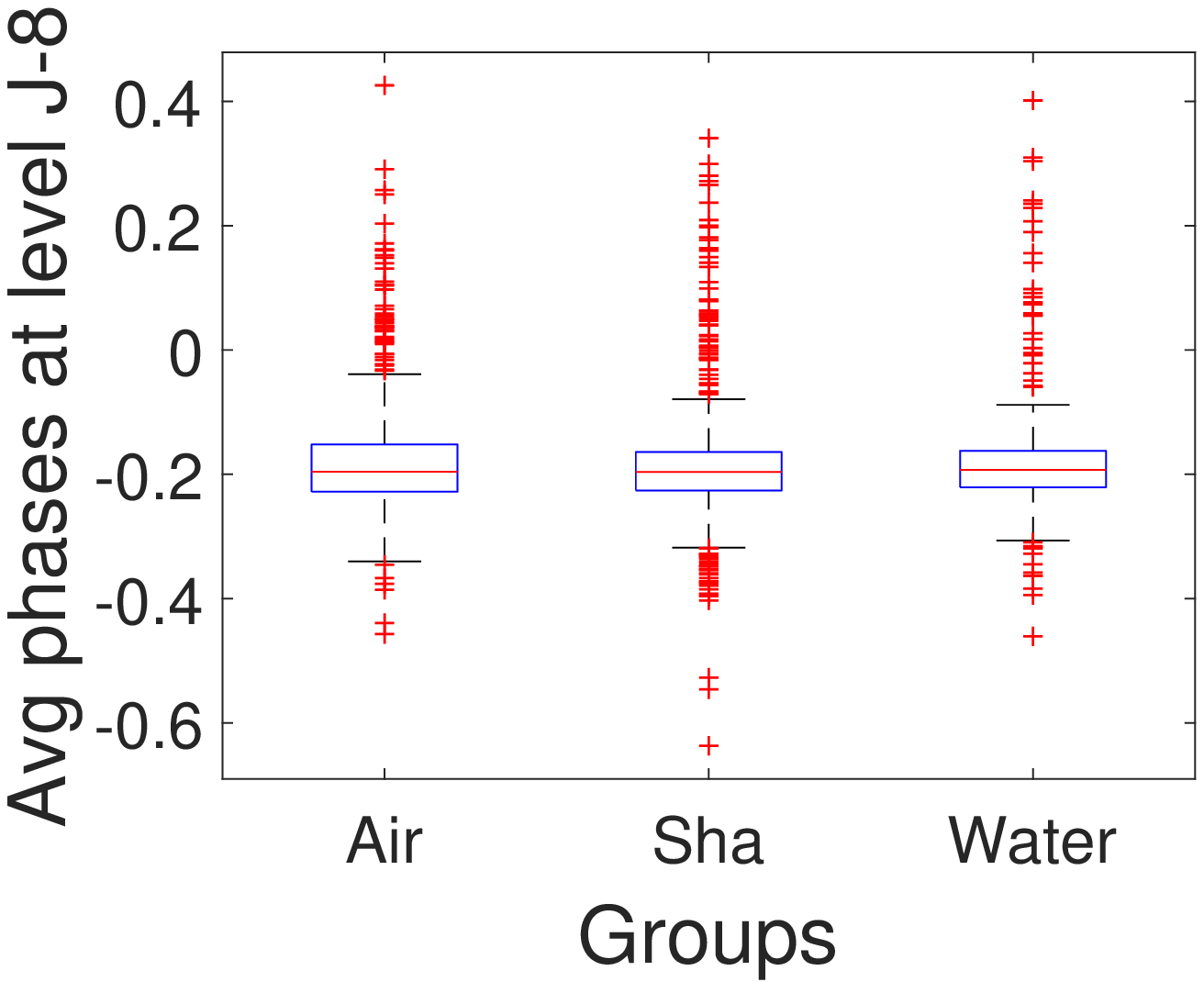}}  \qquad
  \subfigure[]{\includegraphics[width=1.5in, height=1.5in]{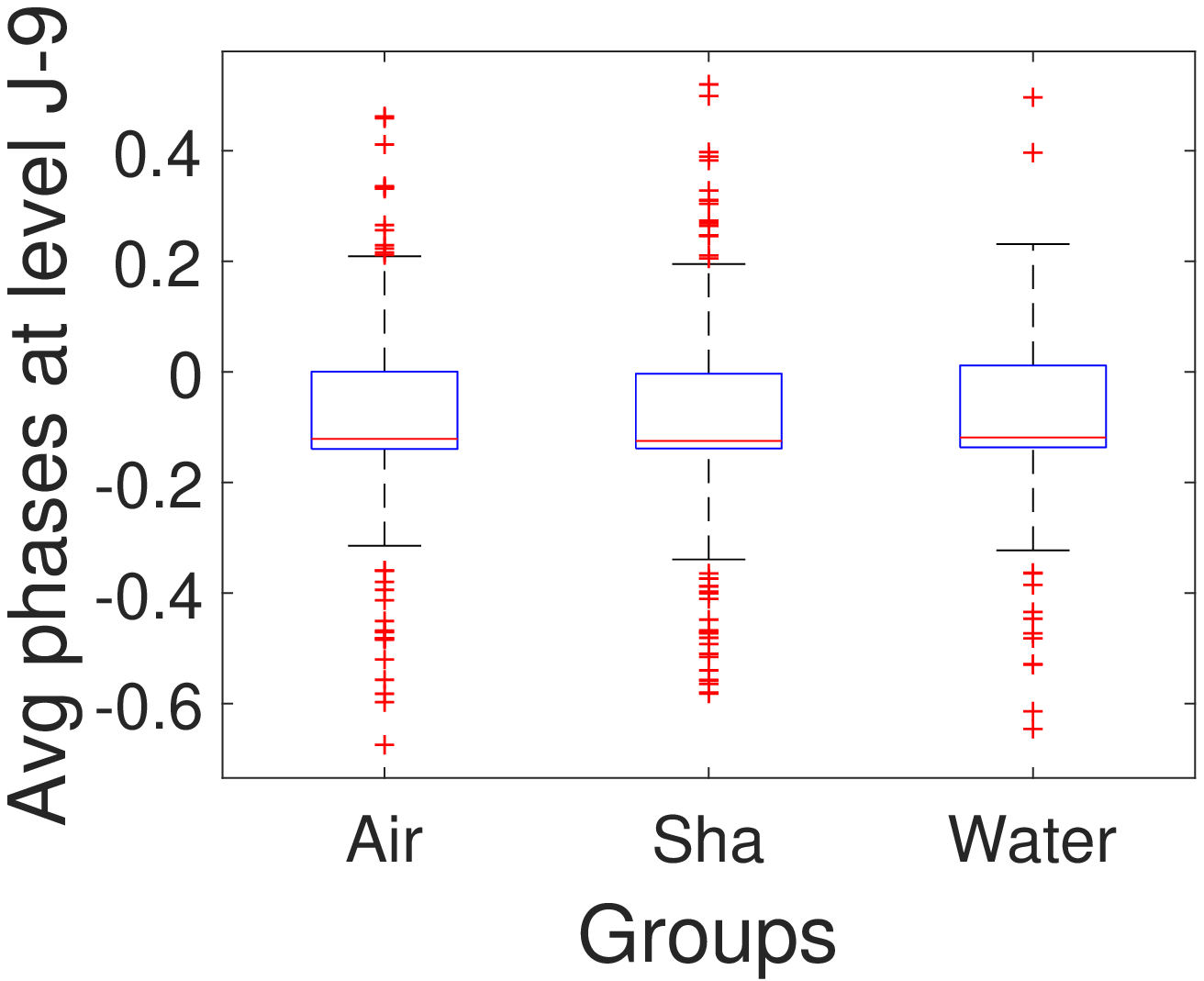}} \\

  \caption{Box plots of averages of phase $\phi$ at all multiresolution levels.}
  \label{boxfig:soundphase1NDQ}
\end{figure}

\begin{figure}[h!tb]
  \centering
  \subfigure[]{\includegraphics[width=1.5in, height=1.5in]{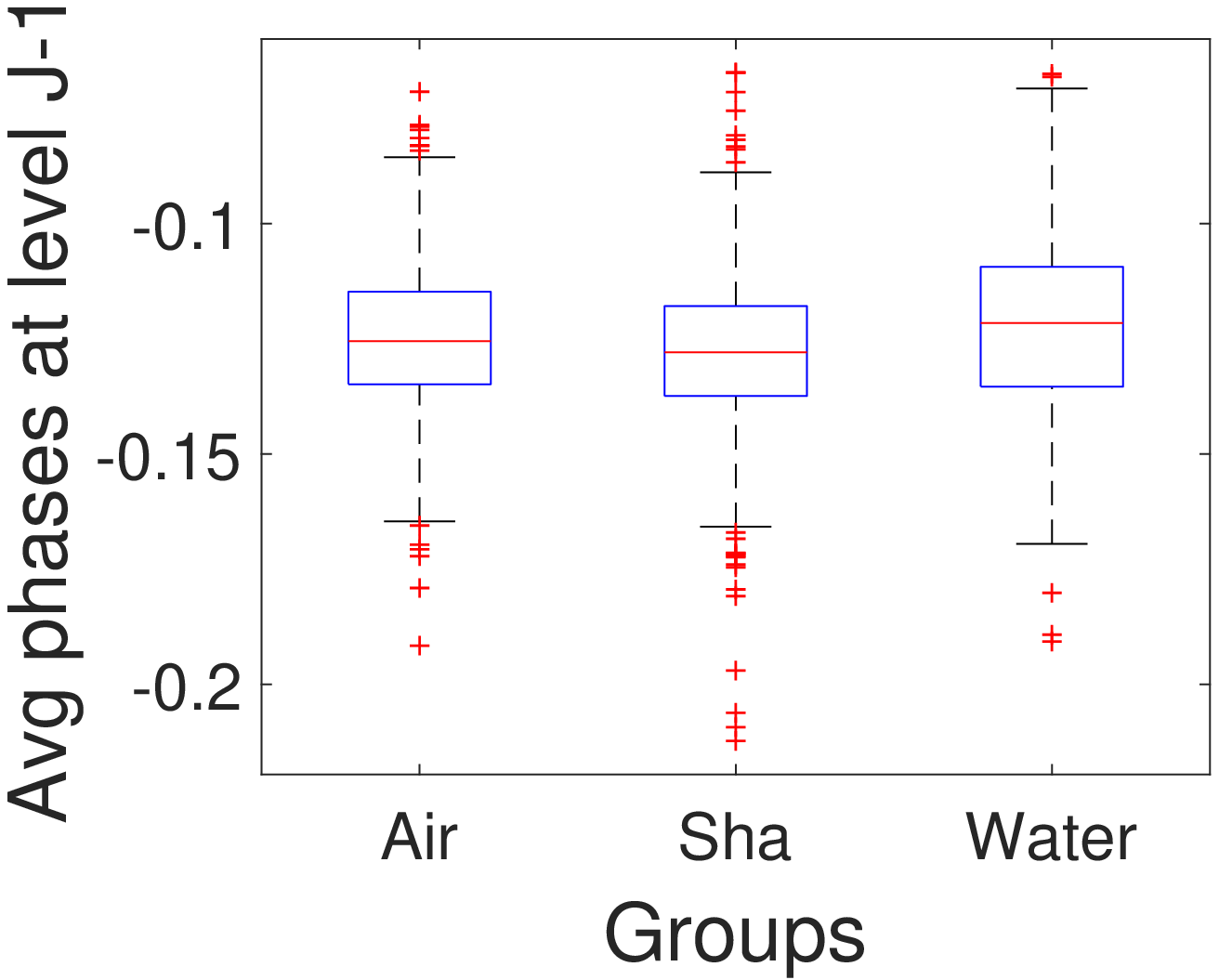}} \qquad
  \subfigure[]{\includegraphics[width=1.5in, height=1.5in]{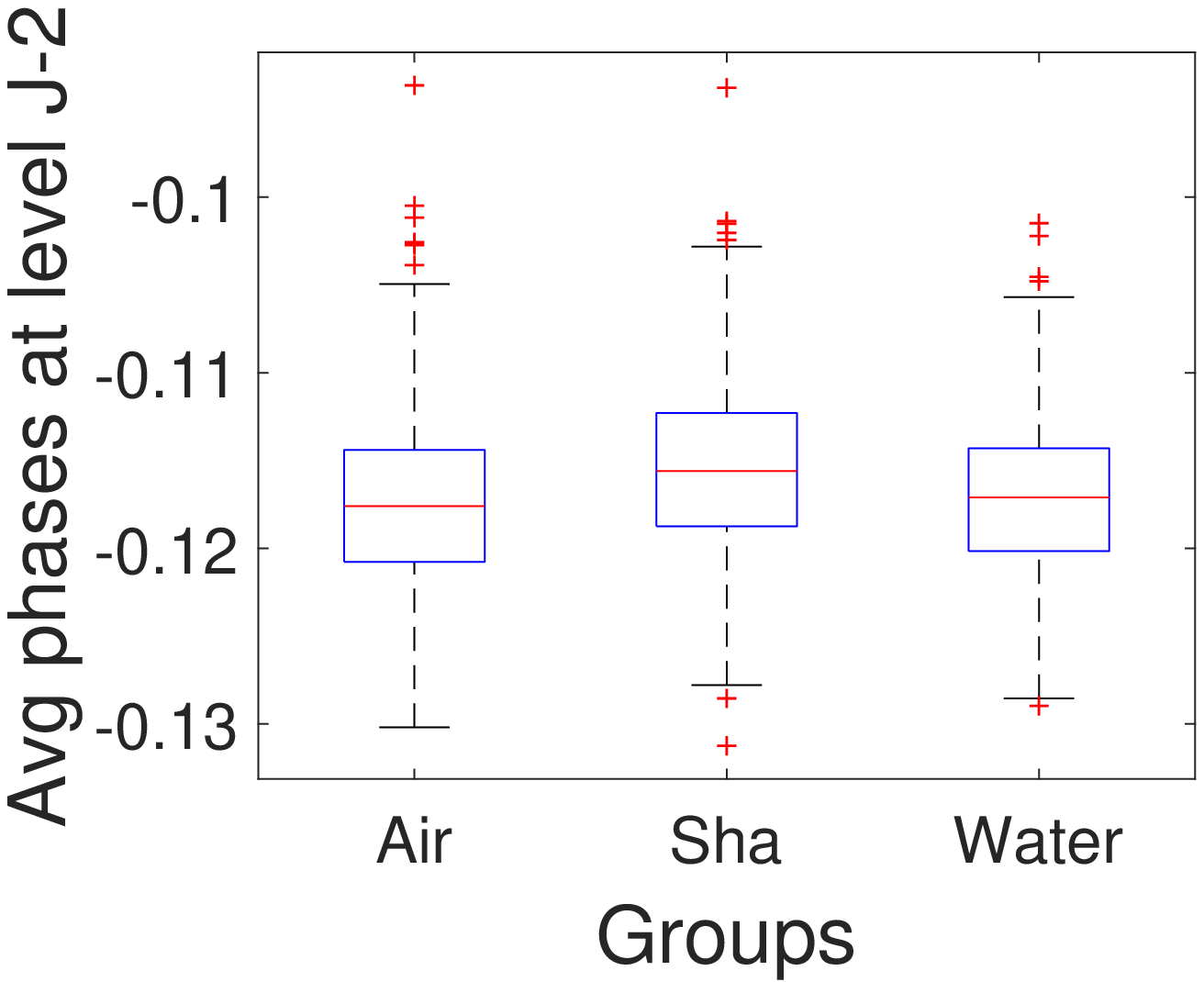}}  \qquad
  \subfigure[]{\includegraphics[width=1.5in, height=1.5in]{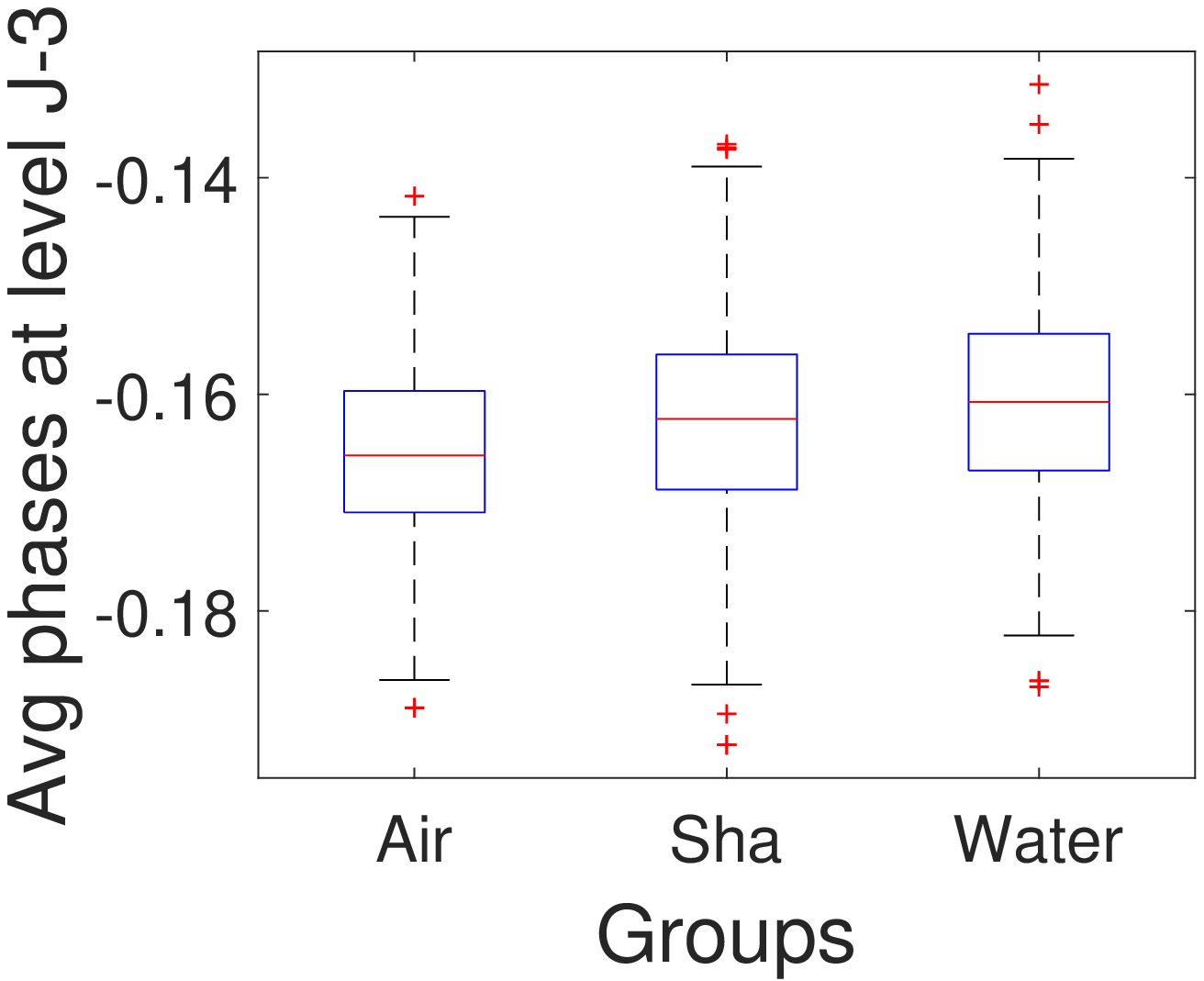}} \\
  \subfigure[]{\includegraphics[width=1.5in, height=1.5in]{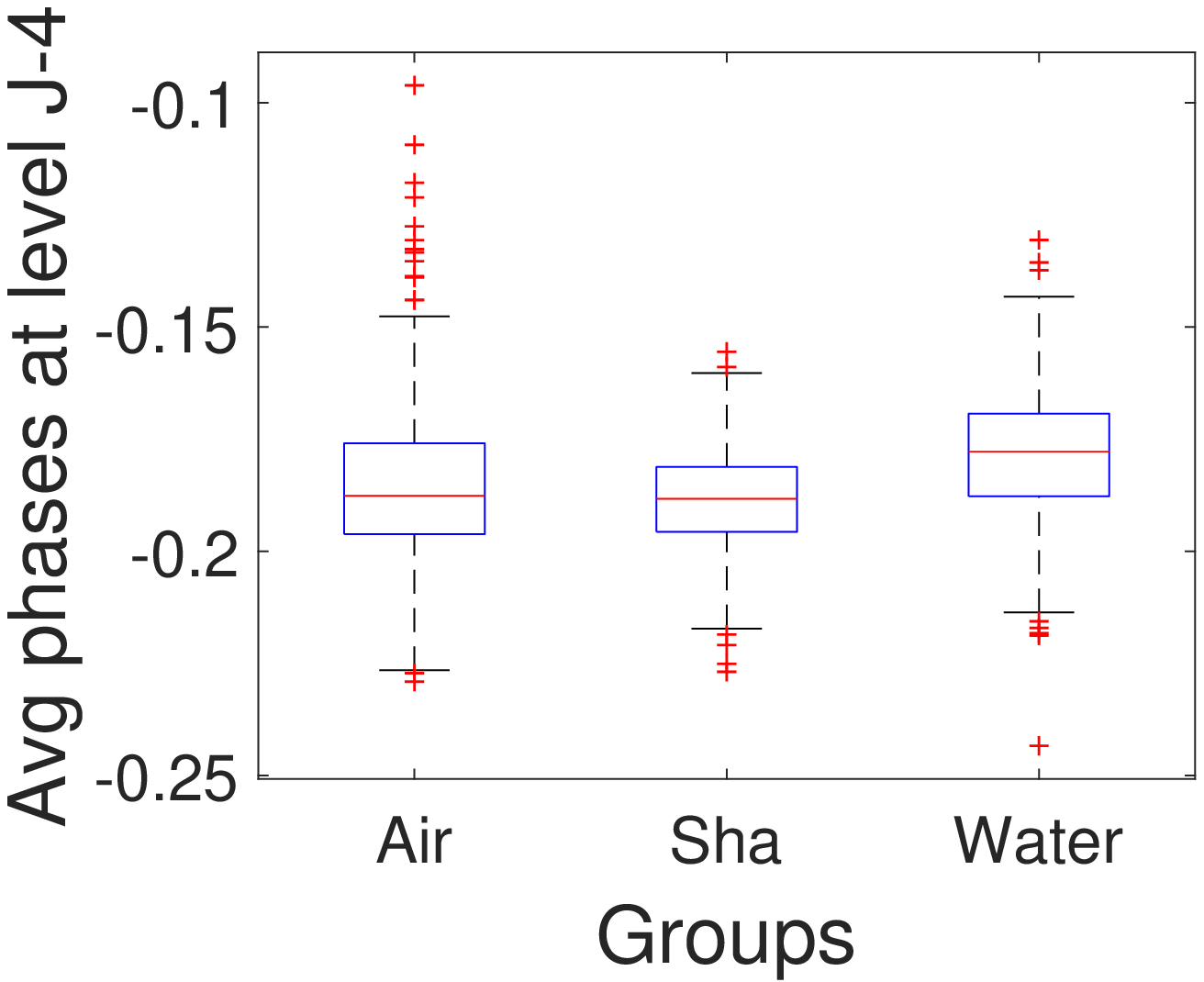}} \qquad
  \subfigure[]{\includegraphics[width=1.5in, height=1.5in]{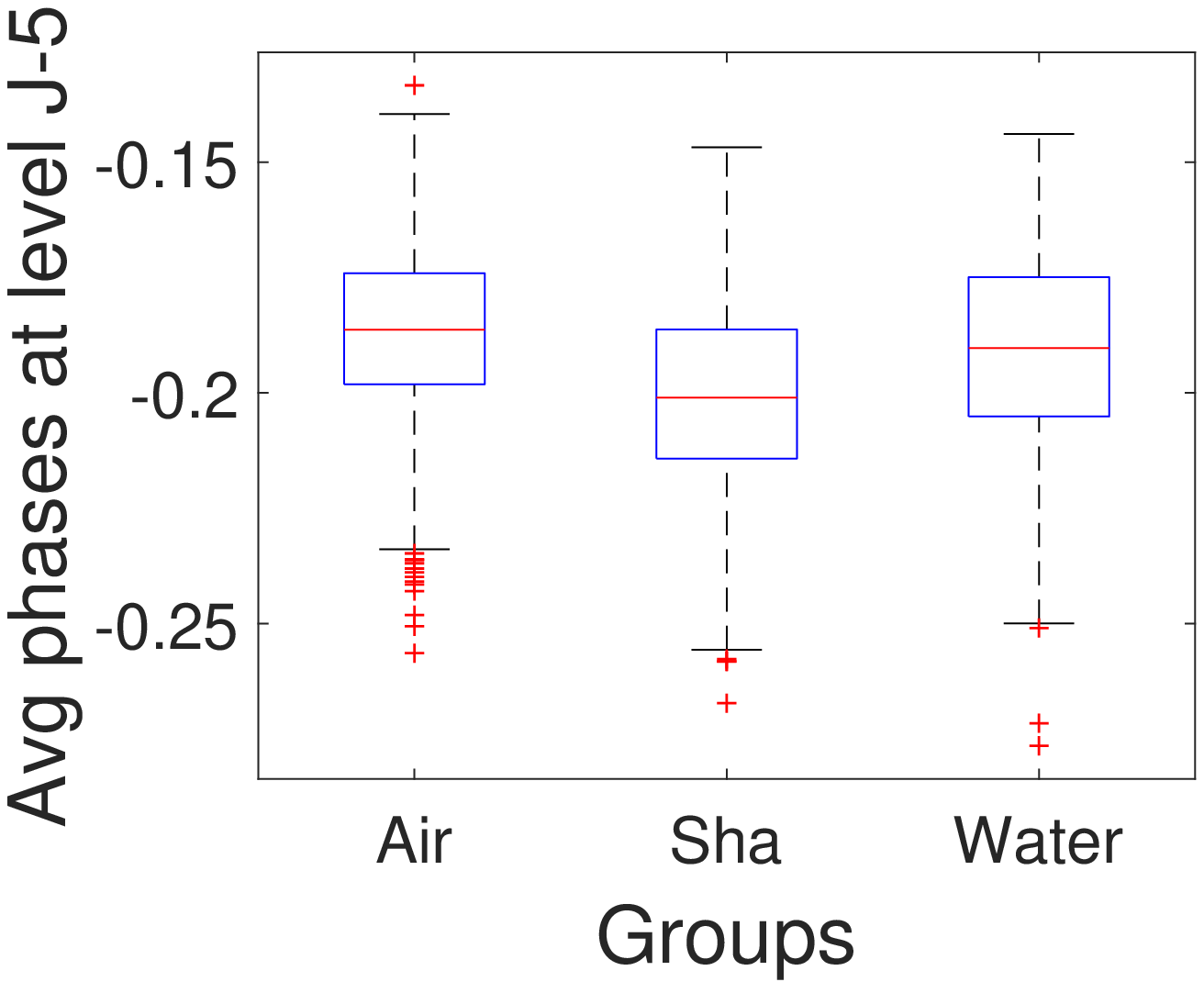}}  \qquad
  \subfigure[]{\includegraphics[width=1.5in, height=1.5in]{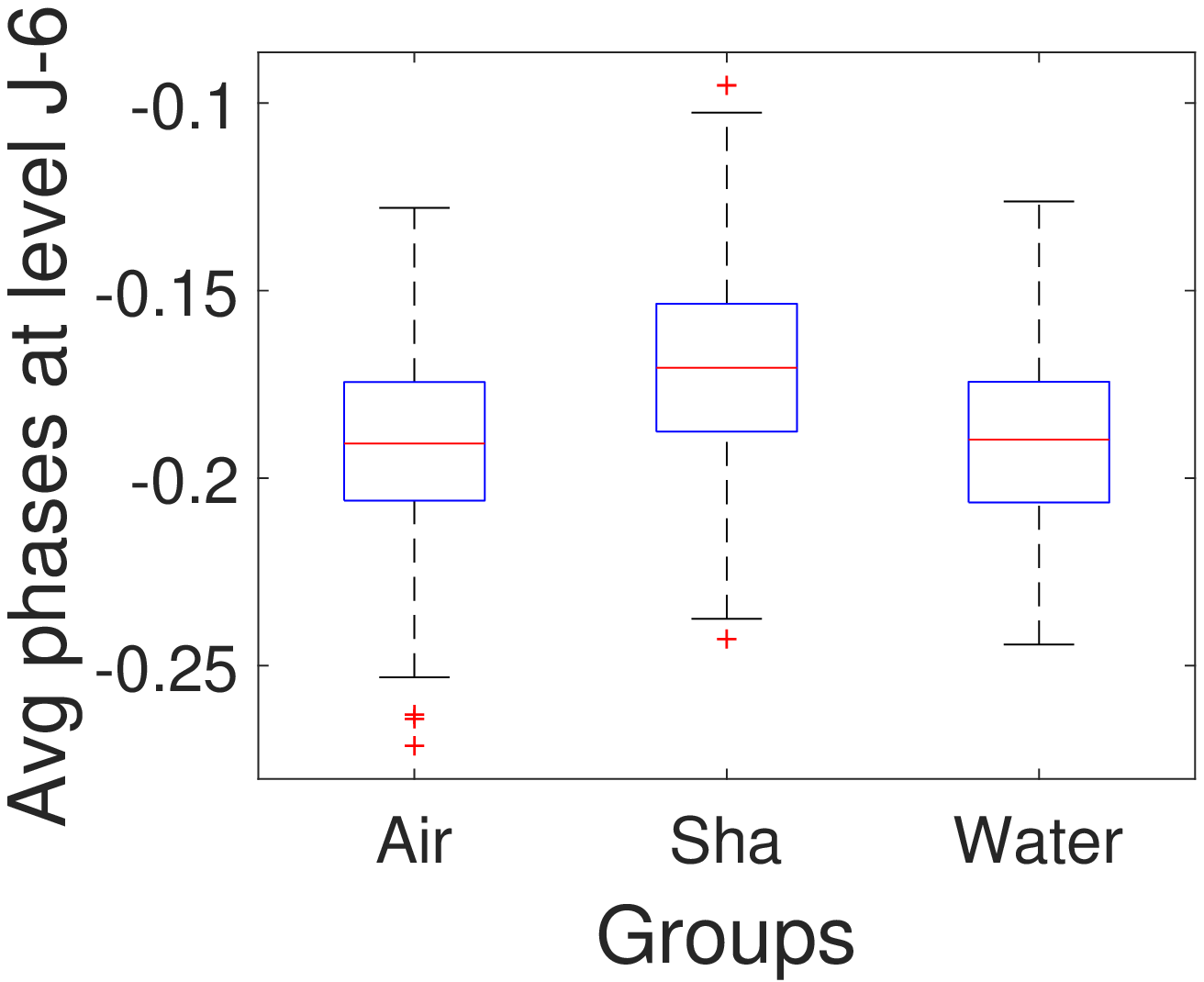}} \\
  \subfigure[]{\includegraphics[width=1.5in, height=1.5in]{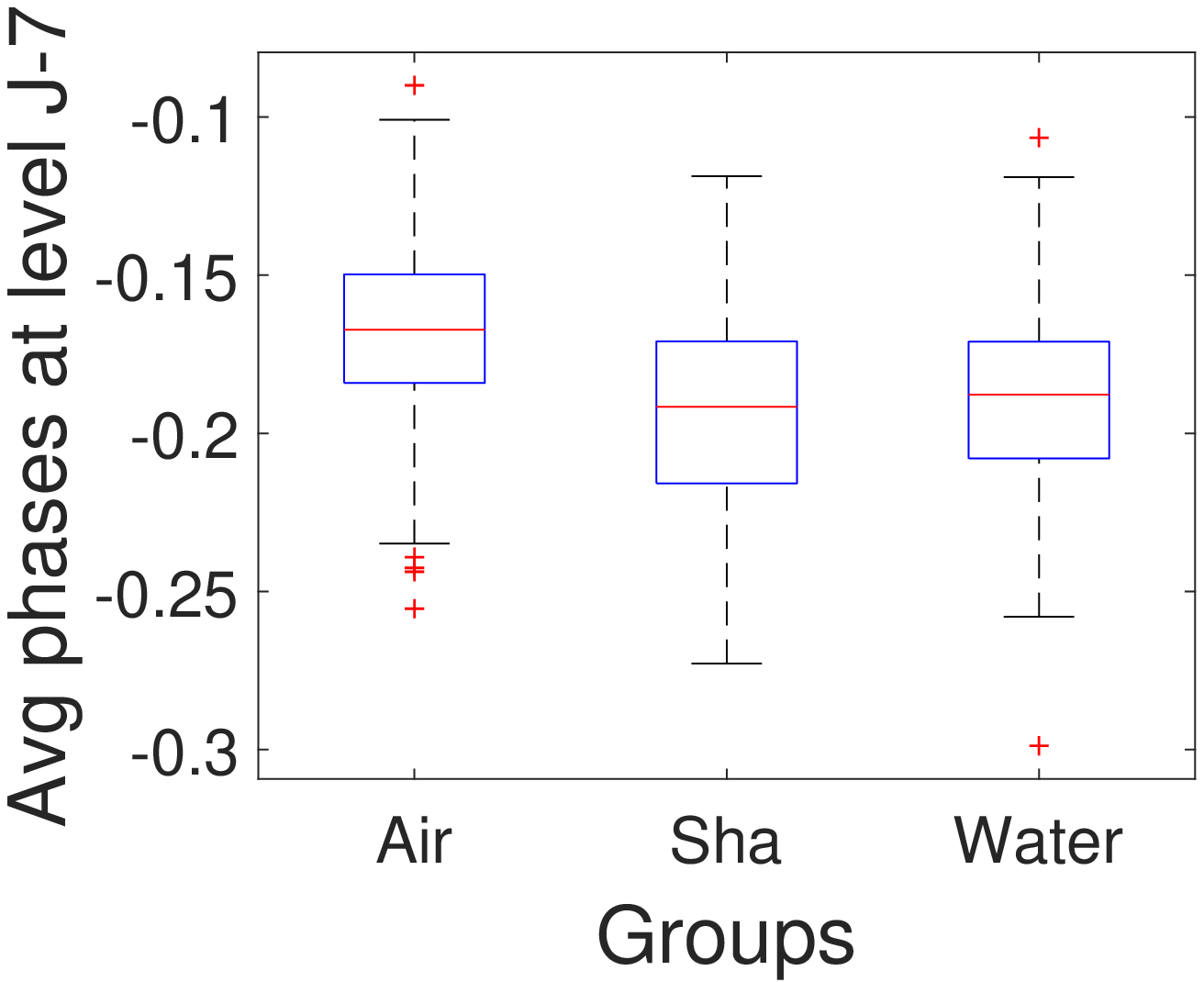}} \qquad
  \subfigure[]{\includegraphics[width=1.5in, height=1.5in]{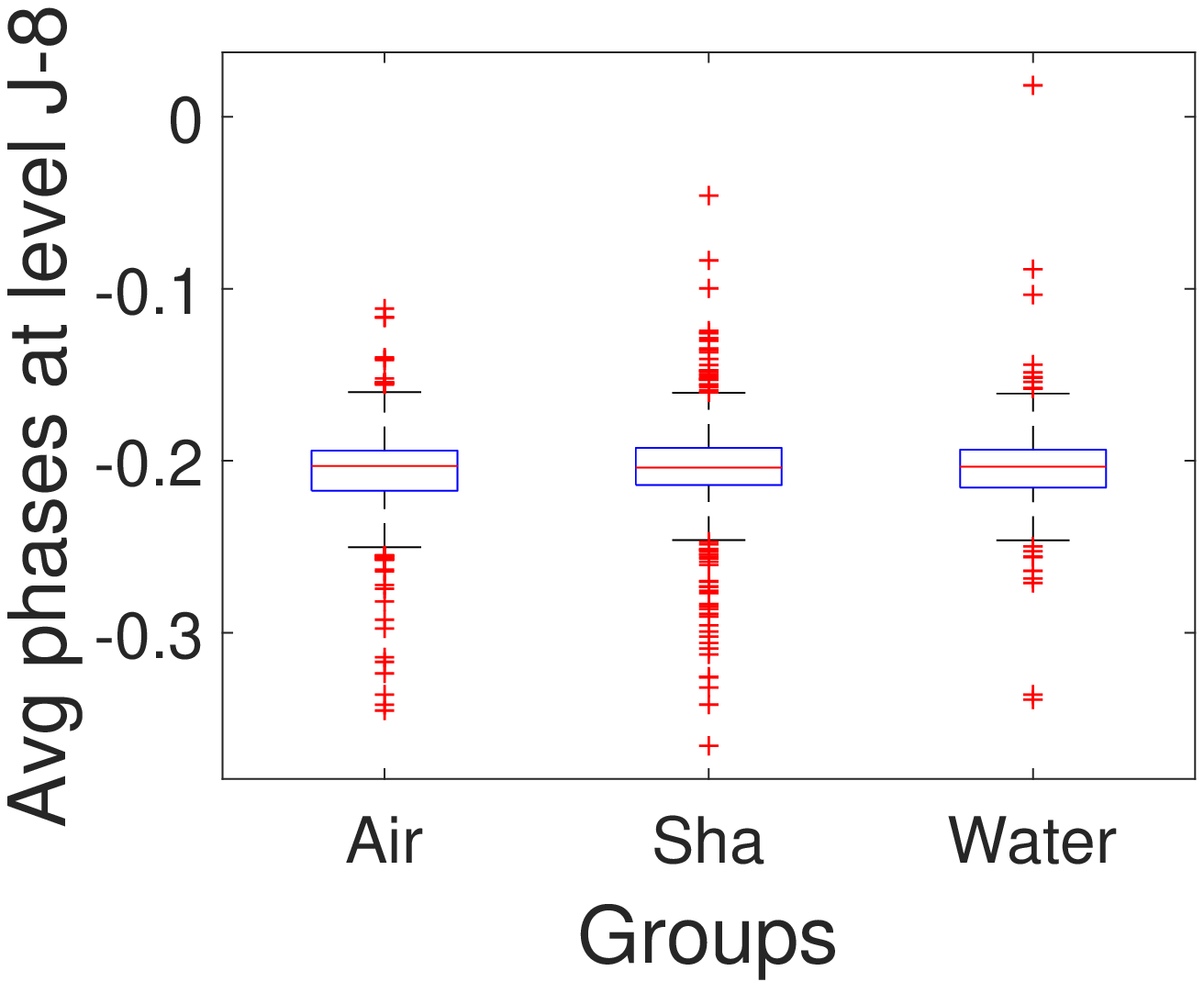}}  \qquad
  \subfigure[]{\includegraphics[width=1.5in, height=1.5in]{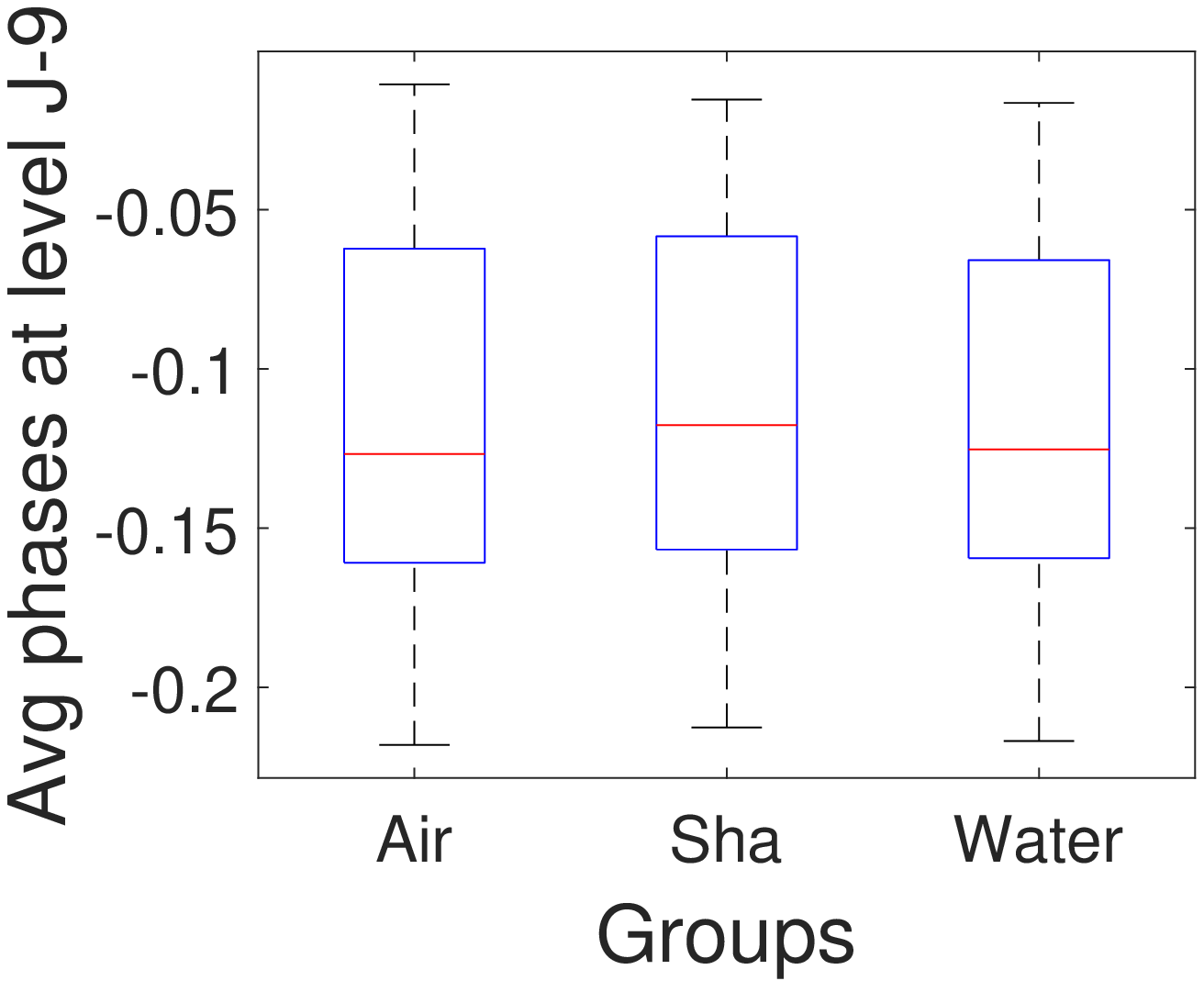}} \\

  \caption{Box plots of averages of phase $\theta$ at all multiresolution levels.}
  \label{boxfig:soundphase2NDQ}
\end{figure}

\begin{figure}[h!tb]
  \centering
  \subfigure[]{\includegraphics[width=1.5in, height=1.5in]{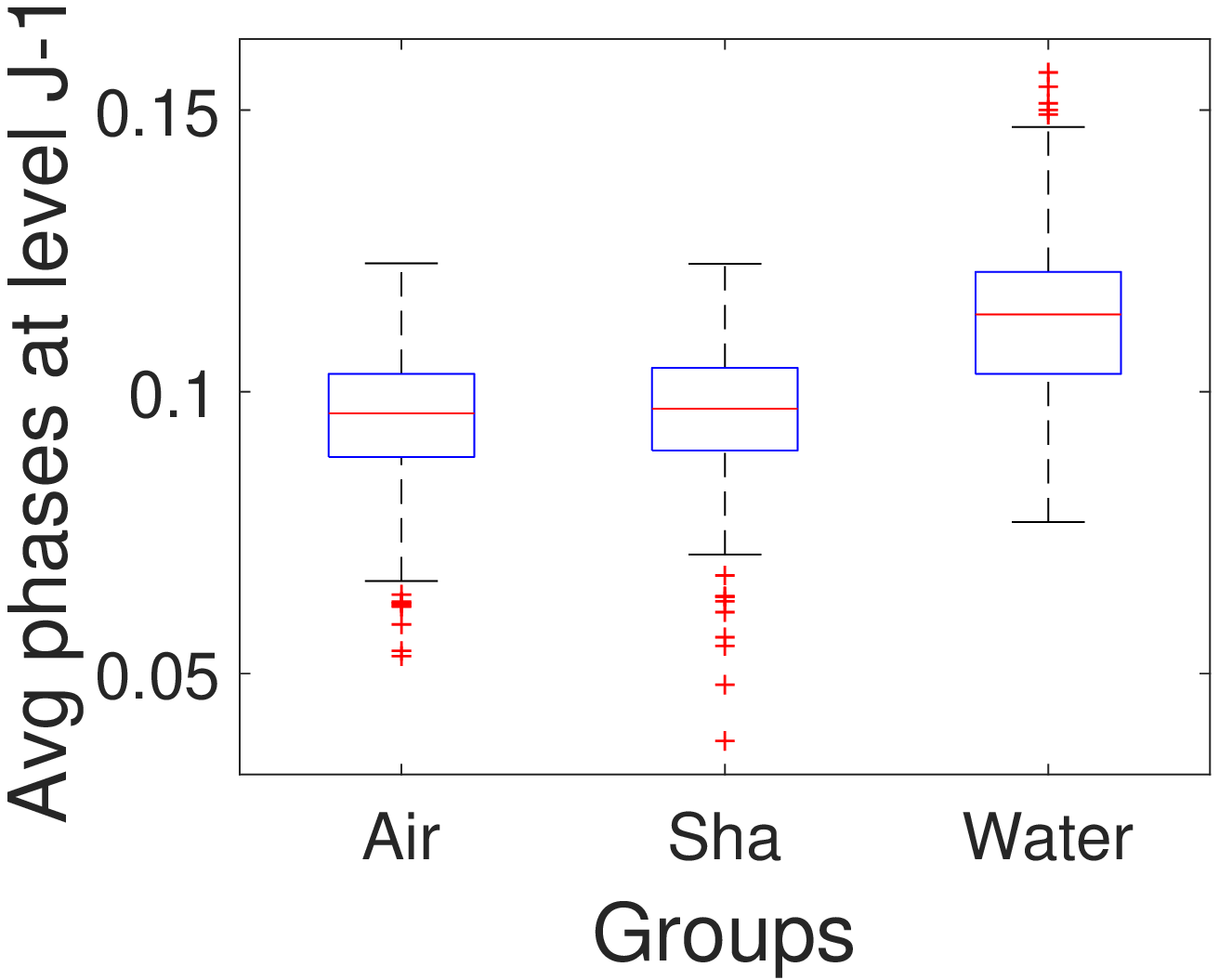}} \qquad
  \subfigure[]{\includegraphics[width=1.5in, height=1.5in]{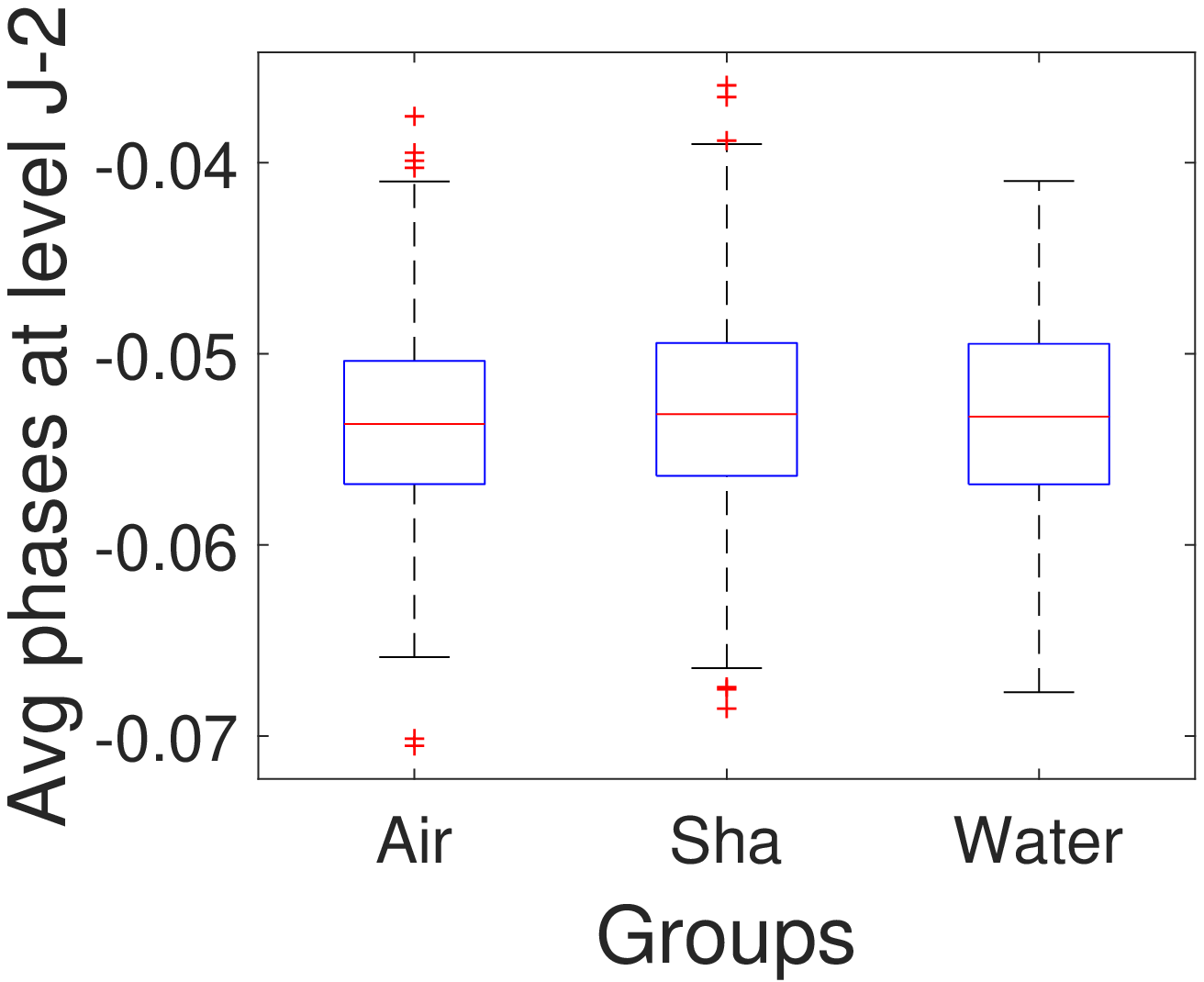}}  \qquad
  \subfigure[]{\includegraphics[width=1.5in, height=1.5in]{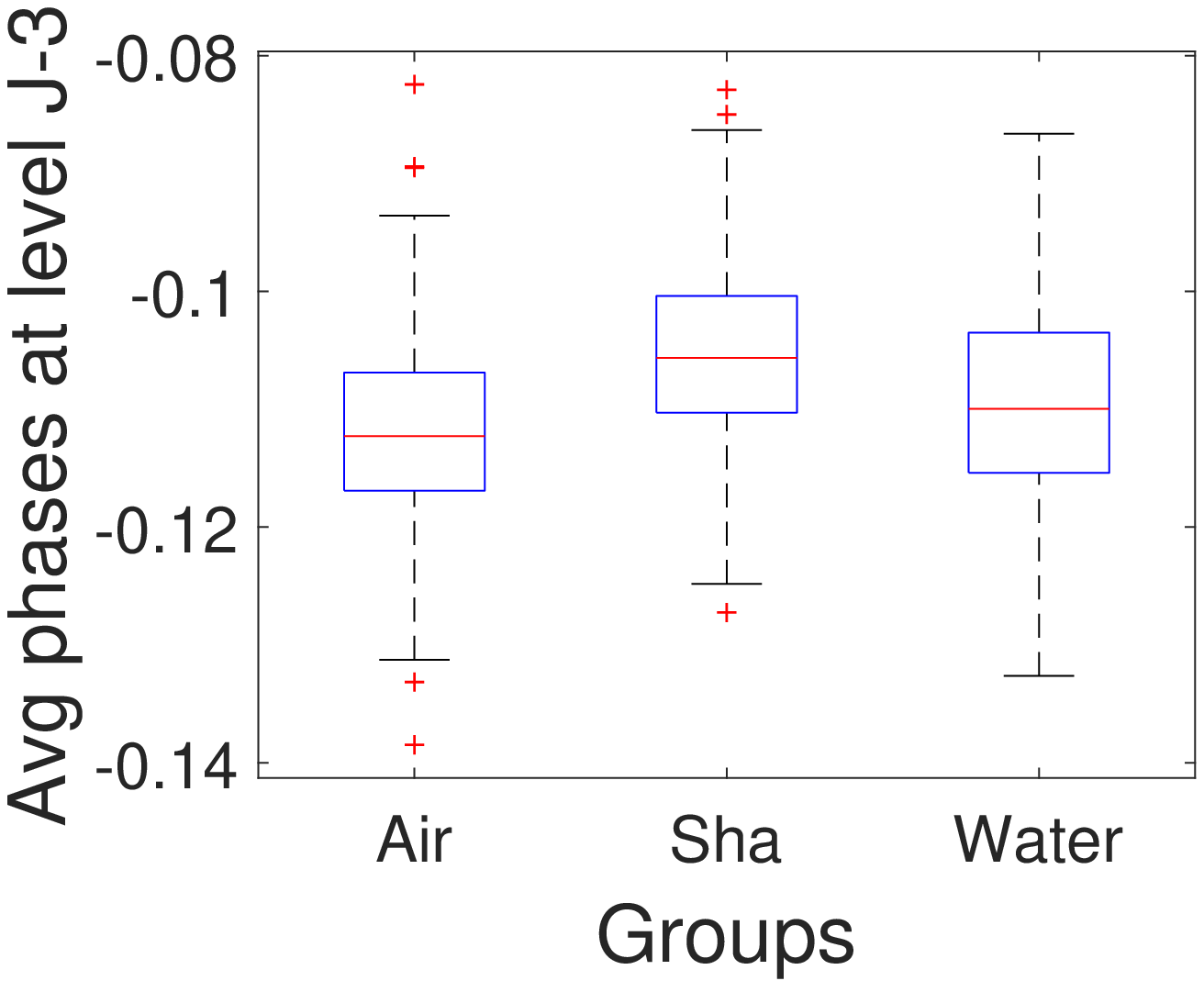}} \\
  \subfigure[]{\includegraphics[width=1.5in, height=1.5in]{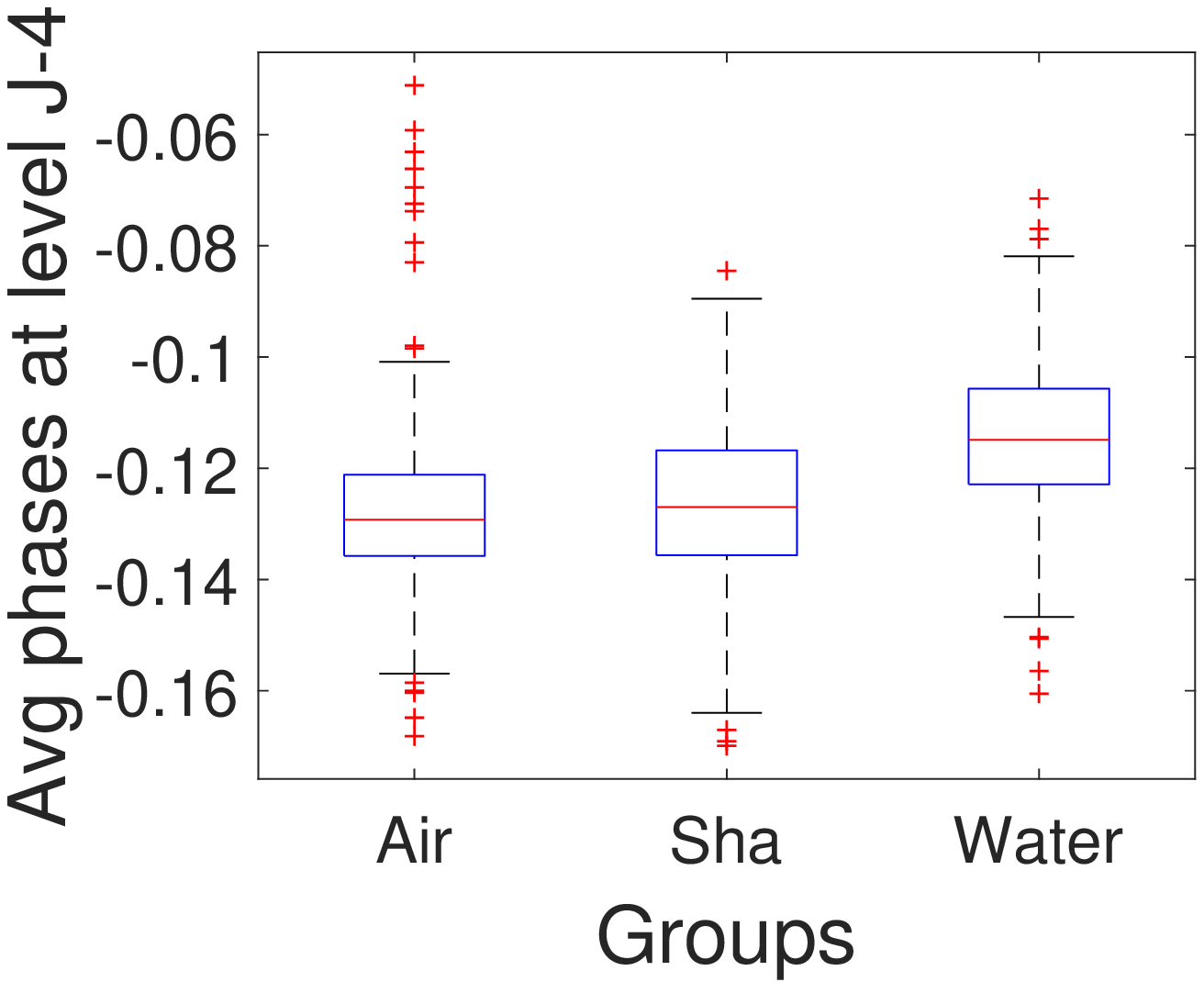}} \qquad
  \subfigure[]{\includegraphics[width=1.5in, height=1.5in]{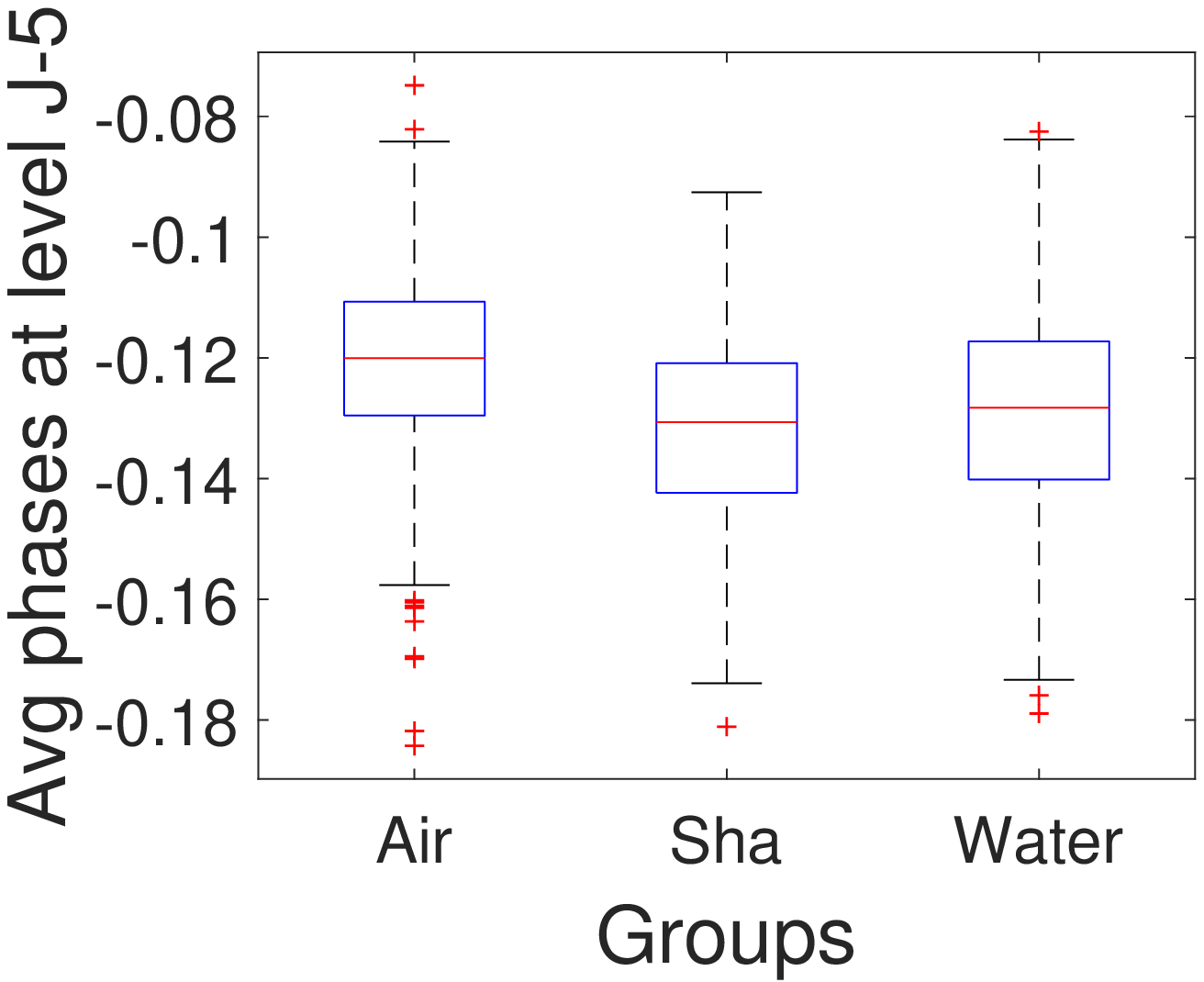}}  \qquad
  \subfigure[]{\includegraphics[width=1.5in, height=1.5in]{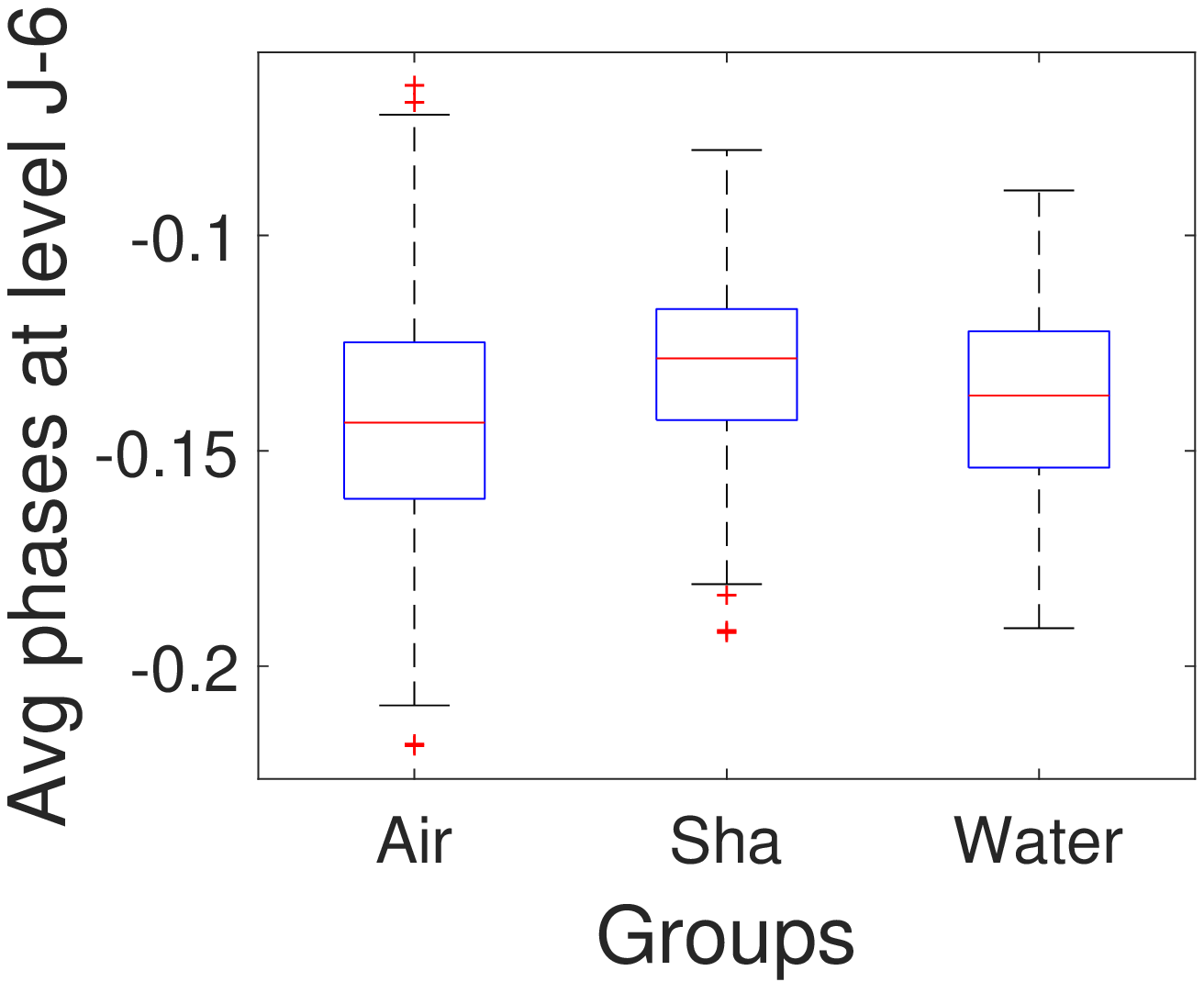}} \\
  \subfigure[]{\includegraphics[width=1.5in, height=1.5in]{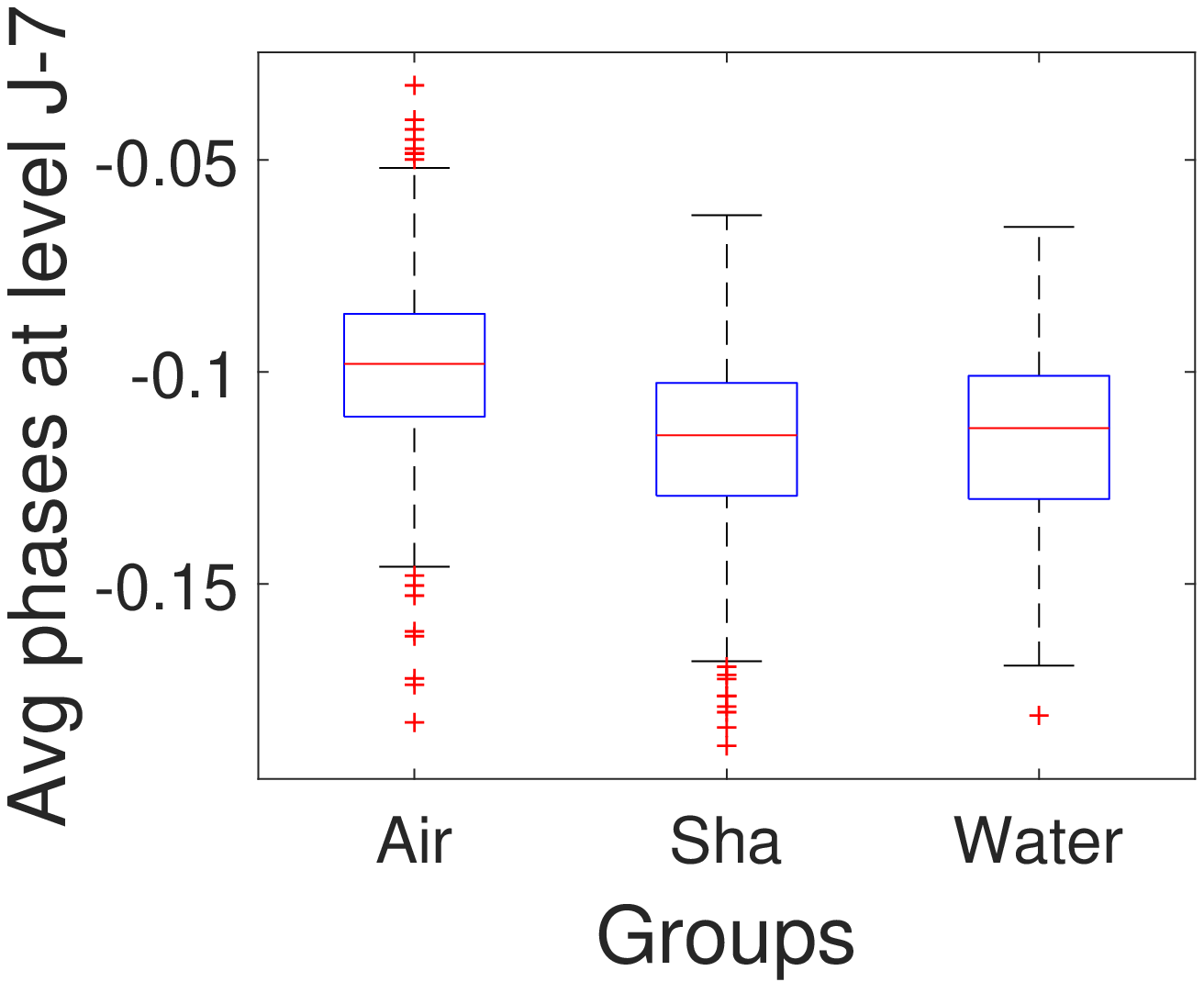}} \qquad
  \subfigure[]{\includegraphics[width=1.5in, height=1.5in]{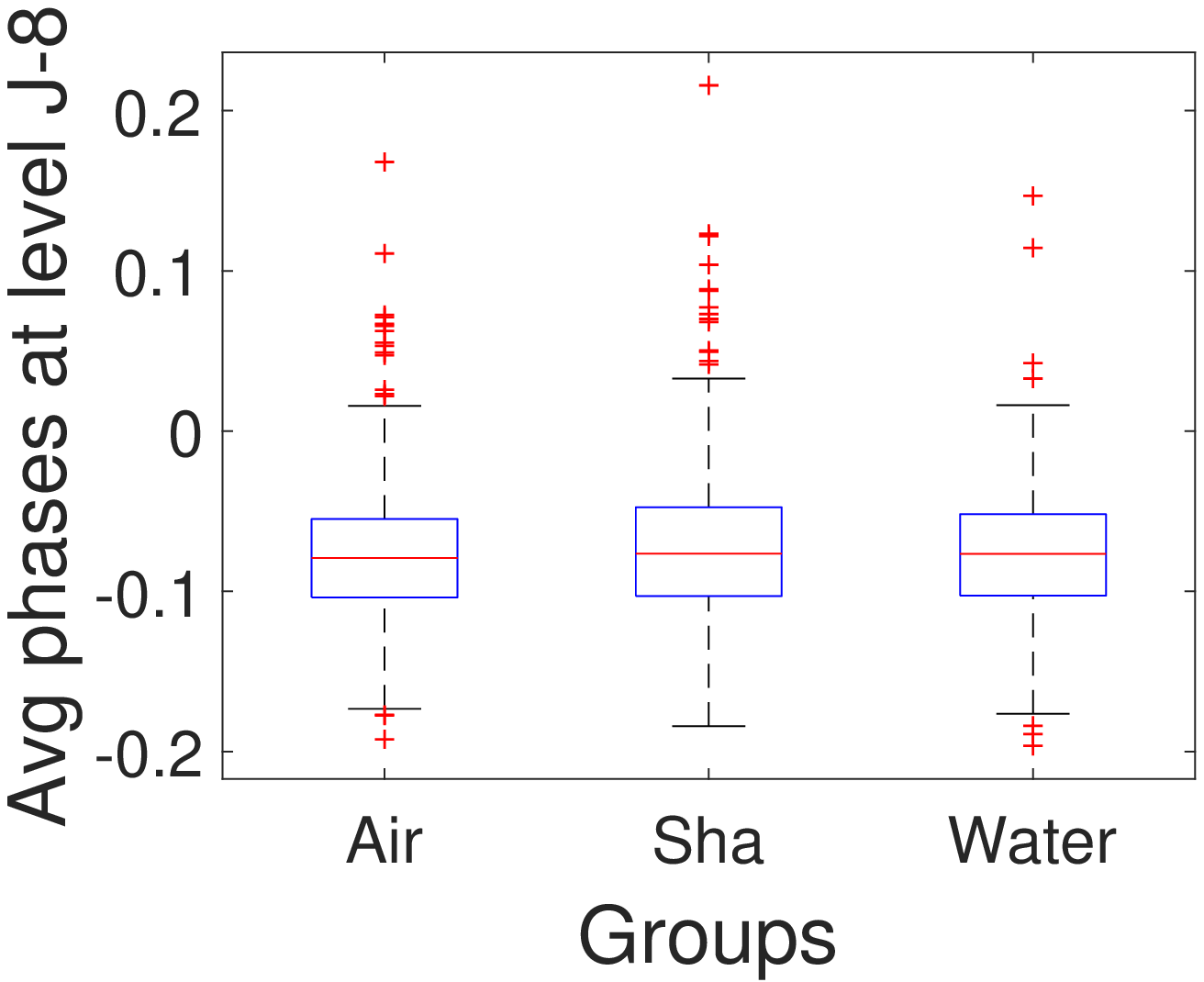}}  \qquad
  \subfigure[]{\includegraphics[width=1.5in, height=1.5in]{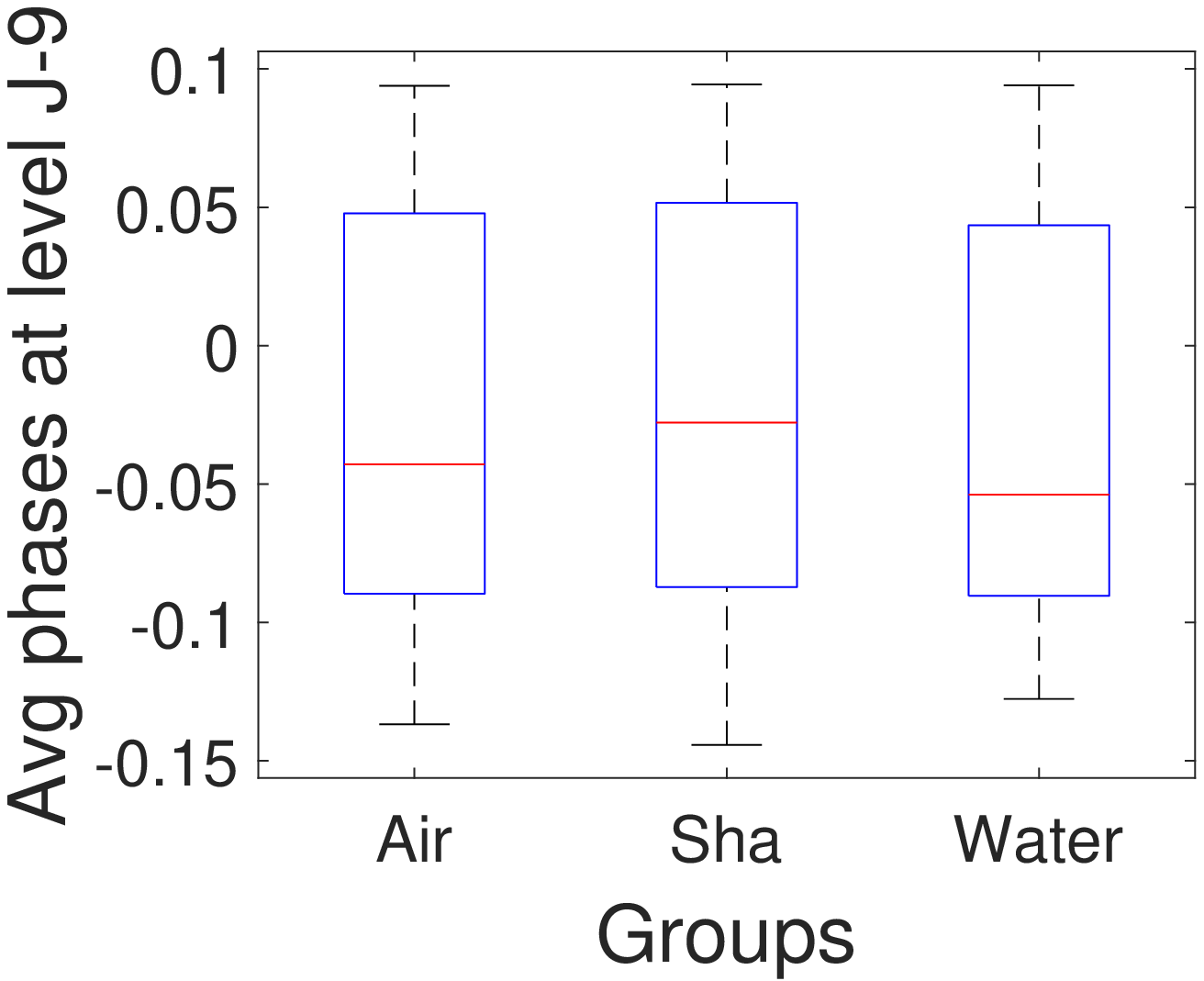}} \\

  \caption{Box plots of averages of phase $\psi$ at all multiresolution levels.}
  \label{boxfig:soundphase3NDQ}
\end{figure}

Next, we calculated a spectral slope of wavelet spectra discussed in Section \ref{sec-NDQwavespec} and found averages of the three phases at all level $j=J_0, \dots, J-1$ defined in Equation (\ref{avgphase}), to use them as supplementary variables in classification analysis.
Box plots of averages of three phases ($\phi, \theta, \psi$) at all multiresolution levels are displayed in Figure \ref{boxfig:soundphase1NDQ}, \ref{boxfig:soundphase2NDQ}, and \ref{boxfig:soundphase3NDQ}.

After distilling the summaries, we chose gradient boosting to classify the sound signals. Random forest, k-NN with $k=1$, and SVM are also considered. The gradient boosting consistently outperformed the rest. In simulations, we randomly split the dataset into $75\%$ part as training set, and take the remaining $25\%$ part as testing set. This random partition to training and testing sets was repeated $1,000$ times, thus, the provided performance measures were averaged over $1,000$ runs.

\subsubsection{Results}{\label{sec-soundresult}}

The performance of NDQWT in terms of overall accuracy and sensitivities of the three groups is shown in Table \ref{soundtable}.
For DWT and NDWT, Haar filter was used.
We denoted the average of phases from $\text{NDWT}_\text{\large{c}}$ as $\angle d_{j},$ and the averages of three phases from NDQWT as $\phi_{j}, \theta_{j}, \psi_{j}$.

For convenience, the considered methods are numbered from $1$ to $15$ according to transformation method and features utilized.
As expected, $\text{WT}_\text{\large{c}}$ ($4$th) is better than DWT ($1$th) and QWT ($7$th) is superior to $\text{WT}_\text{\large{c}}$ ($4$th) because they produce wavelet coefficients in a more redundant domain.
Also, their corresponding non-decimated versions, NDWT ($8$th), $\text{NDWT}_\text{\large{c}}$ ($11$th), and NDQWT ($14$th), show the same trend and better performance compared to decimated versions, except for the NDWT because of the compound effects of componential and structural redundancies.
Generally, all methods using slopes show similar and inadequate performance while the methods that in addition use phase-based statistics perform better.
In particular, classifiers without quaternion-based phase information tend to show low performance in classifying the water sound.
In this context, we can conclude that discriminatory information characteristic for the water sound is located in the phases of quaternion-valued wavelet coefficients.
This, of course, may not be case for arbitrary data, but the results in this case validated the use of quaternion-based phase information.

In conclusion, comparing the $7$th and $8$th with $14$th methods, we can observe that the proposed NDQWT dominates both QWT and NDWT by its compounding positive effect of the componential and structural redundancies.
Note that the best performance is achieved if we utilize all features from the $\text{NDWT}_\text{\large{c}}$ and NDQWT together in one integrated model as $15$th, which is slightly better than the NDQWT alone.

To compare computational costs, we recorded computing times for all considered versions of wavelets (DWT, $\text{WT}_\text{\large{c}}$, QWT, NDWT, $\text{NDWT}_\text{\large{c}}$, and NDQWT)
needed to transform a single signal of length 1024.
As expected, the computation times are proportional to the overall accuracies: more accurate results take longer to calculate.

\begin{table}[h!tb]
\begin{center}
\begin{adjustbox}{max width=\textwidth}

\begin{tabular}{c|c|c|c|ccc|c}
  \specialrule{1.3pt}{1pt}{1pt}
   Order & Transform & Features & \begin{tabular}{@{}c@{}} Overall \\ Accuracy rate\end{tabular}
 & \begin{tabular}{@{}c@{}}Sensitivity \\ Air\end{tabular}
 & \begin{tabular}{@{}c@{}}Sensitivity \\ Sha\end{tabular} & \begin{tabular}{@{}c@{}}Sensitivity \\ Water\end{tabular} & \begin{tabular}{@{}c@{}}Computing \\ Time  \end{tabular} \\\hline \hline
   $1$st & DWT & Slope  &   0.3841	&0.3324	&0.4918&	0.1846 & 0.015s \\ \hline
  $2$nd & $\text{WT}_\text{\large{c}}$ & Slope  & 0.3993&	0.3624	&0.5221	&0.1387  &\\
   $3$rd & & $\angle d_j$ &  0.4014	&0.3767	&0.5002	&0.1787  & 0.021s \\
   $4$th & &  Slope + $\angle d_j$ & 0.4236	&0.3969	&0.5337	&0.1641  &\\ \hline
  $5$th & QWT & Slope  &   0.3947	&0.3416	&0.5214&	0.1423 &\\
  $6$th &  & $\phi_{j}$ + $\theta_{j}$ + $\psi_{j}$  & 0.6539	&0.6208&	0.7377	&0.4881  & 0.030s \\
  $7$th & & Slope + $\phi_{j}$ + $\theta_{j}$ + $\psi_{j}$ & 0.6580	&0.6294&	0.7348&	0.5037 & \\ \hline
  $8$th & NDWT & Slope  &  0.3968	&0.3565&	0.5040	&0.1781 & 0.019s \\ \hline
  $9$th & $\text{NDWT}_\text{\large{c}}$ & Slope  &  0.4026	&0.3618	&0.5085&	0.1962  &\\
  $10$th &  & $\angle d_j$ &0.6536	&0.7088	&0.7422	&0.2898  & 0.027s \\
  $11$th & & Slope + $\angle d_j$ & 0.6614&	0.7006	&0.7504	&0.3213 &\\ \hline
  $12$th & NDQWT & Slope  & 0.3789&	0.3442&	0.4829	&0.1606 &  \\
  $13$th &  & $\phi_{j}$ + $\theta_{j}$ + $\psi_{j}$  & 0.7558&	0.7585	&0.8096	&0.5957   &0.041s\\
  $14$th & & Slope + $\phi_{j}$ + $\theta_{j}$ + $\psi_{j}$ & 0.7564&	0.7541	&0.8199	&0.5860   &\\ \hline
  $15$th &  & $11$th + $14$th & \textbf{0.7822}&	\textbf{0.7895}	&\textbf{0.8407}	&\textbf{0.5997} & 0.068s \\
  \specialrule{1.3pt}{1pt}{1pt}
\end{tabular}

\end{adjustbox}
\end{center}
\caption{Gradient boosting classification results. Total 15 methods are compared and the best result is obtained by the $15$th method.}\label{soundtable}
\end{table}

\subsection{Application in Seam Detection in Steel Rolling Process}{\label{sec-rolling}}

In various manufacturing fields, fault diagnosis or anomaly detection in 2-D images have been widely deployed thanks to their low implementation cost and the rich information from a high acquisition rate of image sensors. Thus, considerable research  has been conducted on inspection systems for rolling process \citep{Jin2004}, composite material fabrication \citep{Sohn2004}, liquid crystal display manufacturing \citep{Jiang2005}, fabric and textile manufacturing \citep{Kumar2008}, and so on.
In these systems, snapshots of  particular products or parts are obtained during production process, to be analyzed as digital images for detecting defects or anomalies.

\begin{figure}[!ht]
\centering
\includegraphics [width=4.5in, height=1.5in]{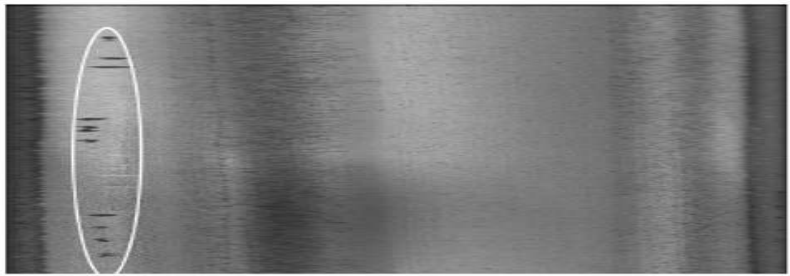}
\caption{An example of surface image of steel rolling bar. The white ellipse indicates seem defects in rolling process.}
\label{exrolling}
\end{figure}

A representative example is an inspection system for the subsequent rolling process, continuous casting manufacturing, in which a semi-finished billet is solidified from molten metal. Specifically, rolling process is a high-speed deformation by consistent diameters  between sets of rollers to decrease the cross-sectional size in a long steel bar by applying compressive forces. One of surface defects caused during the rolling process is a seam defect that leads to stress concentration on the bulk material, which further may create failures when a steel bar is used. Thus, timely detection of such anomaly is significant to preventing such failures and for reducing overall manufacturing costs.
For this purpose, advanced vision sensing systems have been developed in rolling processes to obtain high-resolution snapshots of the product surface with a high data acquisition rate of billets at short time intervals.
An example of the bar surface image with seam defects of rolling process is shown in Figure \ref{exrolling}.

Until recently a quality inspection or anomaly detection had been performed manually.
However, automatic inspection systems with high speed and accuracy have been developed, since machine learning is utilized to analyze images.
Since the seam defects are typically sparse as shown in Figure \ref{exrolling}, the inherent self-similarity of the background is disturbed by the anomaly, that can be sensed well by wavelet tools.
The proposed wavelet-based method considered here should not be used in practice alone.
This should be added to a battery of other standard image recognition tools based on the image features in the original domain.

To illustrate the power of the proposed method, we set a classification problem with a test set of surface images and compare classification results in following sections.

\subsubsection{Description of Data}{\label{sec-data3}}
One hundred surface images of size of $128 \times 512$ pixels of a rolling bar are sequentially collected.
In this data set, the images generally appear smooth in the rolling (vertical) direction. 
Seam defects that occur typically towards the end of the rolling bar started to appear from 76th image, so 71th to 80th images are omitted for an objective experiment. We define the first 70 images as controls without seam defects and the last 20 images as case samples with seam defects.

\subsubsection{Classification}
\begin{figure}[h!tb]
  \centering
  \subfigure[]{\includegraphics[width=1.5in, height=1.5in]{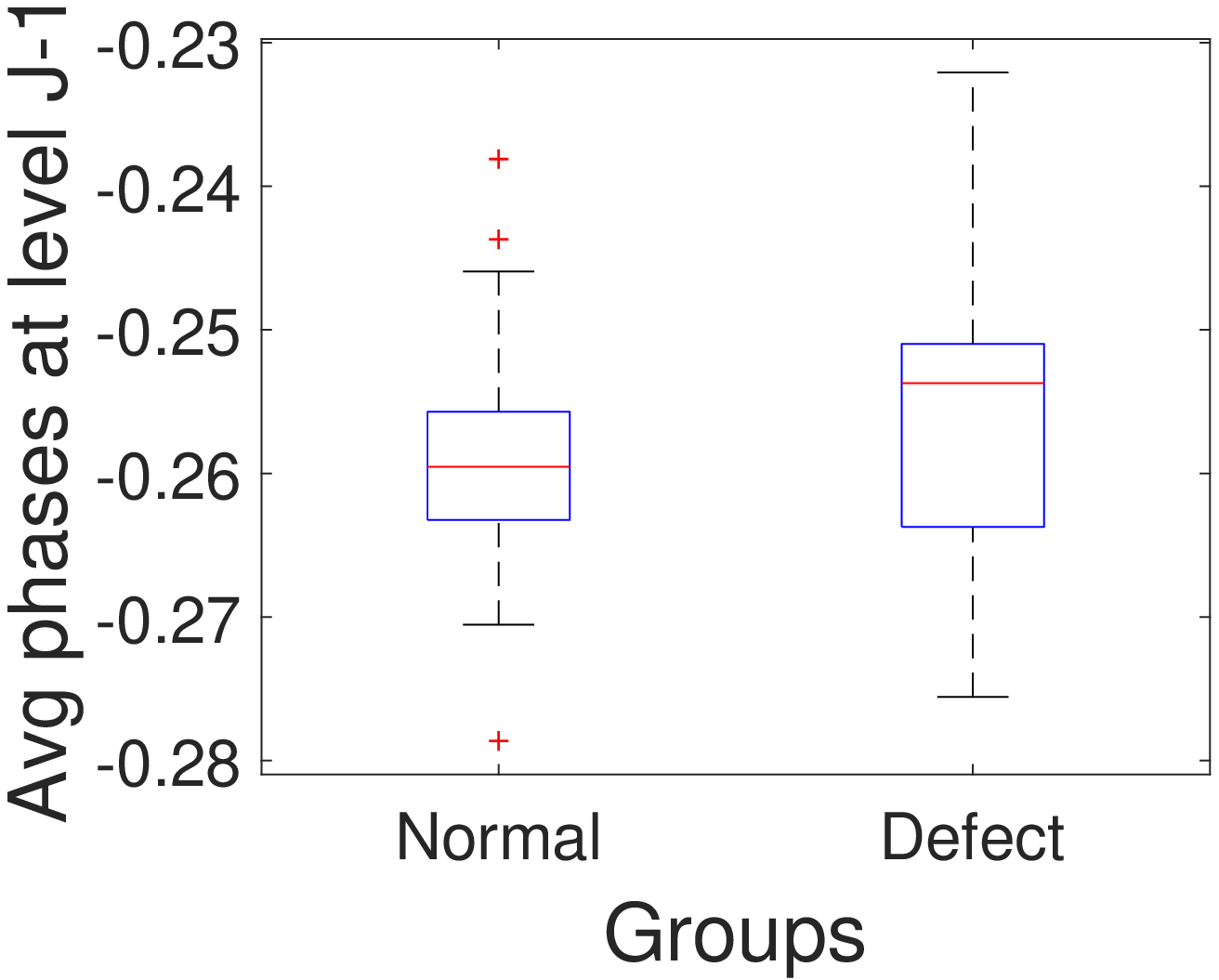}} \qquad
  \subfigure[]{\includegraphics[width=1.5in, height=1.5in]{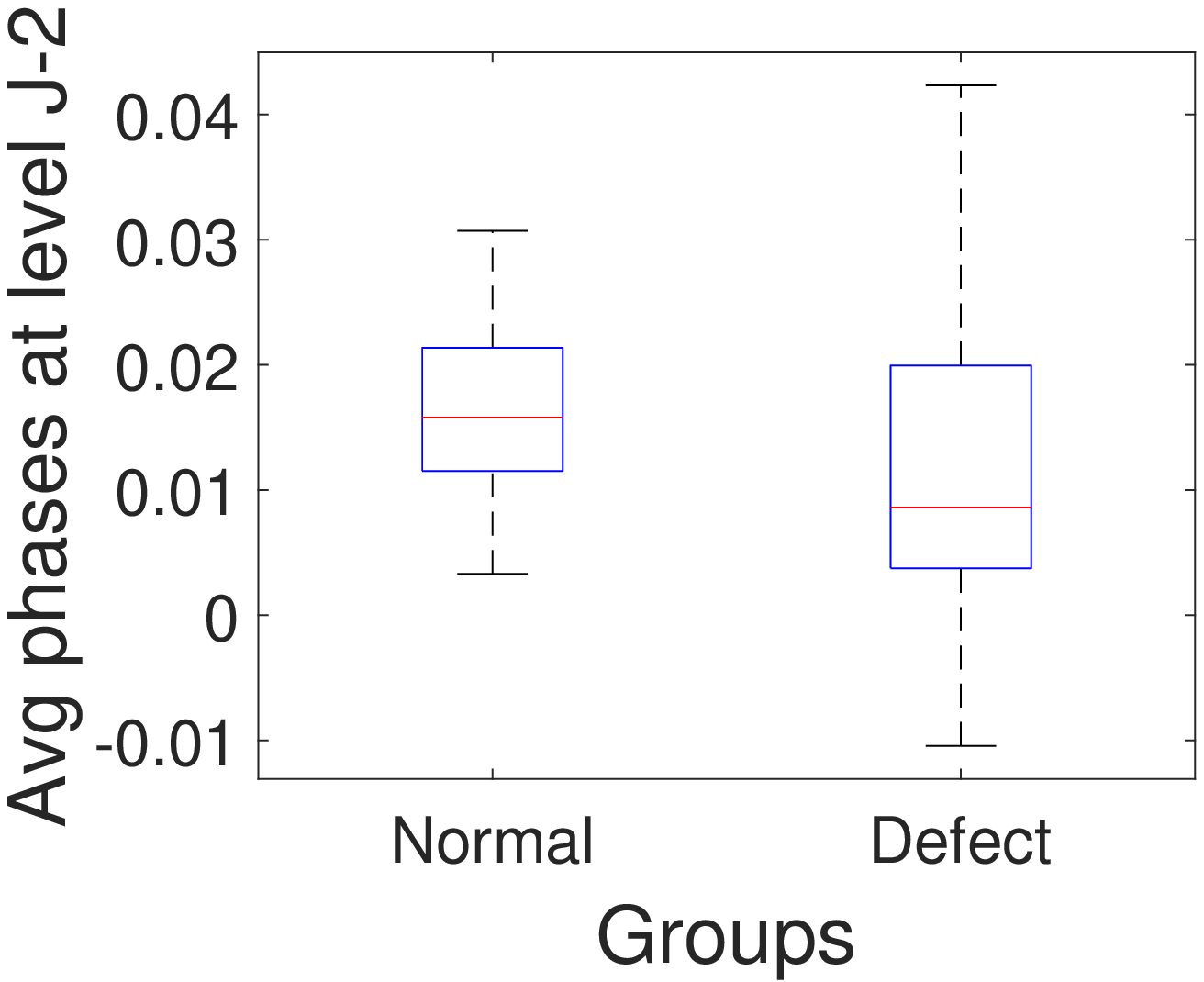}}  \qquad
  \subfigure[]{\includegraphics[width=1.5in, height=1.5in]{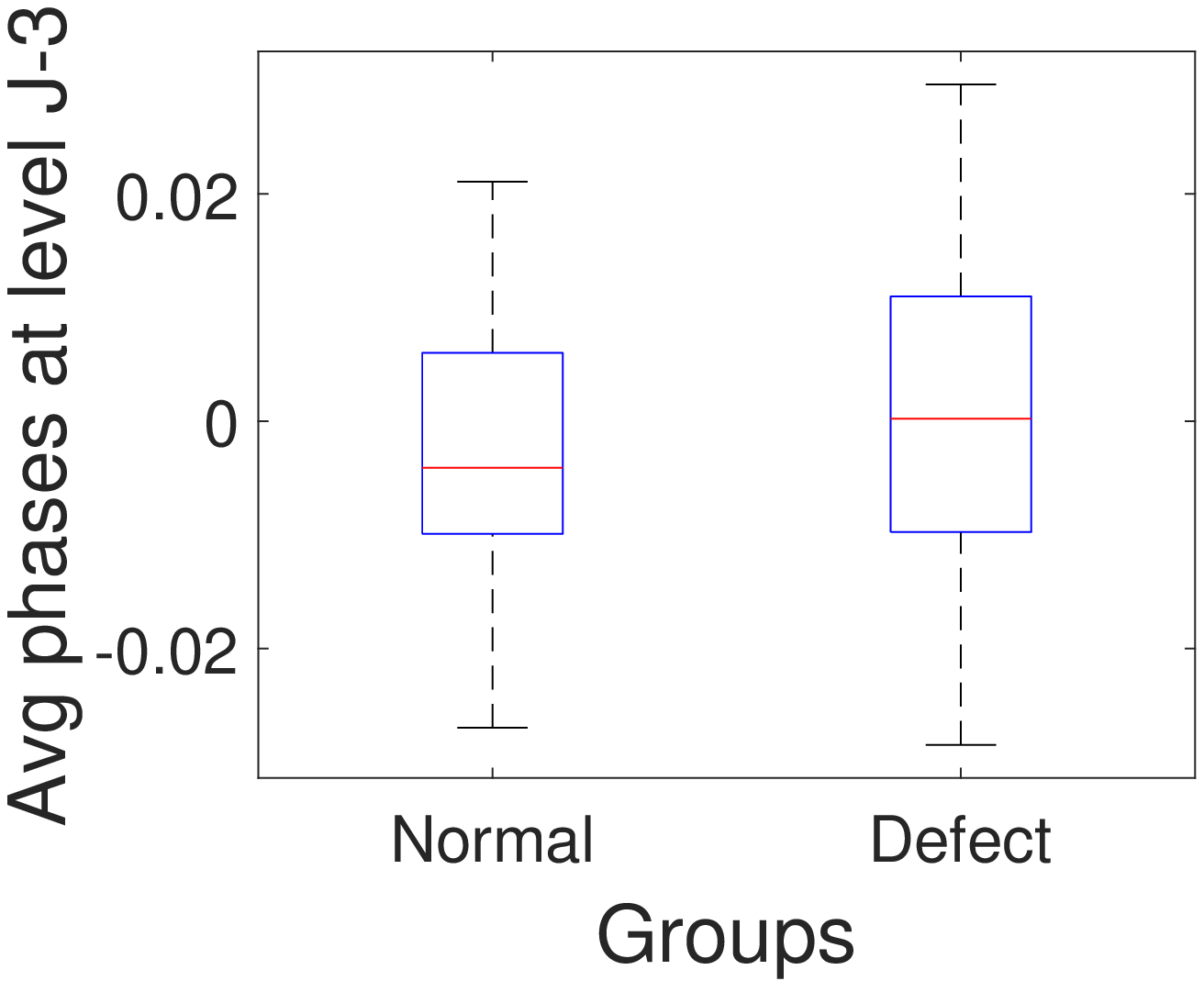}} \\
  \subfigure[]{\includegraphics[width=1.5in, height=1.5in]{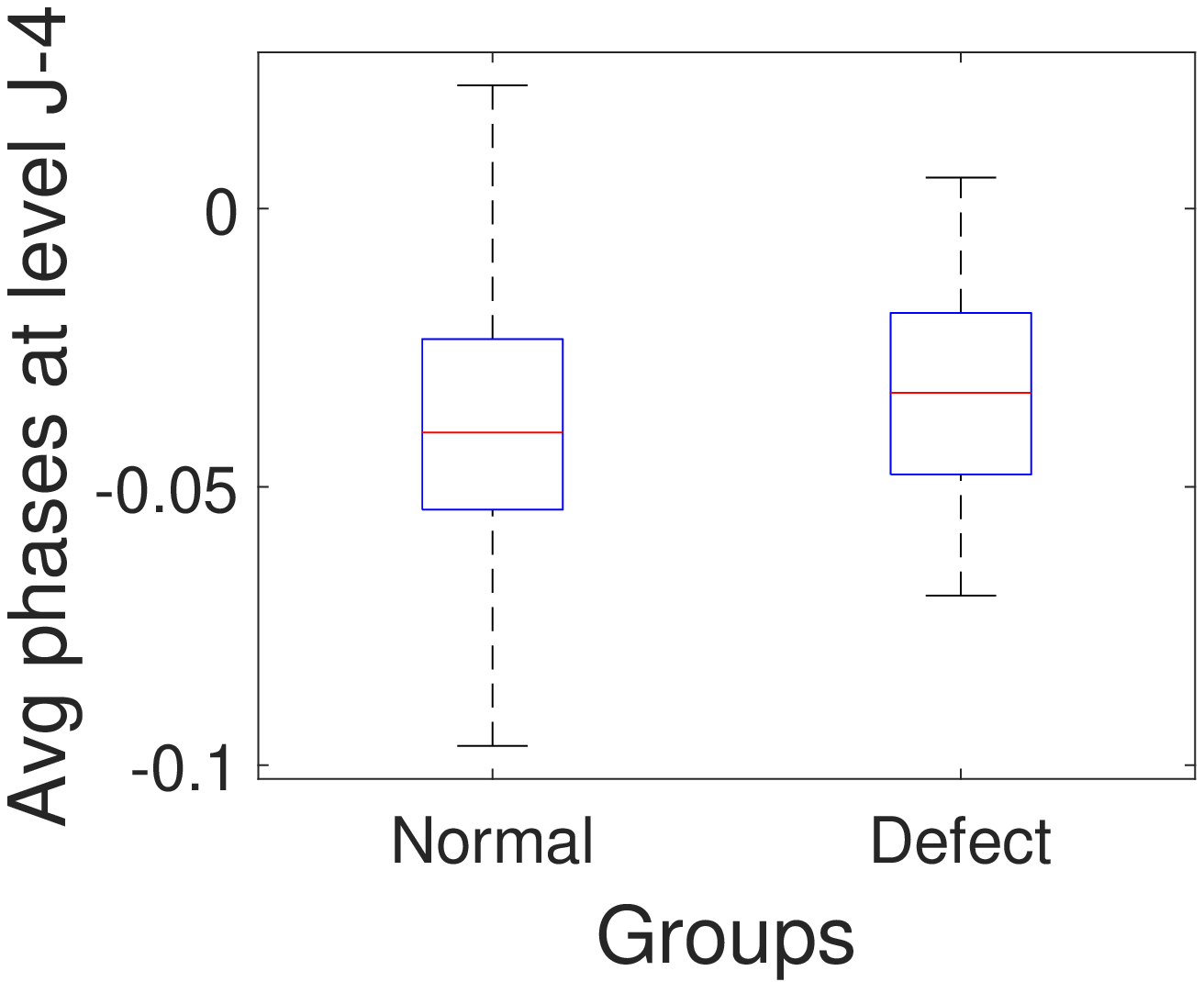}} \qquad
  \subfigure[]{\includegraphics[width=1.5in, height=1.5in]{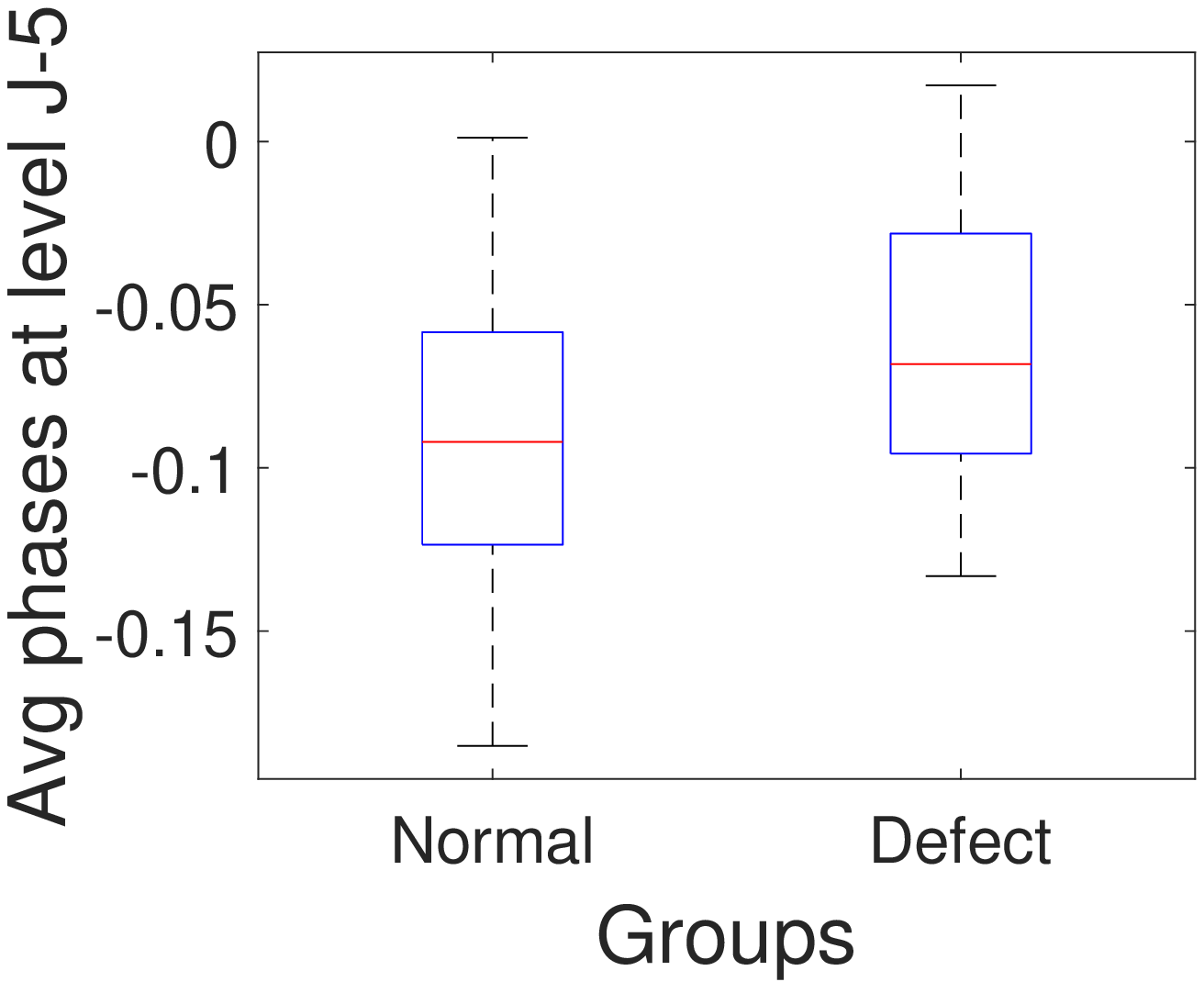}}  \qquad
  \subfigure[]{\includegraphics[width=1.5in, height=1.5in]{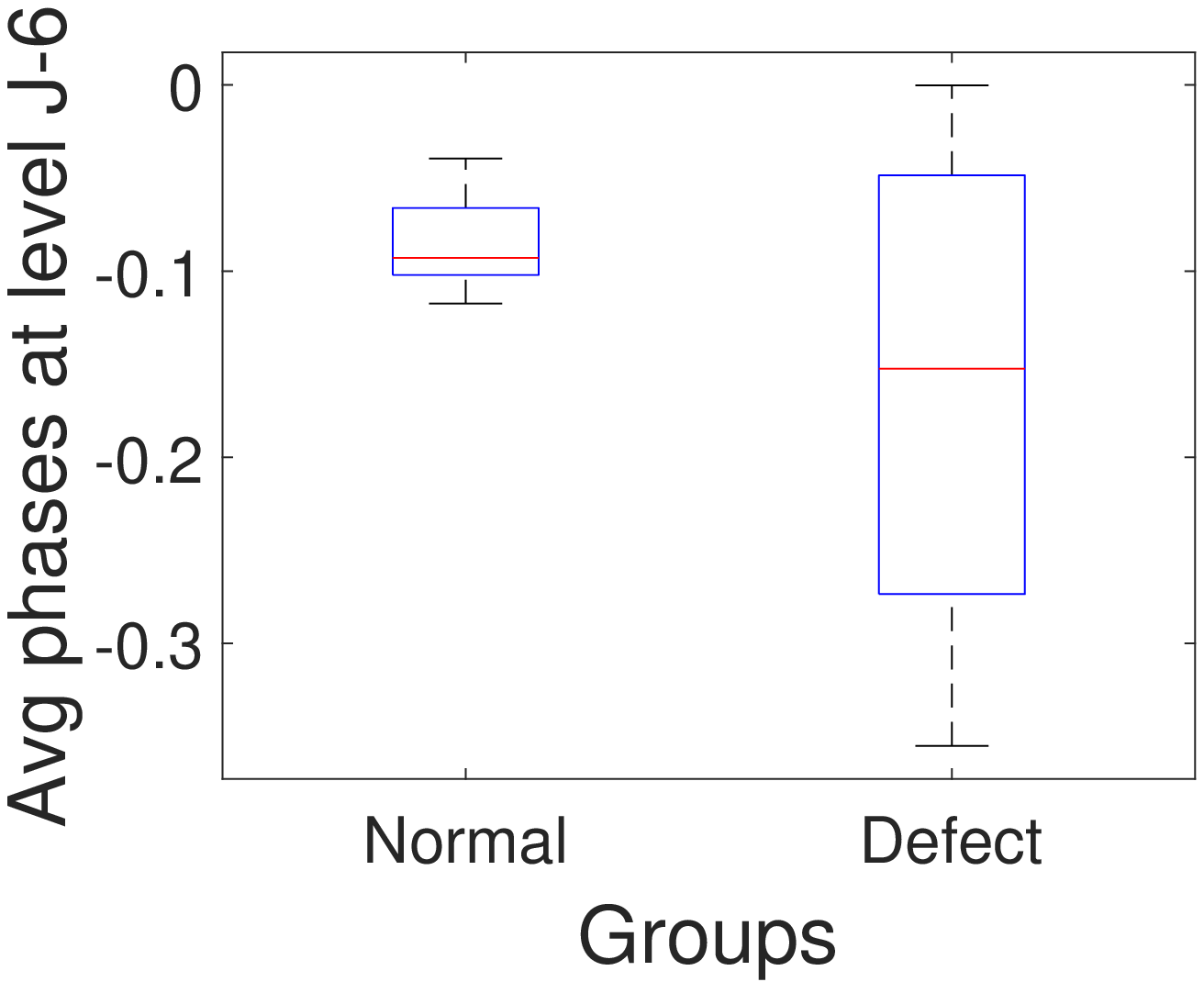}} \\

  \caption{Box plots of averages of phase $\phi$ at all multiresolution levels.}
  \label{boxfig:rollingphase1NDQ}
\end{figure}

\begin{figure}[h!tb]
  \centering
  \subfigure[]{\includegraphics[width=1.5in, height=1.5in]{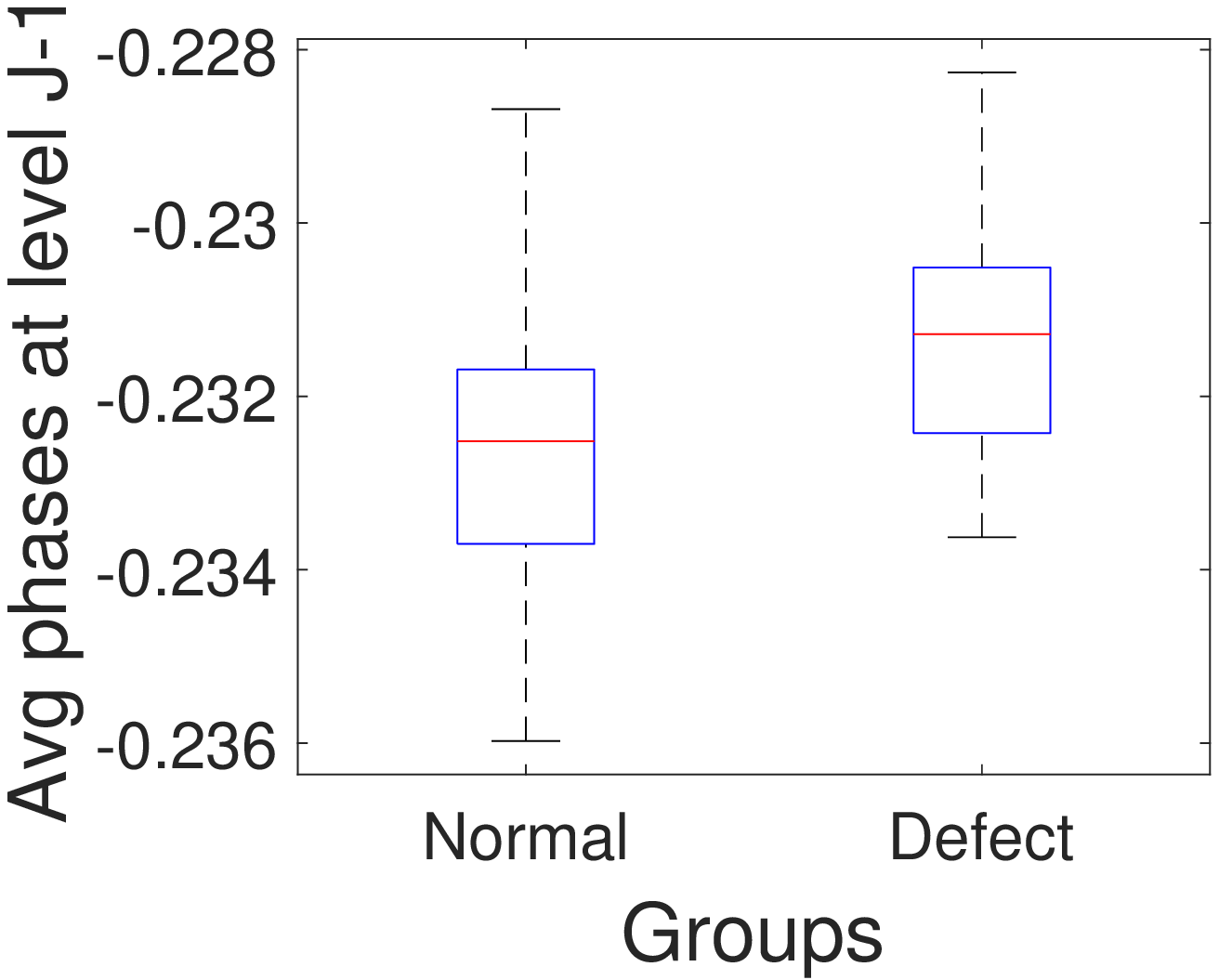}} \qquad
  \subfigure[]{\includegraphics[width=1.5in, height=1.5in]{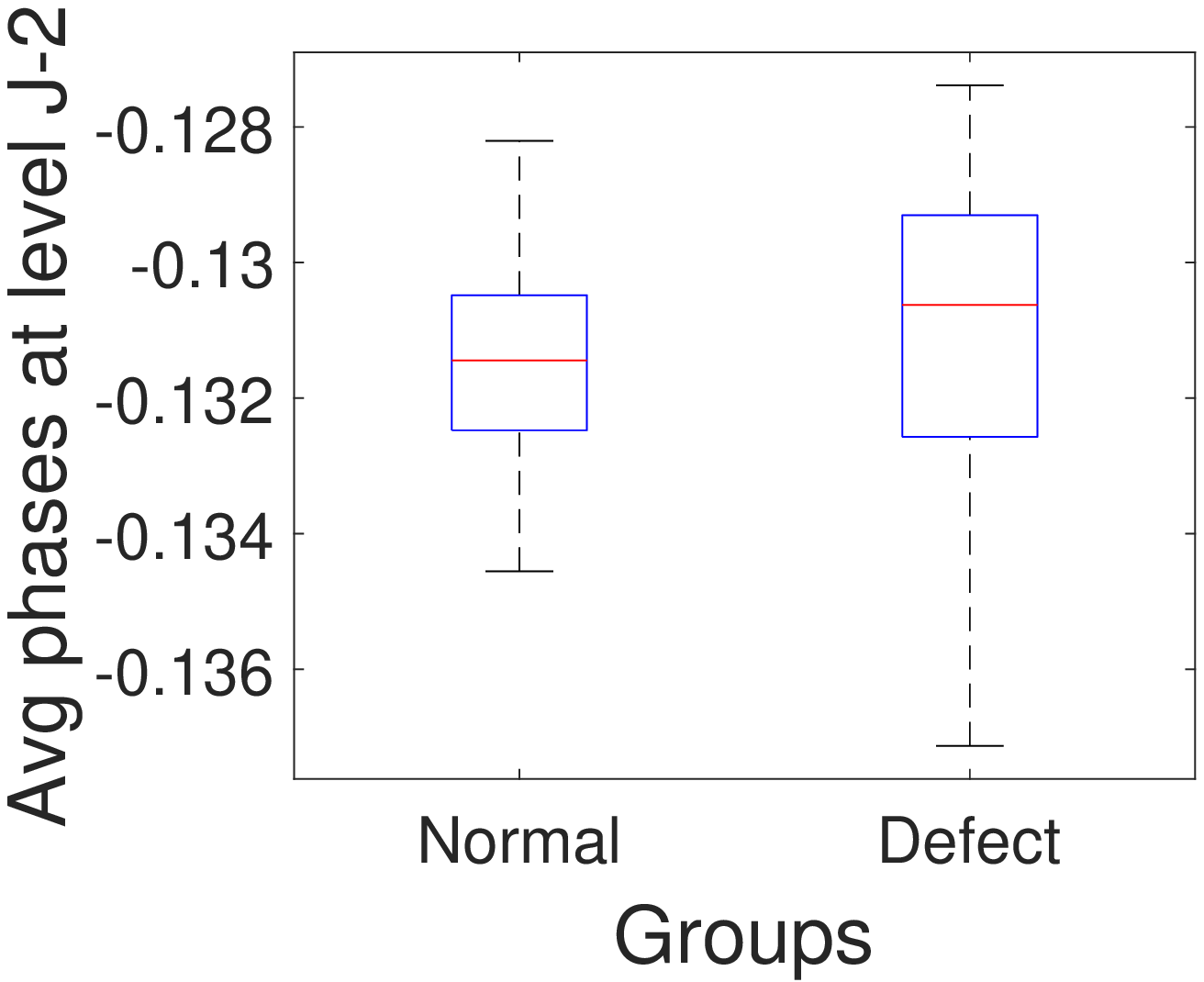}}  \qquad
  \subfigure[]{\includegraphics[width=1.5in, height=1.5in]{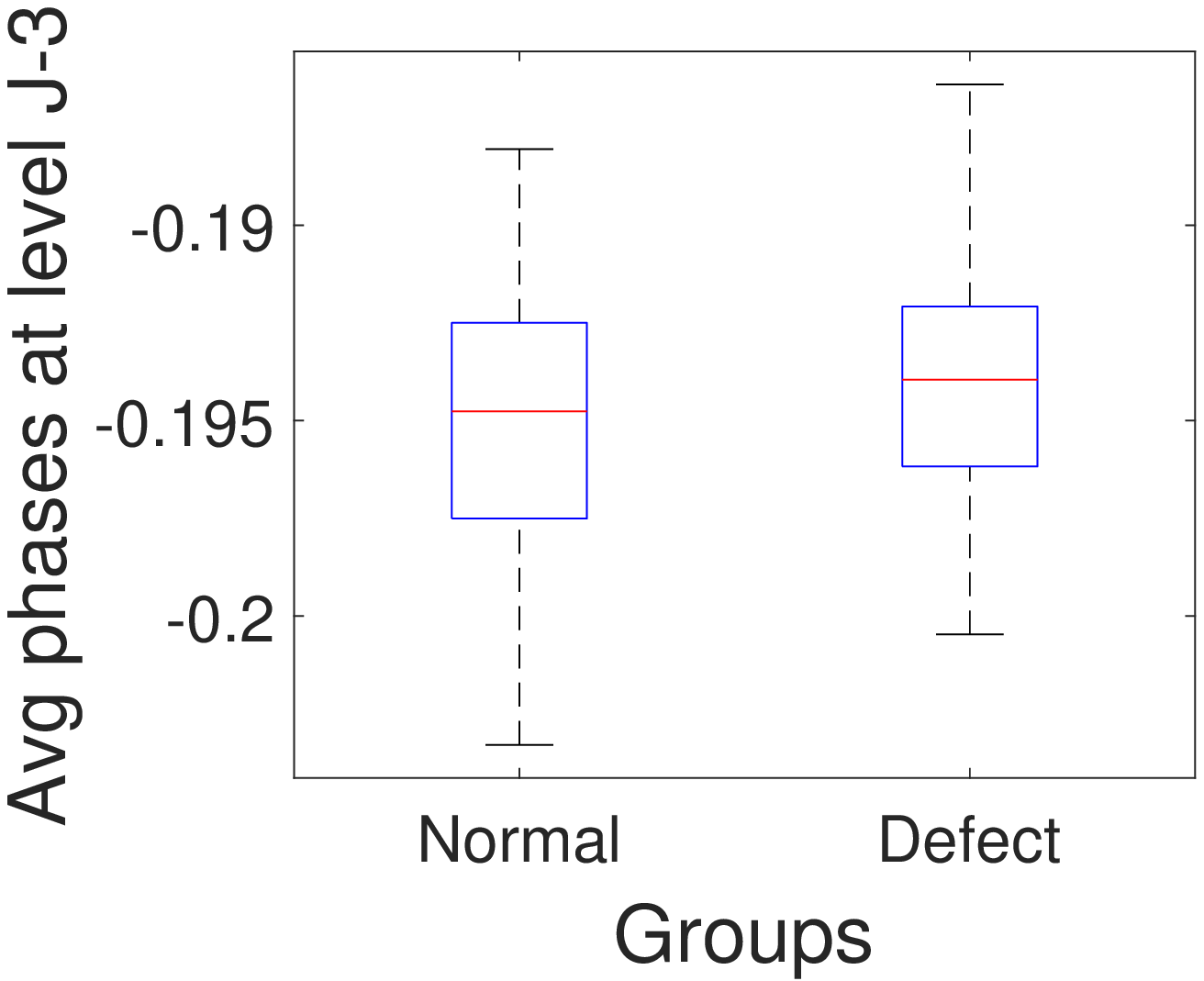}} \\
  \subfigure[]{\includegraphics[width=1.5in, height=1.5in]{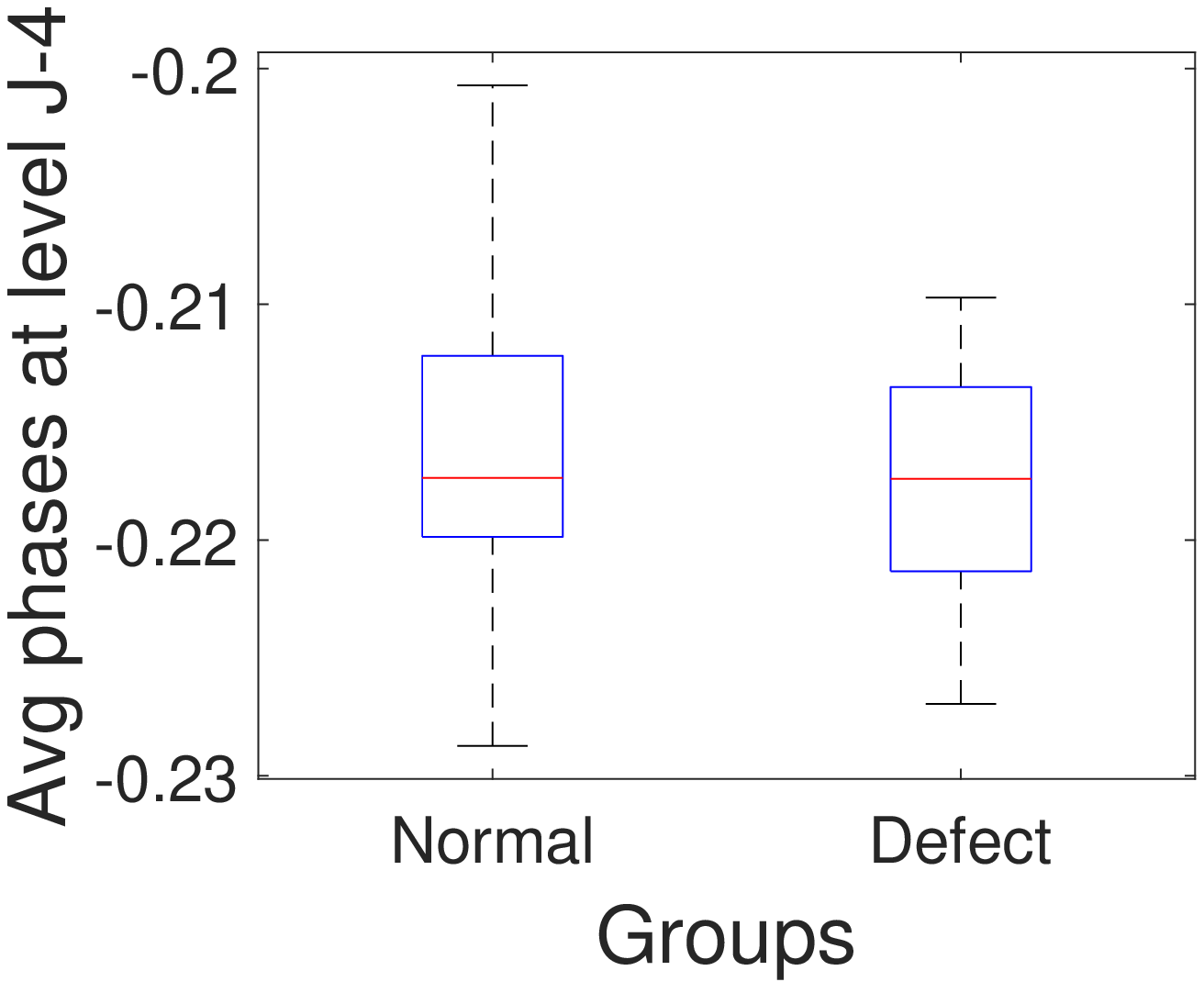}} \qquad
  \subfigure[]{\includegraphics[width=1.5in, height=1.5in]{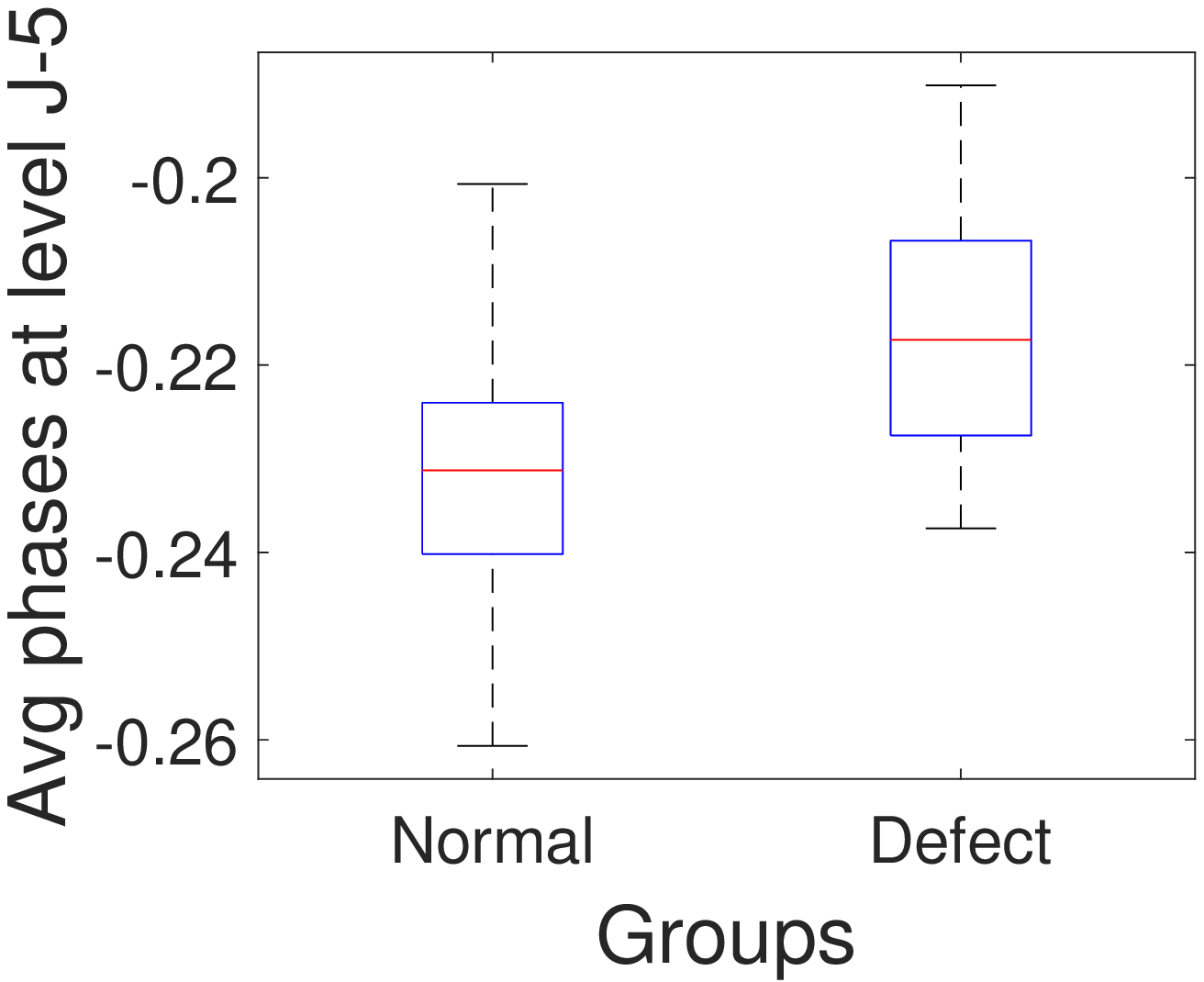}}  \qquad
  \subfigure[]{\includegraphics[width=1.5in, height=1.5in]{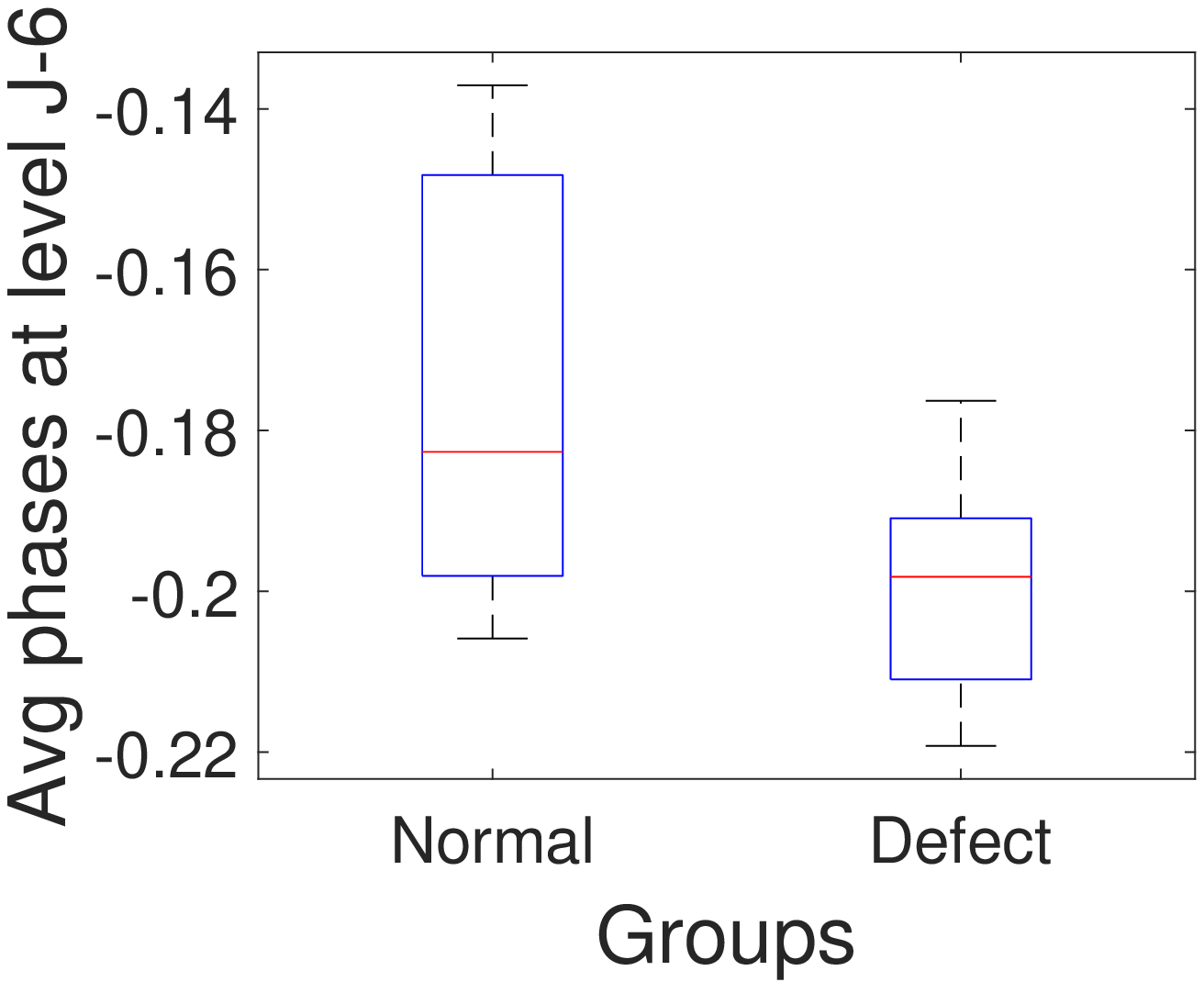}} \\

  \caption{Box plots of averages of phase $\theta$ at all multiresolution levels.}
  \label{boxfig:rollingphase2NDQ}
\end{figure}

\begin{figure}[h!tb]
  \centering
  \subfigure[]{\includegraphics[width=1.5in, height=1.5in]{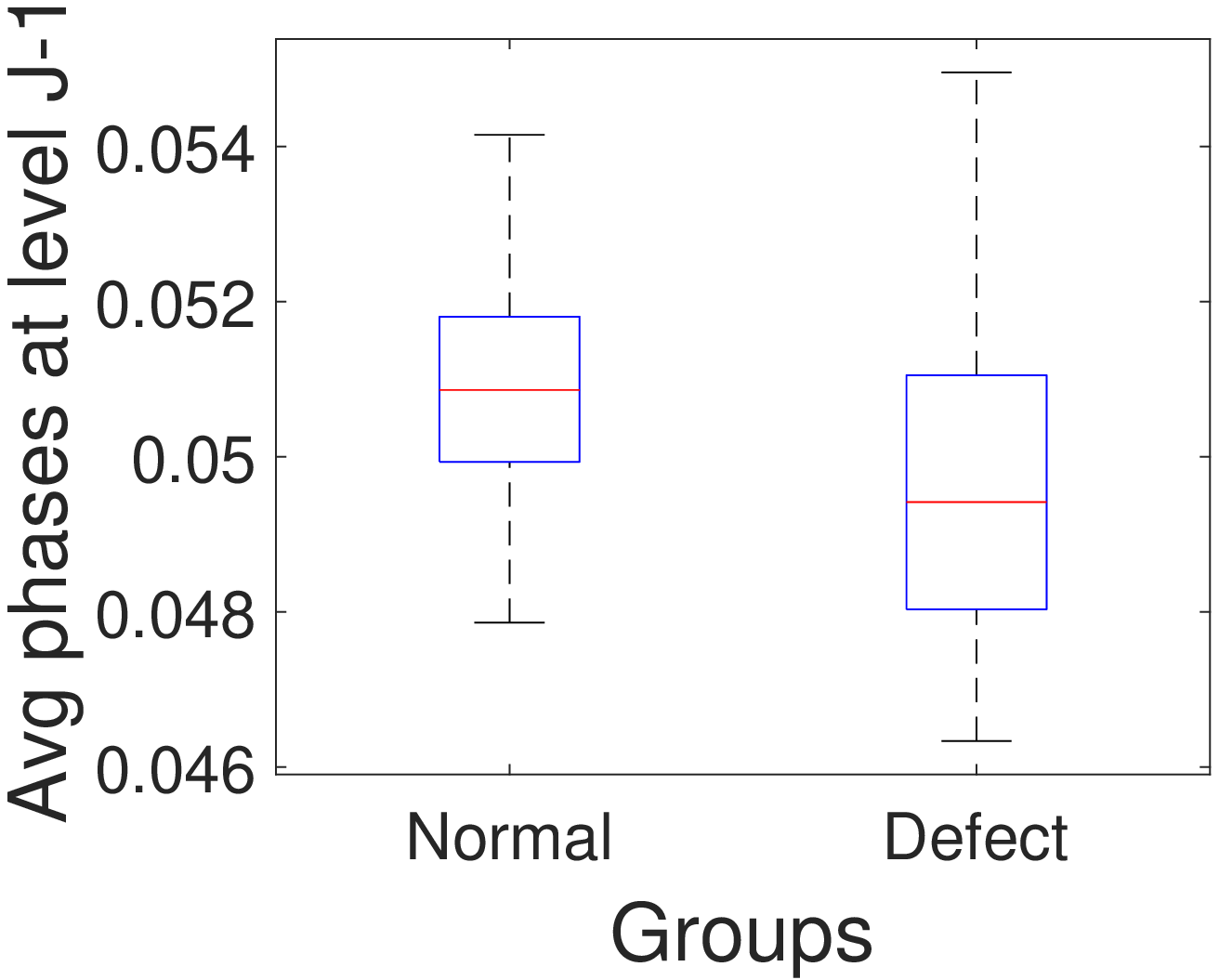}} \qquad
  \subfigure[]{\includegraphics[width=1.5in, height=1.5in]{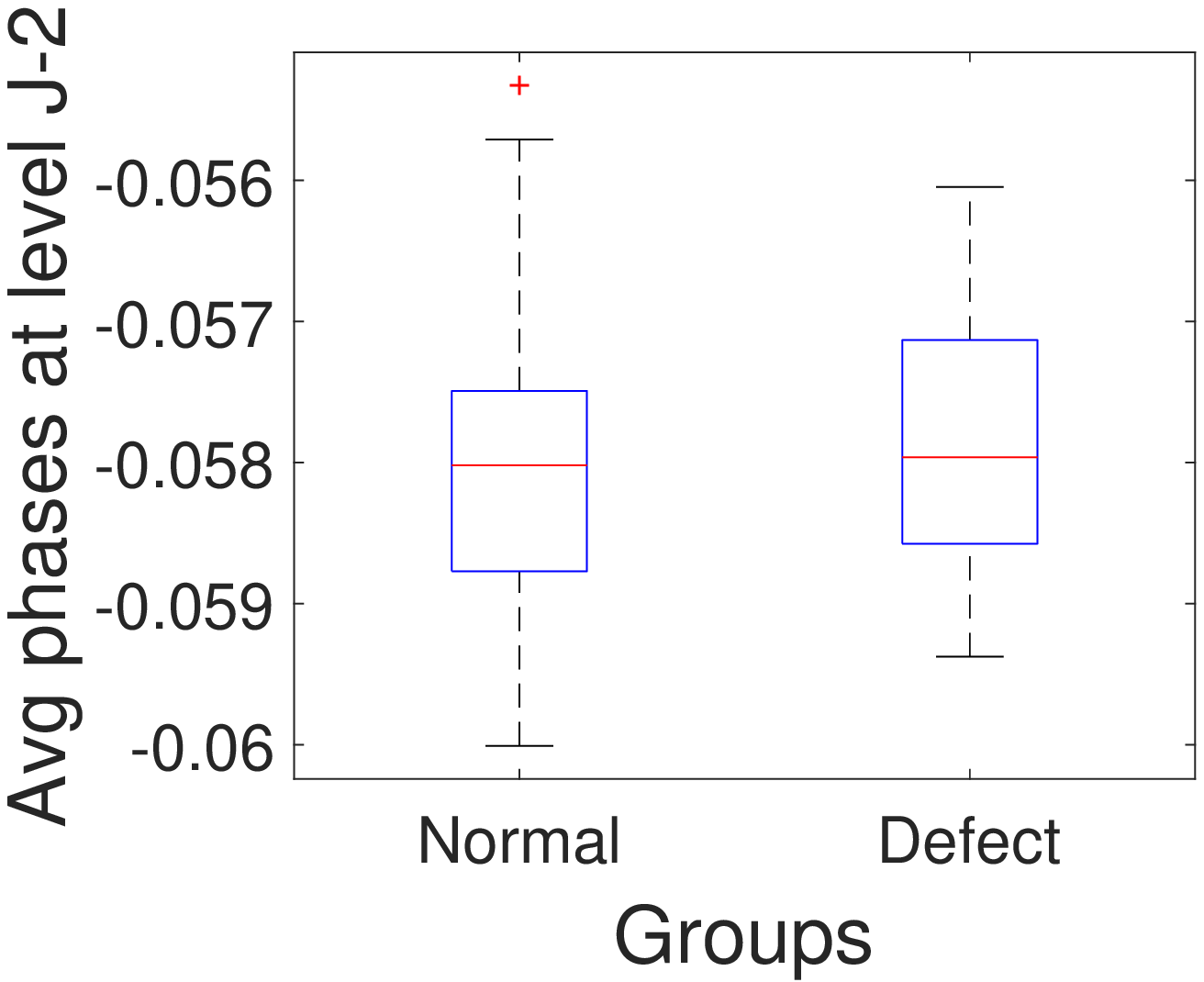}}  \qquad
  \subfigure[]{\includegraphics[width=1.5in, height=1.5in]{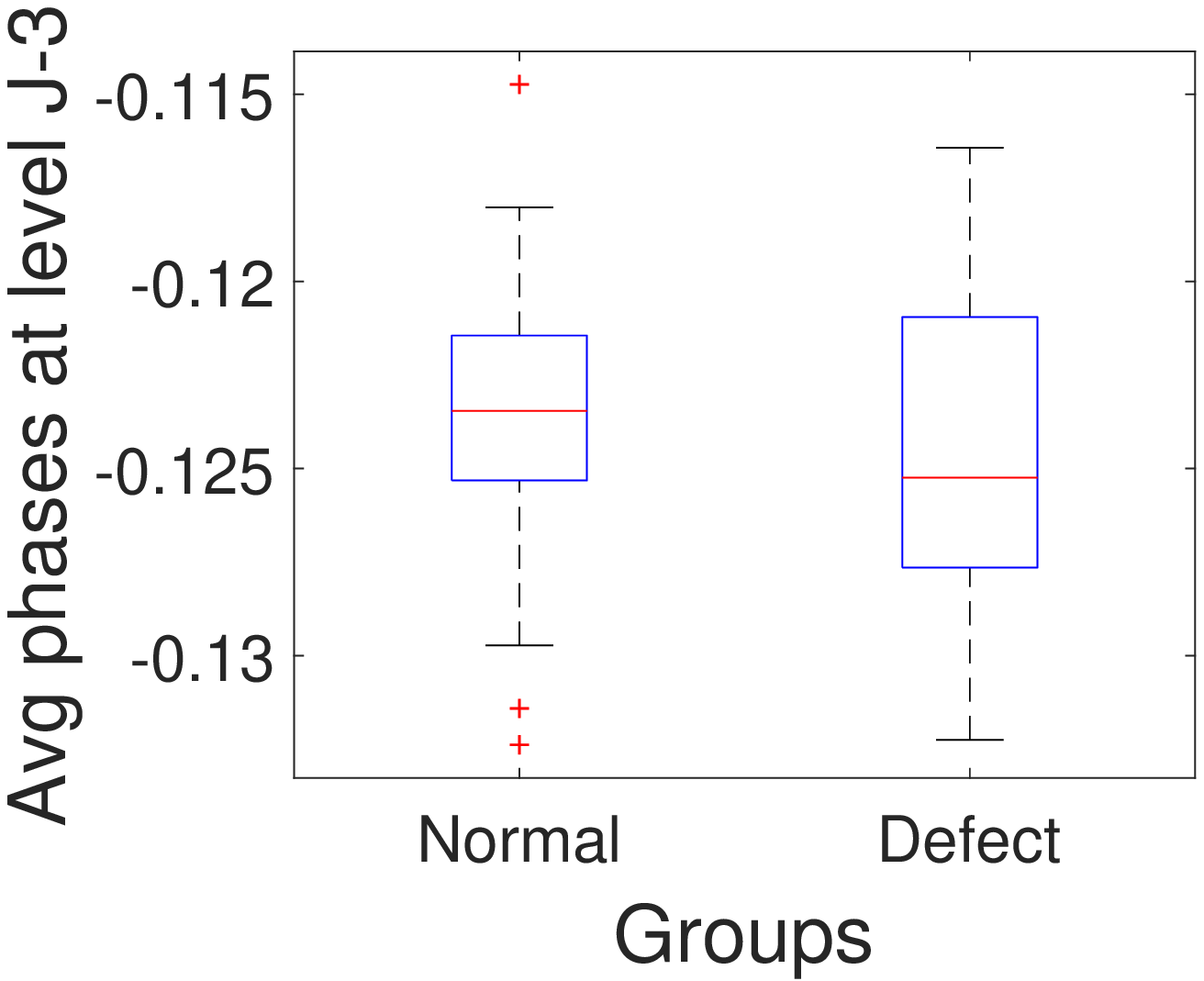}} \\
  \subfigure[]{\includegraphics[width=1.5in, height=1.5in]{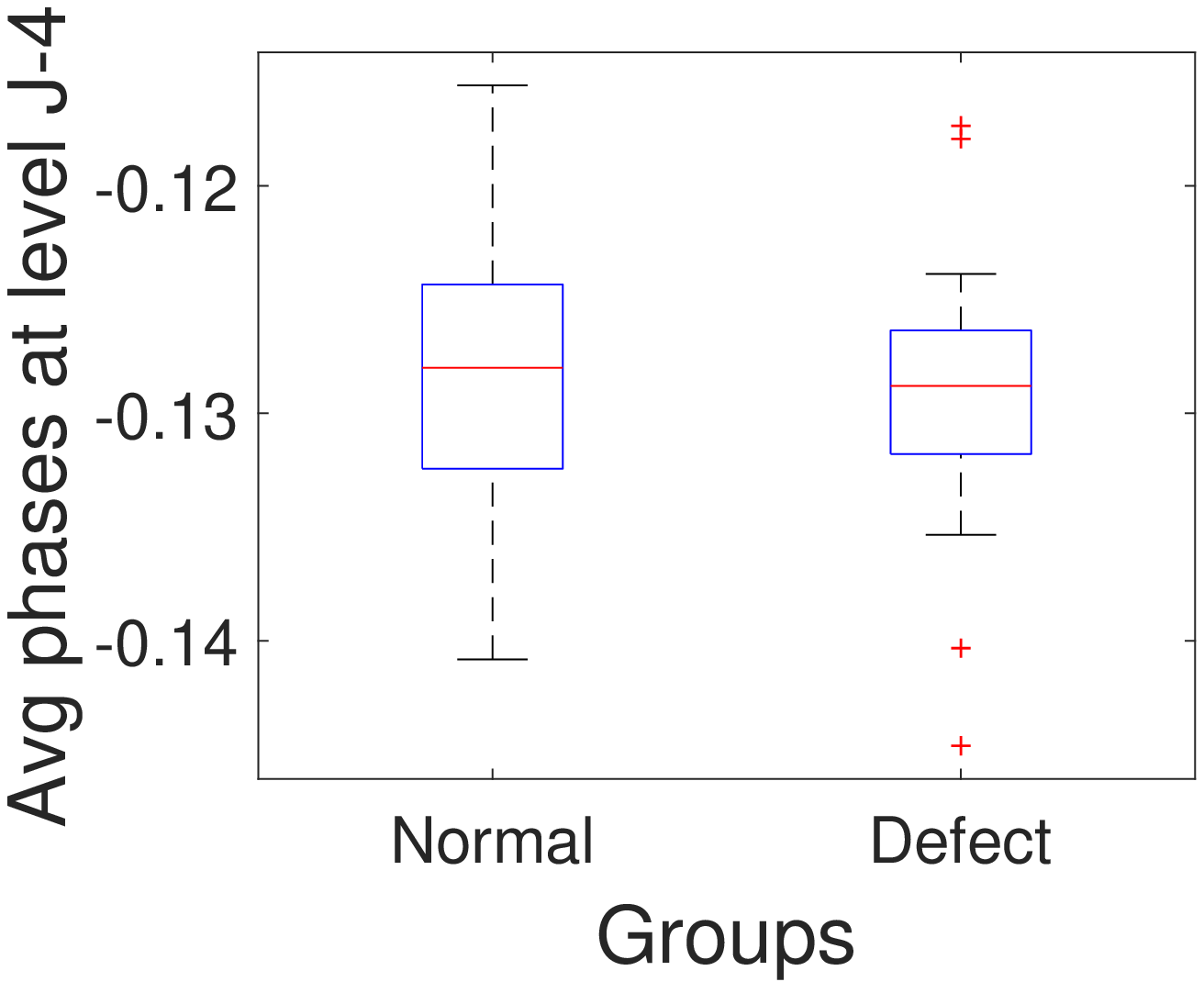}} \qquad
  \subfigure[]{\includegraphics[width=1.5in, height=1.5in]{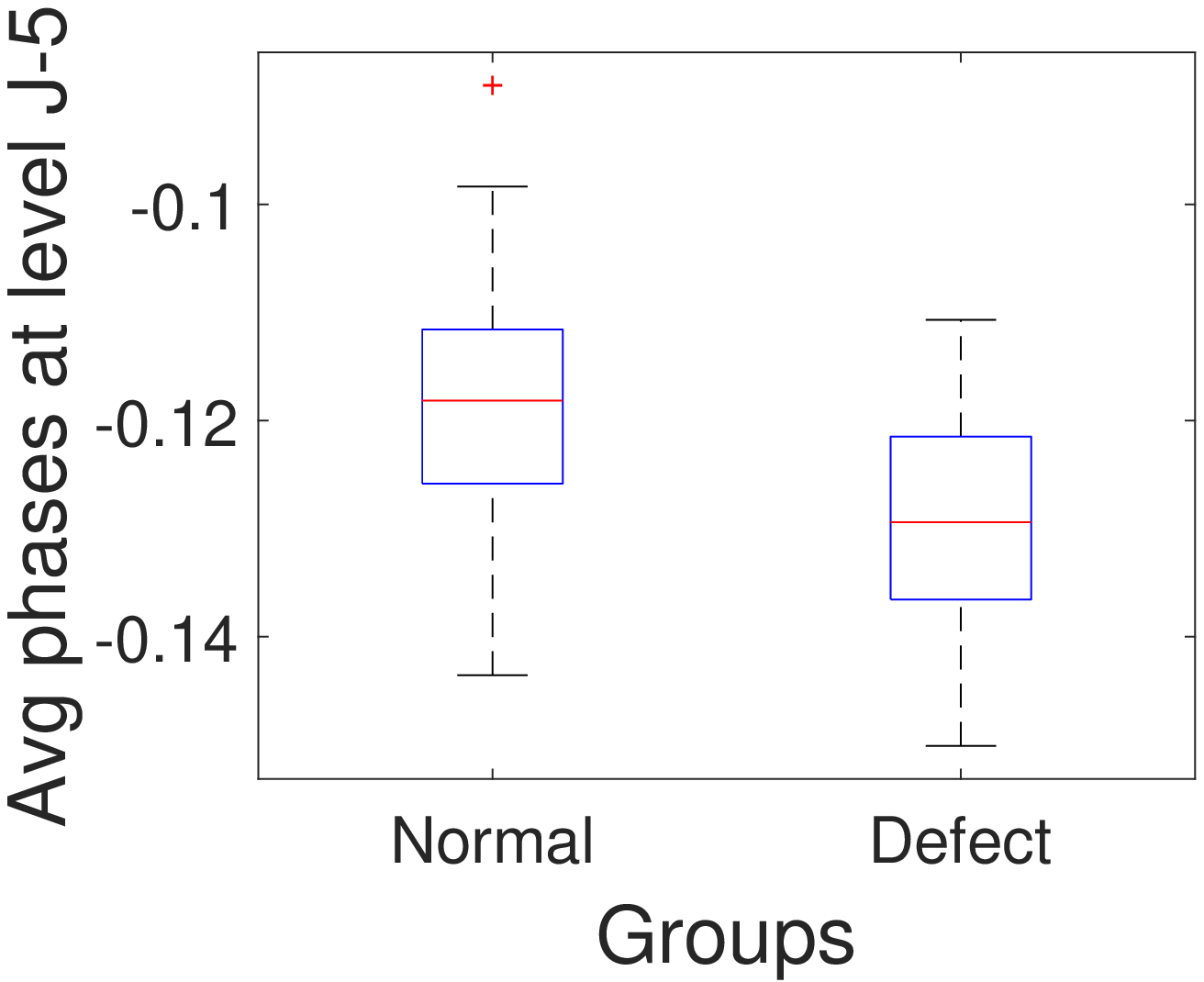}}  \qquad
  \subfigure[]{\includegraphics[width=1.5in, height=1.5in]{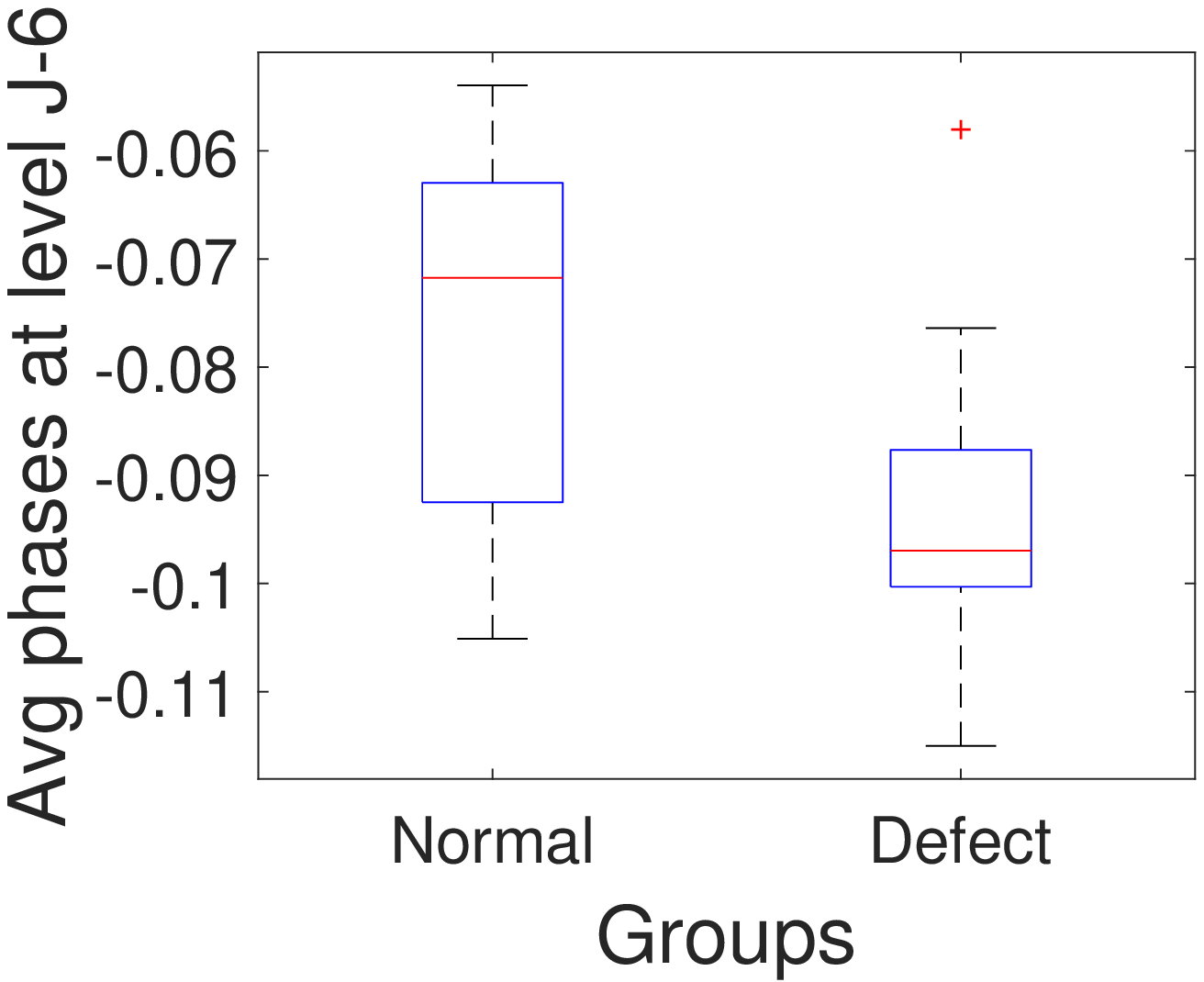}} \\

  \caption{Box plots of averages of phase $\psi$ at all multiresolution levels.}
  \label{boxfig:rollingphase3NDQ}
\end{figure}

{\label{sec-classification3}}
In this section, we describe a way of classifying the rolling surface images focusing on the NDQWT.
We implemented the proposed scale-mixing 2-D NDQWT with $s=0$ as well as DWT, NDWT, $\text{WT}_\text{\large{c}}$, $\text{NDWT}_\text{\large{c}}$, and QWT for comparisons.
As in the previous application in sound signals, we obtained a spectral slope as in Section \ref{sec-scNDQwavespec} and three phase averages at all level $j=J_0, \dots, J-1$ defined in Equation (\ref{avgphase}) as descriptors used in classification analysis.
Box plots of three phase averages ($\phi, \theta, \psi$) at all decomposition levels are displayed in Figure \ref{boxfig:rollingphase1NDQ}, \ref{boxfig:rollingphase2NDQ}, and \ref{boxfig:rollingphase3NDQ}.

Finally, we decided to use random forests methodology to classify the rolling surface images. 
In addition, we considered logistic regression, k-NN, SVM, and gradient boosting, however, the random forests consistently outperformed the rest. 
In simulations, since the dataset is imbalanced and of a relatively small size, we separately selected 75\% for training and 25\% for testing for both control and case samples to repeatedly measure performances 1,000 times. Finally, the prediction measures were obtained by averaging the repeated 1,000 runs.

\subsubsection{Results}{\label{sec-result3}}

We compared classification performances in the context of sensitivity, specificity, and overall accuracy rate, which are shown in Table \ref{rollingtable}. Haar filter is chosen for DWT and NDWT.  
We denoted the phase average from $\text{NDWT}_\text{\large{c}}$ as $\angle d_{j}$ and the three phase averages from NDQWT as $\phi_{j}, \theta_{j}, \psi_{j}$.

\begin{table}[h!tb]
\begin{center}
\begin{adjustbox}{max width=\textwidth}

\begin{tabular}{c|c|c|ccc|c}
  \specialrule{1.3pt}{1pt}{1pt}
   Order & Transform & Features & Accuracy rate & Specificity & Sensitivity & Computing Time \\\hline \hline
   $1$st & DWT & Slope  &  0.6057&	0.7372	&0.2110  & 0.0222 \\\hline
  $2$nd & $\text{WT}_\text{\large{c}}$ & Slope  & 0.7449	&0.8324&	0.4822   &\\
   $3$rd & & $\angle d_{j}$ &  0.6156&	0.7127&	0.3240  & 0.0937 \\
   $4$th & &  Slope + $\angle d_{j}$ & 0.7811	&0.9050&	0.4094 & \\\hline
  $5$th & QWT & Slope  &  0.7222&	0.8254&	0.4124 &\\
  $6$th & & $\phi_{j}$ + $\theta_{j}$ + $\psi_{j}$  & 0.8359&	0.9749	&0.4188 & 0.4206 \\
  $7$th & & Slope + $\phi_{j}$ + $\theta_{j}$ + $\psi_{j}$ & 0.8447	&0.9823&	0.4320  &\\ \hline
  $8$th & NDWT & Slope  &  0.6594	&0.7569	&0.3666  &0.0696 \\\hline
  $9$th & $\text{NDWT}_\text{\large{c}}$ & Slope  &  0.7844&	0.8755	&0.5108 &\\
  $10$th & & $\angle d_{j}$  &0.8600	&0.9664	&0.5408 & 0.2538 \\
  $11$th & & Slope + $\angle d_{j}$ &  0.8938	&0.9717	&0.6598  &\\\hline
  $12$th & NDQWT & Slope  &  0.7483	&0.8227&	0.5250 &\\
  $13$th & & $\phi_{j}$ + $\theta_{j}$ + $\psi_{j}$  &0.9139&	\textbf{0.9868}	&0.6950   &1.1777 \\
  $14$th & & Slope + $\phi_{j}$ + $\theta_{j}$ + $\psi_{j}$ &  0.9220&	0.9831	&0.7384  &\\ \hline
  $15$th &  & $11$th + $14$th & \textbf{0.9310}	&0.9819&	\textbf{0.7529}  & 1.4315\\
  \specialrule{1.3pt}{1pt}{1pt}
\end{tabular}

\end{adjustbox}
\end{center}
\caption{Random forest classification results. Total 15 methods are compared and the best result is achieved by the $15$th method.}\label{rollingtable}
\end{table}

First, we enumerated the methods from $1$ to $15$ depending on the transform and features used.
The methods show almost same trend comparing to the counterparts in Section \ref{sec-soundresult}; the only difference is that in this case the slopes are also informative features.
Contrasting $7$th and $8$th to $14$th, we see that the NDQWT dominates both QWT and NDWT.
In particular, the sensitivity of NDQWT ($14$th) significantly increased compared to that of QWT ($7$th), which means that redundancy is beneficial in capturing information on defects.
It is also notable that the three phase averages as descriptors outperform the slope, when comparing $12$th and $13$th cases.
In conclusion, we find that the best performance is achieved by $15$th method which is based on all descriptors from $\text{NDWT}_\text{\large{c}}$ and NDQWT.
This indicates that the performance can be better if we apply the $\text{NDWT}_\text{\large{c}}$ and NDQWT together in one integrated model.

As in the 1-D application, we recorded computation times for all considered versions of wavelets (DWT, $\text{WT}_\text{\large{c}}$, QWT, NDWT, $\text{NDWT}_\text{\large{c}}$, and NDQWT) needed to transform a single $128 \times 512$ image.
The computing times also increase with the increase of the overall accuracies, as in with 1-D case, however, the rate of increase is much steeper. This is because 2-D wavelet transform requires double matrix multiplications, as explained in section \ref{sec-NDQWT}.
Although the times rapidly increase, they are still in the reasonable range; the NDQWT takes approximately 1.17 seconds per image.

As a final comparison, we applied CNN (Convolutional Neural Network) which is the  state-of-the-art image analyzing tool nowadays.
The goal of this additional experiment is to compare CNN with the proposed method in terms of accuracy and computing time.
Tensorflow 1.5.0 in Python 3.5.2 is used for CNN with 5 layers, 0.001 learning rate, 70 batch size, and 100 training epochs and MATLAB 9.1.0 is for NDQWT on Intel(R) Core(TM) i7-6500U CPU at 2.50GHz with 12GM RAM.
Surprisingly, we found that computing times are notably different.
For NDQWT, its time for extracting features was 1.76 mins and then 1000 iteration of training and testing took about 1.26 mins.
Thus, total processing time was approximately 3.02 mins.
In comparison, CNN showed 0.8987 average accuracy for 100 iterations with 0.9139 specificity and 0.8457 sensitivity, which seems to be competitive with NDQWT.
Surprisingly, the CNN took 22 mins 48 secs on average for its one-time training and testing.
For dataset of large size of training data, CNN does not need multiple training because a large size of testing data would be available as well.
However, due to a small size of dataset, multiple training and validation runs are desired here, which will take whooping $23 \times 1000$ mins for 1000 iterations.
This is the reason why the proposed method should be favored to the CNN in terms of both accuracy and computing time in this application.

\section{Conclusions and Future Studies}\label{sec-Conc}
In this paper, we suggested a non-decimated quaternion wavelet transform (NDQWT) for both 1-D and 2-D cases.
We demonstrated that the proposed wavelet spectra works well in classification problems with standard spectrum based on the magnitudes is enhanced by the three quaternionic phase-based statistics.
Through comparative investigation in two real-life applications, we found that the classification procedure by NDQWT outperforms the QWT and NDWT.
This indicated that combination of two different redundancies, structural one from NDWT and componential one from QWT benefited the performance.
The NDQWT can appeal to researchers seeking more efficient wavelet-based classification methodology for signals or images with intrinsic self-similarity.

There are several directions for possible future research.
The performance could be robustified if we calculate the spectral slopes in different ways as done in \citet{Hamilton2011} or \citet{Feng2018} where robust Theil-type regressions and trimean estimators have been proposed.
For the scale-mixing 2-D NDQWT, the diagonal hierarchy of coefficients $d^{(d)}$ itself is provided good discriminatory descriptors such as spectral slopes and phase-based statistics. 
By using scale mixing hierarchies in addition to $d^{(d)}$ may likely further improve the performance of classification.
Finally, performing a scale-mixing 2-D NDQWT with different wavelet filters for rows and columns of pixels enables more modeling freedom.
For instance, the left-hand side can be a matrix based on the quaternion-valued filter while the right-hand side can be based on real-valued filters such as Haar, Symmlet, Coiflet, and so on.

In the spirit of reproducible research, we prepared an illustrative demo as a stand alone MATLAB software with solved examples.
The demo is posted on the repository Jacket Wavelets \url{http://gtwavelet.bme.gatech.edu/}.

\section{Acknowledgement}\label{sec-Ack}
We are grateful to Prof. Andrew T. Walden, Imperial College London, and Dr. Paul Ginzberg  for their kind permission  to use software on quaternion wavelets they developed.
Also, we acknowledge developers of Quaternion toolbox for MATLAB \citep{qtfm} that facilitated calculation with matrices of quaternions in almost the same manner as with the matrices of complex numbers.
Finally, we thank Prof. Kamran Paynabar and Ana Maria Estrada Gomez for providing the steel rolling bar image data and useful discussions.

Part of this work was supported by NSF grant DMS-1613258 at Georgia Institute of Technology.
%Prof. Bae does not want to reveal his identity for providing the sound data.

\newpage
\vspace{20cm}

\bibliographystyle{spbasic}      % basic style, author-year citations

%\bibliographystyle{plain}
%\bibliographystyle{wileyj}
\bibliography{QuaternionND_Bib}

%\newpage
%
%\vspace{2cm}
%\begin{center}
%\Huge Appendix
%\end{center}
%\vspace{.5cm}

%\appendix
%\setcounter{section}{26}
%\begin{appendices}

\end{document}